\documentclass[12pt,thmsa, a4paper]{article}%
\usepackage{amssymb}
\usepackage{amsfonts}
\usepackage{amsmath}
\usepackage{color}
\usepackage{graphicx}
\usepackage{geometry}
\usepackage{probsoln}%
\providecommand{\U}[1]{\protect\rule{.1in}{.1in}}
\newtheorem{theorem}{Theorem}

\newtheorem{fact}[theorem]{Fact}
\newtheorem{definition}{Definition}
\newtheorem{corollary}[theorem]{Corollary}
\newtheorem{proposition}[theorem]{Proposition}
\newtheorem{lemma}[theorem]{Lemma}

\definecolor{indigo}{rgb}{0.0, 0.25, 0.42}

\newenvironment{super-boxer}
{\begin{center}
		\begin{tabular}{||p{0.9\columnwidth}||}
			\hline \hline \\
		}
		{
			\\ \\ \hline \hline
		\end{tabular}
	\end{center}
}

\newenvironment{Hidden}{\onlysolution \color{indigo}
}
{
	\endonlysolution}
\showanswers
\hideanswers
\ifx\pdfoutput\relax\let\pdfoutput=\undefined\fi
\newcount\msipdfoutput
\ifx\pdfoutput\undefined\else
\ifcase\pdfoutput\else
\ifx\paperwidth\undefined\else
\ifdim\paperheight=0pt\relax\else\pdfpageheight\paperheight\fi
\ifdim\paperwidth=0pt\relax\else\pdfpagewidth\paperwidth\fi
\fi\fi\fi
\begin{document}

\title{Algorithmic Decision Processes}
\author{C. Baldassi\quad F. Maccheroni\quad M. Marinacci\quad M. Pirazzini}
\maketitle

\begin{abstract}
We develop a full-fledged analysis of an algorithmic decision process that, in
a multialternative choice problem, produces computable choice probabilities
and expected decision times.

\end{abstract}

\section{Introduction}

\paragraph{An algorithmic decision procedure}

In a multialternative choice problem, decision units aim to find the best
alternatives within a finite choice set $A$ of available ones. Had they
unlimited resources allowing them to make an unconstrained number of exact
judgments between alternatives, they could proceed by standard revision. This
brute force comparison-and-elimination algorithm sequentially analyzes pairs
of alternatives and permanently discards the inferior ones. If the unit
preferences are complete and transitive, after $\left\vert A\right\vert -1$
comparisons the incumbent solution of this algorithm is an optimal choice.
Implicit in traditional choice theory is an underlying algorithm of this kind.

Yet, decision units' resources are typically too limited to implement a
standard revision procedure. Indeed, binary comparisons are typically costly,
time-consuming and subject to error (so inexact). This happens because the
decision unit may only imperfectly know the relative desirability of the
competing alternatives. As a result, deliberation over them consumes resources
(economic or physiological), requires time and is subject to
error.\footnote{Here we are abstracting from inescapable non-decision times.}
In choice episodes involving the same pair of alternatives, different decision
times and choices may be then observed. Binary choice behavior can be
described only in probabilistic terms, with stochastic decision times and
choice probabilities.

The well-known limits of working memory suggest a sequential structure for the
multialternative choice problem. Costly, time-consuming and inexact binary
comparisons unfold one after the other through a stochastic exploration
process, with alternatives playing the roles of proposals and incumbents in
each binary comparison. An iterative choice process then operates. The
decision unit limited resources constrain this process by limiting the number
of executable binary comparisons, with a (possibly random) cap on the number
of affordable iterations. When the process is terminated, an alternative is
selected. The inexact nature of comparisons and the stochasticity of the
exploration process make this selection random.

We formalize this schema through a decision procedure, the Neural Metropolis
Algorithm, that parsimoniously adapt standard revision by building on
sequential binary comparisons between proposals and incumbents, explicitly
modelled as time-consuming and subject to errors (so stochastic), that unfold
through a Markovian exploration of the choice set $A$. A stopping number,
determined by the decision unit resources, terminates the iterations of the
algorithm and thus makes the algorithm select an alternative. Different
iterations of the algorithm may result in different selections of alternatives
because of the stochastic nature of binary comparisons and of Markovian exploration.

To the best of our knowledge, this is the first full-fledged analysis of an
algorithmic decision process over a choice set. We are able to derive both the
choice probabilities and expected decision times that the Neural Metropolis
Algorithm generates, with closed forms in some noteworthy cases (Section
\ref{sect:nma}). We are also able to provide a value representation for the
algorithm, which proceeds as if governed by some underlying utility judgements
over alternatives (Section \ref{sect:rev}). In so doing, we generalize and
extend to pairs of choice probabilities \ and decision times some basic
results of traditional stochastic choice.

This value foundation makes it possible to interpret our algorithmic decision
unit as the neural system of a decision maker that confronts a decision
problem. In particular, traditional choice analysis can be implemented by the
Neural Metropolis Algorithm when the resource constraint is relaxed, as it is
the case in the traditional choice analysis sketched at the outset. On the
other hand, our algorithm may incorporate neuroscience binary choice models,
like the Drift Diffusion Model (DDM), which thus get embedded in a sequential
multialternative decision process.

\paragraph{Outline of the analysis}

We begin the analysis by generalizing traditional stochastic choice by
introducing binary choice probabilities $\rho\left(  i\mid j\right)  $ to
model binary comparisons that, in a sequential setting, involve alternatives
$i$ and $j$ playing the distinct roles of proposals and incumbents,
respectively (a distinction that stochastic choice does not make). In our
first main result, Theorem \ref{thm:value}, we show that binary choice
probabilities have a value representation through a Fechnerian scaling
$v:A\rightarrow\mathbb{R}$ when they satisfy a basic transitivity property,
thus extending to our setting classic results of traditional deterministic
choice theory and of traditional stochastic choice theory. Indeed, our
analysis includes as special cases both deterministic traditional choice,
where $\rho$ is $0$-$1$ valued, and traditional stochastic choice, where
$\rho$ is strictly positive.

Besides choice probabilities, the other key element of the analysis are the
expected decision times $\tau\left(  i\mid j\right)  $ that account for the
average duration of comparisons between proposal $i$ and incumbent $j$. We
introduce them formally and consider pairs $\left(  \rho,\tau\right)  $ to
study their interplay with binary choice probabilities. We propose a value
representation also for these pairs. Such a representation is behaviorally
characterized in our Theorem \ref{prop:chrono}. Theorem \ref{prop:psycho}
captures the special case of symmetric expected decision times which result
from a classical speed/accuracy relation: faster decisons corresponding to
smaller error rates.

With this, we move to the analysis of the Neural Metropolis Algorithm. It
sequentially compares pairs of alternatives, playing the roles of incumbents
and proposals. These comparisons use a binary choice model $\left(
\mathrm{C},\mathrm{RT}\right)  $ consisting of choice variables and response
times that determine the frequency $\rho_{\mathrm{C}}\left(  i\mid j\right)  $
with which proposal $i$ is accepted over incumbent $j$ and the mean response
time $\tau_{\mathrm{RT}}\left(  i\mid j\right)  $ required by the comparison.
To the pair $\left(  \rho_{\mathrm{C}},\tau_{\mathrm{RT}}\right)  $ we can
apply the value analysis previously developed, with a Fechnerian scaling
$v_{\mathrm{C}}:A\rightarrow\mathbb{R}$ for the stochastic binary comparisons
(featuring a positive $\rho_{\mathrm{C}}$) and a Paretian utility function
$w_{\mathrm{C}}:A\rightarrow\mathbb{R}$ for the deterministic ones (featuring,
instead, a $0$-$1$ valued $\rho$). An initial random condition $\mu$ and an
exploration matrix $Q$ complete the description of the constituent elements of
the Neural Metropolis Algorithm. A stopping time $N$ terminates the algorithm,
which thus selects an alternatives.

The algorithm thus generated a choice probability $p_{N}$ over alternatives,
where $p_{N}\left(  i,A\right)  $ is the probability that the algorithm
selects alternative $i$ from the choice set $A$, as well as a mean response
time $\tau_{N}\geq0$, the average time that the algorithm takes to select an
alternative from $A$. We obtain closed forms for both $p_{N}$ and $\tau_{N}$
in the important case of negative binomial stopping times, which includes the
geometric ones.

Our value analysis shows that, as the stopping number allows more and more
iterations, the Neural Metropolis Algorithm has noteworthy optimality
properties. It selects optimal alternatives when all binary comparisons are
deterministic, thus implementing traditional choice behavior. When, instead,
deterministic and stochastic binary comparisons coexist, the algorithm first
selects alternatives $i$ that are best across the deterministic comparisons,
so belong to $\arg\max_{A}w_{\mathrm{C}}$, and then choose over them according
to a logit a rule $\medskip$%
\[
\frac{e^{v_{\mathrm{C}}\left(  i\right)  }}{\sum_{j\in\arg\max_{A}%
w_{\mathrm{C}}}e^{v_{\mathrm{C}}\left(  j\right)  }}%
\]
where $v_{\mathrm{C}}$ and $w_{\mathrm{C}}$ are the Fechnerian scaling and
Paretian utility previously mentioned.

\paragraph{Limitations of the analysis}

In our analysis the stopping number, which accounts for the decision unit
limited resources, is exogenous. It is a convenient assumption in a first
analysis of an algorithmic decision process, but a topic for future research
is the study of a decision problem that would endogenously deliver it.

Relatedly, we take the Markovian stochastic exploration as exogenous, though
the decision unit may want to adjust it as exploration progresses. The study
of more sophisticated exploration strategies is another topic for future research.

The sequential structure of the Neural Metropolis Algorithm appears to be
natural in view of the limitations of working memory. It would interesting,
however, to understand its optimality status by making explicit the working
memory constraints that impede the parallel, rather than sequential,
consideration of all competing alternatives in the choice set.

\paragraph{Related literature}

The Neural Metropolis Algorithm has the Metropolis DDM Algorithm of Baldassi
et al. (2020) and Cerreia-Vioglio et al. (2022) as special cases in which
binary comparisons are performed according the DDM of Ratcliff (1978), as
adapted by Krajbich et al. (2010) and Milosavljevic et al. (2010) to
value-based binary choices. The generalization is significant, moving from a
specific binary comparison model to virtually all of them. Moreover, our
results are novel even when binary comparisons are DDM based.

The Neural Metropolis Algorithm differs from most neuro-computational models
of neuroscience that typically consider simultaneous evidence accumulation for
all the alternatives in the choice set $A$. See, e.g., Roe et al. (2001),
Anderson et al. (2004), McMillen and Holmes (2006), Bogacz et al. (2007),
Ditterich (2010), and Krajbich and Rangel (2011). This simultaneity
assumption, although efficient per se, is at odds with the known limits of
attention and working memory, as previously mentioned.

An important exception is Reutskaja et al. (2011), who present three two-stage
models in which subjects randomly search through the feasible set during an
initial search phase and, when this phase is concluded, select the best item
that was encountered during the search (up to some noise). This approach can
be called quasi-exhaustive search in that time pressure may terminate the
search phase before all alternatives have been evaluated and introduces an
error probability.

Although different from the models considered by Krajbich and Rangel (2011)
and Reutskaja et al. (2011), our model is consistent with some of their
experimental findings about the exploration process of choice sets and with
the conclusions of the seminal eye fixation study of Russo and Rosen (1975).

\section{Preliminaries}

\subsection{Mathematics}

\paragraph{Stochastic matrices}

A square matrix $B$ is (\emph{left}) \emph{stochastic} if the sum of the
entries of each column is $1$. Its powers are the stochastic matrices
$B^{0}=I$ and $B^{n}=BB^{n-1}$, with entry $b_{ij}^{(n)}$ for each $i,j$ in
the index set of $B$. A stochastic matrix $B$ is:

\begin{enumerate}
\item[(i)] \emph{positive} if its entries are strictly positive;

\item[(ii)] \emph{quasi-positive} if its off diagonal entries are strictly positive;

\item[(iii)] \emph{primitive} if there exists $n\geq1$ such that $B^{n}$ is positive;

\item[(iv)] \emph{irreducible} if, for each $i,j$ in its index set, there
exists $n\geq1$ such that $b_{ij}^{(n)}>0$;

\item[(v)] \emph{non-traceless} if, it has at least one strictly positive
element on its diagonal;

\item[(vi)] \emph{nice} if it is symmetric and quasi-positive;

\item[(vii)] \emph{reversible} if there exists a probability vector
$p\gg\mathbf{0}$ such that%
\begin{equation}
b_{ij}p_{j}=b_{ji}p_{i}\label{eq:balance-pre}%
\end{equation}
for each off diagonal entry $b_{ij}$.
\end{enumerate}

This terminology is standard, except (vi). Clearly, a positive matrix is
quasi-positive, a quasi-positive matrix is primitive if at least of order $3$.
An irreducible and non-traceless matrix is primitive. Given a stochastic
matrix $B$, the matrix $I-\zeta B$ is invertible when $\zeta\in\left(
-1,1\right)  $ because $\left\Vert \zeta B\right\Vert _{1}=\left\vert
\zeta\right\vert <1$. Instead, the matrix $I-B$ is not invertible because, by
Markov's Theorem, there exists a probability vector $p$ such that
\[
Bp=p
\]
Such a vector is called a \emph{stationary distribution} of $B$.

\paragraph{Stopping number}

A \emph{stopping number} (or \emph{rule}) is a $\mathbb{N}$-valued random
variable with finite mean $\mathbb{E}\left[  N\right]  $ defined on an
underlying probability space featuring a probability measure $\mathbb{P}$. Two
important functions are associated with a stopping number $N$. The
\emph{probability generating function} $f_{N}:\left[  0,1\right]
\rightarrow\mathbb{R}$ is defined by%
\[
f_{N}\left(  z\right)  =%
{\displaystyle\sum_{n=0}^{\infty}}
\mathbb{P}\left[  N=n\right]  z^{n}%
\]
while the \emph{survival generating function} $g_{N}:\left[  0,1\right]
\rightarrow\mathbb{R}$ is defined by%
\[
g_{N}\left(  z\right)  =%
{\displaystyle\sum\limits_{n=0}^{\infty}}
\mathbb{P}\left[  N>n\right]  z^{n}%
\]
These two functions are related as follows (Feller, 1968, p. 265):%
\[
g_{N}\left(  z\right)  =\dfrac{1-f_{N}\left(  z\right)  }{1-z}\qquad\forall
z\in\left[  0,1\right]
\]
under the limit convention
\begin{equation}
g_{N}\left(  1\right)  =\lim_{z\rightarrow1^{-}}\dfrac{1-f_{N}\left(
z\right)  }{1-z}=\mathbb{E}\left[  N\right]  \label{eq:limit-bis}%
\end{equation}
Probability generating functions are widely used and several formulas are
available for them (see, e.g., Johnson et al. 2005).

Let $\mathcal{B}$ be the collection of all stochastic matrices. Given a
stochastic matrix $B\in\mathcal{B}$, we denote by $f_{N}\left(  B\right)  $
and $g_{N}\left(  B\right)  $ the square matrices of the same order of $B$
defined by%
\begin{equation}
f_{N}\left(  B\right)  =%
{\displaystyle\sum_{n=0}^{\infty}}
\mathbb{P}\left[  N=n\right]  B^{n}\quad\text{and\quad}g_{N}\left(  B\right)
=%
{\displaystyle\sum\limits_{n=0}^{\infty}}
\mathbb{P}\left[  N>n\right]  B^{n} \label{eq:matrix-gen-fct}%
\end{equation}
It is easy to check that the matrix power series on the r.h.s. converges entry
by entry, and so the matrix $f_{N}\left(  B\right)  $ is well
defined.\footnote{See Section \ref{sect:more-stoch-matrices} for more
details.} As $B\in\mathcal{B}$ varies, via (\ref{eq:matrix-gen-fct}) one
defines the matrix generating function on $\mathcal{B}$, still denoted by
$f_{N}$, induced by a probability generating function $f_{N}$.\footnote{On
matrix functions see, e.g., Rinehart (1955) and Higham (2008).}

There is a natural partial order on stopping numbers: we say that stopping
number $N$ \emph{stochastically dominates} stopping number $N^{\prime}$,
written $N\geq N^{\prime}$, if
\[
\mathbb{P}\left[  N>n\right]  \geq\mathbb{P}\left[  N^{\prime}>n\right]
\qquad\forall n\geq0
\]
Intuitively, $N$ is a less tight stopping number than $N^{\prime}$. At the
limit, we say that a sequence $N_{k}$ of stopping numbers \emph{diverges},
written $N_{k}\rightarrow\infty$, if%
\[
\lim_{k\rightarrow\infty}\mathbb{P}\left(  N_{k}>n\right)  =1\qquad\forall
n\geq0
\]
This means, as easily checked, that the probability of stopping at any finite
$n$ vanishes as $k\rightarrow\infty$, that is, $\lim_{k\rightarrow\infty
}\mathbb{P}\left(  N_{k}=n\right)  =0$ for each $n\geq0$.

\subsection{Stochastic choice}

Let $A$ be a finite choice set, with at least three alternatives,\footnote{In
this paper we do not carry out comparative statics exercises across menus. For
this reason, we develop the analysis in terms of an arbitrarily fixed menu
$A$.} called \emph{menu}. Its typical elements are $i$, $j$, $h$ and $k$. We
denote by $\Delta\left(  A\right)  $ the set of all probability distributions
on $A$, viewed as $\left\vert A\right\vert $-dimensional vectors. In other
words, $\Delta\left(  A\right)  $ is the standard simplex in the Euclidean
space $\mathbb{R}^{\left\vert A\right\vert }$.

A \emph{choice probability} $p\left(  \cdot,A\right)  \in\Delta\left(
A\right)  $ assigns to each alternative $i$ the probability $p\left(
i,A\right)  $ that the decision unit chooses $i$ within $A$. Formally,
$p\left(  i,A\right)  $ is the component $i$ of the $\left\vert A\right\vert
$-dimensional vector $p\left(  \cdot,A\right)  $.

\section{Binary choice\label{sect:bin-choice}}

As discussed in the Introduction, our algorithmic decision process considers a
sequence of binary choices between an incumbent and a proposal. To model these
binary choices, in this section we generalize traditional stochastic choice to
account for the distinction between the roles of incumbents and proposals that
alternatives may play in binary choices. In the next section we will apply
this generalized framework to observed binary choice behavior.

\subsection{Binary choice probabilities}

A neural system, our decision unit,\footnote{Throughout we use the terms
\textquotedblleft decision unit\textquotedblright\ and \textquotedblleft
neural system\textquotedblright\ interchangeably.} compares two alternatives
$i$ and $j$ in a menu $A$ of alternatives through a \emph{probability kernel}
a function $\rho:A^{2}\rightarrow\left[  0,1\right]  $. For distinct $i$ and
$j$,%
\[
\rho\left(  i\mid j\right)
\]
denotes the probability with which \emph{proposal} $i$ is accepted when $j$ is
the \emph{incumbent} (or \emph{status quo}). So, $1-\rho\left(  i\mid
j\right)  $ is the probability with which the proposal is rejected and the
incumbent maintained. Next we introduce the class of kernels that we will study.

\begin{definition}
A probability kernel $\rho:A^{2}\rightarrow\left[  0,1\right]  $ is a
\emph{binary choice probability }if%
\begin{equation}
\rho\left(  i\mid j\right)  =1\Longleftrightarrow\rho\left(  j\mid i\right)
=0 \label{eq:pcb}%
\end{equation}
with the convention $\rho\left(  i\mid i\right)  =\varepsilon>0$.
\end{definition}

We thus assume throughout that when an alternative is chosen for sure over
another alternative, this happens regardless of the roles that they play. With
this, we now introduce some basic properties.

\begin{definition}
A binary choice probability $\rho$ is:

\begin{itemize}
\item \emph{(status-quo) unbiased} if
\[
\underset{\text{prob. }i\text{ if proposal}}{\underbrace{\rho\left(  i\mid
j\right)  }}=\underset{\text{prob. }i\text{ if incumbent}}{\underbrace
{1-\rho\left(  j\mid i\right)  }}%
\]
for all distinct alternatives $i\ $and $j$;

\item \emph{positive} if $\rho\left(  i\mid j\right)  >0$ for all distinct
alternatives $i\ $and $j$;

\item \emph{Dirac} if $\rho\left(  i\mid j\right)  \in\left\{  0,1\right\}  $
for all distinct alternatives $i\ $and $j$.
\end{itemize}
\end{definition}

These properties have a simple interpretation: a binary choice probability is
unbiased when gives the incumbent alternative no special status, it is
positive when selects either alternative with strictly positive probability,
and it is Dirac when selects either alternative deterministically.\footnote{If
a binary choice probability is positive, then for all distinct $i$ and $j$,
$\rho\left(  i\mid j\right)  >0$ and $\rho\left(  j\mid i\right)  >0$, then
neither of them can be $1$ (otherwise the other would be $0$). Thus we have
$0<\rho\left(  i\mid j\right)  <1$ for all $i$ and \thinspace$j$.}

Traditional stochastic choice usually considers unbiased (and often positive)
binary choice probabilities, where $\rho\left(  i\mid j\right)  =p\left(
i,\left\{  i,j\right\}  \right)  $ describes the probability of choosing $i$
from the doubleton $\left\{  i,j\right\}  $. General, possibly biased, binary
choice probabilities account for the incumbent and proposal distinct roles
that are peculiar to a sequential analysis.

\begin{definition}
A binary choice probability $\rho$ is \emph{transitive} if
\begin{equation}
\rho\left(  j\mid i\right)  \rho\left(  k\mid j\right)  \rho\left(  i\mid
k\right)  =\rho\left(  k\mid i\right)  \rho\left(  j\mid k\right)  \rho\left(
i\mid j\right)  \label{eq:trans-ddm-bis}%
\end{equation}
for all distinct alternatives $i$, $j$ and $k$.
\end{definition}

In words, a binary choice probability is transitive when violations of
transitivity in the choices that it determines are due only to the presence of
noise.\footnote{Cf. Luce and Suppes (1965) p. 341. Also note that if two
alternatives are not distinct, then condition (\ref{eq:trans-ddm-bis}) is
automatically satisfied.} Indeed, condition (\ref{eq:trans-ddm-bis}) amounts
to require the intransitive cycles
\[
i\rightarrow j\rightarrow k\rightarrow i\text{\quad and\quad}i\rightarrow
k\rightarrow j\rightarrow i
\]
to be equally likely (over independent choices). Transitivity ensures that
systematic intransitivities, a violation of a basic rationality tenet, cannot
occur. We expect that a viable neural system satisfies this property.

\subsection{Binary value analysis}

The binary choice probability $\rho$ induces two binary relations $\succ
^{\ast}$ and $\succsim$ $\ $defined by%
\[
i\succ^{\ast}j\Longleftrightarrow\rho\left(  i\mid j\right)  =1\quad
\text{and\quad}i\succsim j\Longleftrightarrow\rho\left(  i\mid j\right)
\geq\rho\left(  j\mid i\right)
\]
for all alternatives $i$ and $j$. We interpret $\succ^{\ast}$ as a clear-cut,
deterministic, strict preference over the alternatives that the decision unit
is able to perfectly discriminate in value. It is a standard notion in
traditional (non-stochastic) utility theory.\footnote{See Fishburn (1970) and
Kreps (1988).} Since%
\begin{equation}
i\succ^{\ast}j\Longleftrightarrow\rho\left(  i\mid j\right)
=1\Longleftrightarrow\rho\left(  j\mid i\right)  =0 \label{eq:sstr}%
\end{equation}
a strict preference holds irrespective of the alternatives' roles as
incumbents or proposals.\footnote{To further elaborate, $\rho\left(  i\mid
j\right)  =1$ means that $i$ is accepted for sure when proposed, while
$\rho\left(  j\mid i\right)  =0$, that is, $1-\rho\left(  j\mid i\right)  =1$,
means that $i$ is maintained for sure when it is the incumbent.} We write
$i\parallel^{\ast}j$ when there is no strict preference over the two
alternatives, i.e.,%
\[
i\parallel^{\ast}j\Longleftrightarrow i\not \succ ^{\ast}j\text{ and
}j\not \succ ^{\ast}i
\]
by (\ref{eq:sstr}) this is equivalent to $\rho\left(  i\mid j\right)
\in\left(  0,1\right)  $, and also to $\rho\left(  j\mid i\right)  \in\left(
0,1\right)  $. That is when choice is truly stochastic (again irrespective of
the alternatives' roles).

In contrast, we interpret $\succsim$ as a weak notion of preference that
extends the strict preference $\succ^{\ast}$ by allowing for the stochastic
rankings that occur over alternatives that the decision unit only imperfectly
discriminates in value. The consistency property%
\begin{equation}
i\succ^{\ast}j\Longrightarrow i\succ j \label{eq:cons}%
\end{equation}
shows that, as natural, a strict preference is preserved by (the asymmetric
part of $\succsim$). Finally, the binary relation $\succsim^{%
{{}^\circ}%
}$ defined by
\[
i\succsim^{%
{{}^\circ}%
}j\Longleftrightarrow i\succsim j\text{ and }i\parallel^{\ast}j
\]
describes the rankings that are stochastic. Indeed,%
\[
i\succsim^{%
{{}^\circ}%
}j\Longleftrightarrow1>\rho\left(  i\mid j\right)  \geq\rho\left(  j\mid
i\right)  >0
\]

The next lemma substantiates our interpretations.

\begin{lemma}
\label{lm:bcp-prop}If the binary choice probability $\rho$ is transitive, then

\begin{enumerate}
\item[(i)] $\succ^{\ast}$ is asymmetric and negatively transitive;

\item[(ii)] $\parallel^{\ast}$ is an equivalence relation;

\item[(iii)] $\succsim$ is complete and transitive;

\item[(iv)] $\succsim^{%
{{}^\circ}%
}$ is reflexive and transitive as well as complete on each equivalence class
of $\parallel^{\ast}$.
\end{enumerate}
\end{lemma}

The two preferences $\succsim^{%
{{}^\circ}%
}$ and $\succ^{\ast}$ complement each other by accounting for the stochastic
and deterministic comparisons that occur, respectively, with imperfect and
perfect discrimination in value. Jointly, they rank all pairs of
alternatives.\footnote{Set-theoretically, we have $\succ^{\ast}\cap\succsim^{%
{{}^\circ}%
}=\emptyset$ and $\succ^{\ast}\cup\succsim^{%
{{}^\circ}%
}=\succsim$.} In view of this, we focus our analysis on them.

\begin{definition}
A binary choice probability $\rho:A^{2}\rightarrow\left[  0,1\right]  $ has a
\emph{binary} \emph{value} \emph{representation} if there exist
$v,w:A\rightarrow\mathbb{R}$ and a symmetric $s:A^{2}\rightarrow\left(
0,\infty\right)  $ such that%
\begin{equation}
\rho\left(  i\mid j\right)  =\left\{
\begin{array}
[c]{ll}%
1\medskip & \qquad\text{if }w\left(  i\right)  >w\left(  j\right) \\
s\left(  i,j\right)  \dfrac{e^{v\left(  i\right)  }}{e^{v\left(  i\right)
}+e^{v\left(  j\right)  }}\medskip & \qquad\text{if }w\left(  i\right)
=w\left(  j\right) \\
0 & \qquad\text{if }w\left(  i\right)  <w\left(  j\right)
\end{array}
\right.  \label{eq:stoch-utt}%
\end{equation}
for all $i$ and $j$.
\end{definition}

It is readily seen that, in this case%
\begin{equation}
i\succ^{\ast}j\Longleftrightarrow w\left(  i\right)  >w\left(  j\right)
\quad\text{;\quad}i\parallel^{\ast}j\Longleftrightarrow w\left(  i\right)
=w\left(  j\right)  \label{eq:stoch-ut-pre}%
\end{equation}
and, when $w\left(  i\right)  =w\left(  j\right)  $,%
\begin{equation}
i\succsim^{%
{{}^\circ}%
}j\Longleftrightarrow v\left(  i\right)  \geq v\left(  j\right)
\label{eq:stoch-ut}%
\end{equation}
Thus, we interpret $w$ as a utility function for $\succ^{\ast}$ and $v$ as a
utility function for $\succsim^{%
{{}^\circ}%
}$. Moreover, we interpret $s$ as a status quo bias index. These
interpretations are corroborated by the next result. In reading it, keep in
mind that $v$ and $s$ are relevant only for stochastic rankings (as identified
by the equivalence classes of $\parallel^{\ast}$, so by the level sets of $w$).

\begin{lemma}
\label{prop:vb}If a binary choice probability admits a binary value
representation, then,

\begin{enumerate}
\item[(i)] the utility function $w$ is unique up to a strictly increasing transformation;

\item[(ii)] the utility function $v$ is, on each level set of $w$, unique up
to an additive constant;

\item[(iii)] the status quo bias index $s$ is, on each level set of $w$,
unique with%
\begin{align}
\rho\left(  i\mid j\right)   &  <1-\rho\left(  j\mid i\right)
\Longleftrightarrow s\left(  i,j\right)  <1\nonumber\\
\rho\left(  i\mid j\right)   &  =1-\rho\left(  j\mid i\right)
\Longleftrightarrow s\left(  i,j\right)  =1\label{eq:propvb}\\
\rho\left(  i\mid j\right)   &  >1-\rho\left(  j\mid i\right)
\Longleftrightarrow s\left(  i,j\right)  >1\nonumber
\end{align}

\end{enumerate}
\end{lemma}

The relations in the last point of the lemma clarify the interpretation of $s$
as a status quo bias index for the comparison of proposal $i$ and incumbent
$j$. In particular, bias favors the incumbent when $s\left(  i,j\right)  <1$,
the proposal when $s\left(  i,j\right)  >1$, and it is absent otherwise. Thus,
the binary choice probability $\rho$ is unbiased if and only if $s$ is
constant to $1$.

The utility $w$ is a traditional Paretian utility function that, by ranking
alternatives in an ordinal manner, represents the strict preference
$\succ^{\ast}$. This Paretian utility is constant, so irrelevant in
(\ref{eq:stoch-utt}), if and only if $\rho$ is positive, i.e., when all
rankings are stochastic. When $\rho$ is both positive and unbiased, the binary
value representation (\ref{eq:stoch-utt}) reduces to%
\[
\rho\left(  i\mid j\right)  =\dfrac{e^{v\left(  i\right)  }}{e^{v\left(
i\right)  }+e^{v\left(  j\right)  }}\medskip
\]
This is the strict utility\ representation of Marschak (1960) and Luce and
Suppes (1965),\footnote{Luce (1959) and Block and Marschak (1960) study a
stronger non-binary version of strict utility.} which our binary value
representation thus extends to general, possibly biased and partly
deterministic, binary choice probabilities. As well-known, using the logistic
function $\xi$ we can write:%
\begin{equation}
\dfrac{e^{v\left(  i\right)  }}{e^{v\left(  i\right)  }+e^{v\left(  j\right)
}}\medskip=\xi\left(  v\left(  i\right)  -v\left(  j\right)  \right)
\label{eq:fech-diff}%
\end{equation}
The binary choice probability $\rho\left(  i\mid j\right)  $ thus depends, in
a Fechnerian way, on the utility difference $v\left(  i\right)  -v\left(
j\right)  $.\footnote{Cf. Luce and Suppes (1965) p. 334. The logistic function
$\xi:\mathbb{R}\rightarrow\mathbb{R}$ is given by $\xi\left(  x\right)
=1/\left(  1+e^{-x}\right)  $.}

In our extension, $v\ $continues to be a \emph{bona fide} utility function on
the level sets of $w$, as (\ref{eq:stoch-ut}) shows. We call it a
\emph{Fechnerian utility} \emph{function}. When $w\left(  i\right)  =w\left(
j\right)  $, it holds%
\begin{equation}
\rho\left(  i\mid j\right)  \geq\rho\left(  j\mid i\right)
\Longleftrightarrow v\left(  i\right)  \geq v\left(  j\right)  \iff
1-\rho\left(  j\mid i\right)  \geq1-\rho\left(  i\mid j\right)
\label{eq:ut-stoch}%
\end{equation}
Alternatives with a higher Fechnerian utility thus have a higher probability
to be selected, regardless of their roles as proposals or incumbents. When
$\rho$ is both positive and unbiased, (\ref{eq:ut-stoch}) takes the form%
\[
v\left(  i\right)  \geq v\left(  j\right)  \Longleftrightarrow\rho\left(
i\mid j\right)  \geq\frac{1}{2}%
\]
familiar from traditional stochastic choice. The Fechnerian utility function
is immaterial when $\rho$ is Dirac, i.e., when all rankings of distinct
alternatives are deterministic.

\begin{lemma}
\label{lm:dirac}A binary choice probability $\rho$ is Dirac and transitive if
and only if $\succ^{\ast}$ is weakly complete and transitive.\footnote{The
strict preference $\succ^{\ast}$ is \emph{weakly complete} if, for each $i\neq
j$, either $i\succ^{\ast}j$ or $j\succ^{\ast}i$ (cf. Fishburn, 1970). Under
weak completeness, we do not have to worry about indifferences, a notoriously
delicate issue.}
\end{lemma}

The transitivity of a binary choice probability thus generalizes the
transitivity of a strict preference of traditional utility theory. In this
case, the binary value representation (\ref{eq:stoch-utt}) reduces to%
\[
\rho\left(  i\mid j\right)  =\left\{
\begin{array}
[c]{ll}%
1\medskip & \qquad\text{if }w\left(  i\right)  >w\left(  j\right) \\
\dfrac{1}{2}\medskip & \qquad\text{if }w\left(  i\right)  =w\left(  j\right)
\\
0 & \qquad\text{if }w\left(  i\right)  <w\left(  j\right)
\end{array}
\right.
\]

The next representation theorem, our first main result, shows that
transitivity characterizes the binary choice probabilities having a binary
value representation.

\begin{theorem}
\label{thm:value}A binary choice probability has a binary value
representation\emph{\ }if and only if it is transitive.
\end{theorem}

This theorem generalizes standard utility representations in stochastic choice
(e.g., Luce and Suppes, 1965, p. 350) as well as, in view of Lemma
\ref{lm:dirac}, in traditional utility theory.

We conclude by observing that the preference $\succsim$ has, in terms of the
binary value representation (\ref{eq:stoch-utt}), a lexicographic
representation via the Fechnerian utility $v$ and the Paretian utility $w$.
Indeed, it is easy to see that, for each $i$ and $j$,%
\[
i\succsim j\Longleftrightarrow\left(  w\left(  i\right)  ,v\left(  i\right)
\right)  \geq_{lex}\left(  w\left(  j\right)  ,v\left(  j\right)  \right)
\]
where $\geq_{lex}$ is the lexicographic order on the plane.

\subsection{Expected response times}

Besides the choice probability $\rho\left(  i\mid j\right)  $, the other
quantity featured in sequential binary choice is the expected time
\[
\tau\left(  i\mid j\right)
\]
that the decision unit takes to choose between distinct proposal $i$ and
incumbent $j$. We represent it with a function $\tau:A^{2}\rightarrow\left[
0,\infty\right)  $.

\begin{definition}
A pair $\left(  \rho,\tau\right)  $ of a binary choice probability and an
expected response time forms a \emph{tandem} if, for each $i\neq j$,%
\begin{equation}
\rho\left(  i\mid j\right)  \in\left\{  0,1\right\}  \Longleftrightarrow
\tau\left(  i\mid j\right)  =0 \label{eq:tao-rho}%
\end{equation}
and%
\begin{equation}
\tau\left(  i\mid j\right)  =\tau\left(  j\mid i\right)  \Longrightarrow
\rho\left(  i\mid j\right)  =1-\rho\left(  j\mid i\right)  \label{eq:tao-rho2}%
\end{equation}

\end{definition}

A tandem provides a thorough description of the binary choices of our decision
unit in the menu $A$. In such a description, the consistency condition
(\ref{eq:tao-rho}) ensures that deterministic choices are the ones that take
no time (so we abstract from non-decision times). In particular, since $\rho$
is a binary choice probability,
\[
\tau\left(  i\mid j\right)  =0\iff\tau\left(  j\mid i\right)  =0
\]

The consistency condition (\ref{eq:tao-rho2}), instead, requires that the
absence of a status quo bias manifests itself primarily in the symmetry of
response times.

\begin{definition}
A tandem $\left(  \rho,\tau\right)  $ has a \emph{binary} \emph{value}
\emph{representation} if there exist $v,w:A\rightarrow\mathbb{R}$, a symmetric
$s:A^{2}\rightarrow\left(  0,\infty\right)  $, and a strictly quasiconcave and
unimodal $\varphi:\mathbb{R}\rightarrow\left(  0,\infty\right)  $ such that
(\ref{eq:stoch-utt}) holds and
\begin{equation}
\tau\left(  i\mid j\right)  =\left\{
\begin{array}
[c]{ll}%
0\medskip & \qquad\text{if }w\left(  i\right)  \neq w\left(  j\right) \\
\varphi\left(  v\left(  i\right)  -v\left(  j\right)  \right)  &
\qquad\text{if }w\left(  i\right)  =w\left(  j\right)
\end{array}
\right.  \label{eq:response-time-bis}%
\end{equation}
for all $i$ and $j$.
\end{definition}

A strictly quasiconcave and unimodal $\varphi:\mathbb{R}\rightarrow\left[
0,\infty\right)  $ is a function that first strictly increases to a strong
maximum and then strictly decreases. This pattern is motivated by the standard
psychophysical assumption that the stimulus strength determines response times
with stronger stimuli inducing faster responses. Since here the stimulus
strength corresponds to the preference intensity represented by utility
differences $v\left(  i\right)  -v\left(  j\right)  $, this standard
assumption requires that large (positive) differences and small (negative)
differences command short response times. This is exactly what is captured by
the shape of $\varphi$.

To give an observable counterpart of this standard assumption we need to
introduce the following observables:%
\[
\ell_{ij}=\ln\frac{\rho\left(  i\mid j\right)  }{\rho\left(  j\mid i\right)  }%
\]
Using these log-odds we can introduce a class of tandems:

\begin{definition}
A tandem $\left(  \rho,\tau\right)  $ is \emph{chronometric} if $\rho$ is
transitive and there exists a threshold $l$ such that,
\begin{align}
\left.  \ell_{ij}=\ell_{hk}\right.   &  \implies\tau\left(  i\mid j\right)
=\tau\left(  h\mid k\right) \label{eq:qe}\\
\left.  l\leq\ell_{ij}<\ell_{hk}\right.   &  \implies\tau\left(  i\mid
j\right)  >\tau\left(  h\mid k\right) \label{eq:qi}\\
\left.  \ell_{ij}<\ell_{hk}\leq l\right.   &  \implies\tau\left(  i\mid
j\right)  <\tau\left(  h\mid k\right)  \label{eq:qo}%
\end{align}
for all pairs of alternatives $i,j$ and $h,k$ with nonzero response times.
\end{definition}

We can now state our second representation theorem that characterizes tandems
having a binary value representation.

\begin{theorem}
\label{prop:chrono}A tandem has a binary value representation\emph{\ }if and
only if it is chronometric.
\end{theorem}

Another standard psychophysical assumption is that stimulus strength
determines error rates, with stronger stimuli inducing lower error rates.
Consistency of this assumption with the previous one requires shorter response
times to correspond to lower error rates. Surprisingly this leads to unbiased tandems.

Specifically, observe that when comparing a proposal $i$ and an incumbent $j$
(or \textit{viceversa}) we may make errors of two types: we may reject a superior proposal or accept
an inferior one. In analogy with standard terminology, we call them
\emph{first-type }and\emph{\ second-type errors}, respectively. Their
probabilities are%
\begin{equation}
\mathrm{ER}_{i,j}^{\mathrm{I}}=\min\left\{  1-\rho\left(  i\mid j\right)
,1-\rho\left(  j\mid i\right)  \right\}  \quad\text{;}\quad\mathrm{ER}%
_{i,j}^{\mathrm{II}}=\min\left\{  \rho\left(  i\mid j\right)  ,\rho\left(
j\mid i\right)  \right\}  \label{eq:error-prob}%
\end{equation}

Next we introduce a basic error-monotonicity property.

\begin{definition}
A tandem $\left(  \rho,\tau\right)  $ is \emph{psychometric} if $\rho$ is
transitive and
\[
\tau\left(  i\mid j\right)  <\tau\left(  h\mid k\right)  \Longrightarrow
\mathrm{ER}_{i,j}^{\mathrm{I}}<\mathrm{ER}_{h,k}^{\mathrm{I}}\quad
\text{and\quad}\mathrm{ER}_{i,j}^{\mathrm{II}}<\mathrm{ER}_{h,k}^{\mathrm{II}}%
\]
and
\[
\tau\left(  i\mid j\right)  \leq\tau\left(  h\mid k\right)  \Longrightarrow
\mathrm{ER}_{i,j}^{\mathrm{I}}\leq\mathrm{ER}_{h,k}^{\mathrm{I}}%
\quad\text{and\quad}\mathrm{ER}_{i,j}^{\mathrm{II}}\leq\mathrm{ER}%
_{h,k}^{\mathrm{II}}%
\]
for all pairs of alternatives $i,j$ and $h,k$ with nonzero response times.
\end{definition}

In words, shorter binary expected response times correspond to lower errors of
both types, a property that regards as easier to make the choices between
alternatives with larger utility differences. The next representation theorem
shows that psychometricity characterizes chronometric tandems with symmetric
expected response times, thus featuring no status quo biases.

\begin{theorem}
\label{prop:psycho}A tandem has a binary value representation\emph{\ }with an
even $\varphi$ if and only if it is psychometric.
\end{theorem}

It is noteworthy that psychometricity implies chronometricity since the two
definitions are not obviously related. With this theorem, our third main
result, we conclude the general analysis of binary choices.

\section{An algorithmic decision process\label{sect:algo}}

\subsection{Binary choice behavior\label{sect:bin}}

It is time to turn to observed binary choice behavior and apply to it the
general binary choice framework just introduced.

\begin{definition}
A \emph{binary choice model }(\emph{BCM}) is a pair of random matrices
$\left(  \mathrm{C},\mathrm{RT}\right)  $ where:

\begin{enumerate}
\item[(i)] $\mathrm{C}=\left[  \mathrm{C}_{i,j}\right]  $ consists of the
random \emph{choice variables} $\mathrm{C}_{i,j}$ that describe the random
outcome of the comparison between proposal $i$ and status quo $j$, with
\[
\mathrm{C}_{i,j}=\left\{
\begin{array}
[c]{ll}%
i & \text{if }i\text{ accepted}\medskip\\
j & \text{if }i\text{ rejected}%
\end{array}
\right.
\]

\item[(ii)] $\mathrm{RT}=\left[  \mathrm{RT}_{i,j}\right]  $ consists of
random \emph{response times} $\mathrm{RT}_{i,j}$ required by the
comparison.\footnote{Throughout we assume that random response times (say
measured in seconds) have finite mean and variance.}
\end{enumerate}
\end{definition}

The distributions of $\mathrm{C}$ and $\mathrm{RT}$ are, in principle, both
observable in choice behavior. By equating probabilities and frequencies, they
induce a pair $\left(  \rho_{\mathrm{C}},\tau_{\mathrm{RT}}\right)  $ where%
\[
\rho_{\mathrm{C}}\left(  i\mid j\right)  =\mathbb{P}\left[  \mathrm{C}%
_{i,j}=i\right]
\]
is the frequency with which proposal $i$ is accepted over incumbent $j$, and
\[
\tau_{\mathrm{RT}}\left(  i\mid j\right)  =\mathbb{E}\left[  \mathrm{RT}%
_{i,j}\right]
\]
is the mean response time required by the comparison.

When $\left(  \rho_{\mathrm{C}},\tau_{\mathrm{RT}}\right)  $ has a binary
value representation, we denote by
\[
\left(  s_{\mathrm{C}},v_{\mathrm{C}},w_{\mathrm{C}},\varphi_{\mathrm{RT}%
}\right)
\]
its elements. The most basic example of a BCM $\left(  \mathrm{C}%
,\mathrm{RT}\right)  $ occurs in traditional utility theory when the choices
of the decision unit are deterministic. In this case, the pair $\left(
\rho_{\mathrm{C}},\tau_{\mathrm{RT}}\right)  $ has the binary value
representation of the Dirac form:%
\[
\rho_{\mathrm{C}}\left(  i\mid j\right)  =\left\{
\begin{array}
[c]{ll}%
1\medskip & \qquad\text{if }w_{\mathrm{C}}\left(  i\right)  >w_{\mathrm{C}%
}\left(  j\right) \\
\dfrac{1}{2}\medskip & \qquad\text{if }w_{\mathrm{C}}\left(  i\right)
=w_{\mathrm{C}}\left(  j\right) \\
0 & \qquad\text{if }w_{\mathrm{C}}\left(  i\right)  <w_{\mathrm{C}}\left(
j\right)
\end{array}
\right.
\]
and $\tau_{\mathrm{RT}}$ is typically undefined.

A popular stochastic binary choice model is the \emph{Drift Diffusion Model
}(\emph{DDM}) introduced by Ratcliff (1978). In its value version, developed
by Krajbich et al. (2010) and Milosavljevic et al. (2010), the comparison of
two alternatives $i$ and $j$ is governed by their \emph{neural utilities }%
$\nu\left(  i\right)  $ and $\nu\left(  j\right)  $ about which the decision
unit learns, for instance via memory retrieval, during the deliberation that
precedes the choice between the two alternatives.\footnote{To ease the
analysis, we assume that the neural utility $\nu:A\rightarrow\mathbb{R}$ is
injective.} Evidence accumulation in favor of either alternative is
represented by the two Brownian motions with drift $\mathrm{V}_{i}\left(
t\right)  =\nu\left(  i\right)  t+\mathrm{W}_{i}\left(  t\right)  $ and
$\mathrm{V}_{j}\left(  t\right)  =\nu\left(  j\right)  t+\mathrm{W}_{j}\left(
t\right)  $. Each accumulation experiences independent white noise
fluctuations modeled by the uncorrelated Wiener processes $\mathrm{W}_{i}$ and
$\mathrm{W}_{j}$. With this,

\begin{itemize}
\item the net\ evidence in favor of $i$ over $j$ is given, at each $t>0$, by
the difference%
\begin{equation}
\mathrm{Z}_{i,j}\left(  t\right)  =\mathrm{V}_{i}\left(  t\right)
-\mathrm{V}_{j}\left(  t\right)  =\left[  \nu\left(  i\right)  -\nu\left(
j\right)  \right]  t+\sqrt{2}~\mathrm{W}\left(  t\right)  \label{eq:ddm}%
\end{equation}
where $\mathrm{W}$ is the Wiener difference process $\left(  \mathrm{W}%
_{i}-\mathrm{W}_{j}\right)  /\sqrt{2}$;

\item comparison ends when $\mathrm{Z}_{i,j}\left(  t\right)  $ reaches either
the barrier $\lambda>0$ or $-\beta<0$; so the response time is%
\[
\mathrm{RT}_{i,j}=\min\left\{  t:\mathrm{Z}_{i,j}\left(  t\right)
=\lambda\text{ or }\mathrm{Z}_{i,j}\left(  t\right)  =-\beta\right\}
\]

\item proposal $i$ is accepted when the upper barrier $\lambda$ is reached,
while incumbent $j$ is maintained (so proposal $i$ is rejected) when the lower
barrier $-\beta$ is reached; so the choice variable is%
\[
\mathrm{C}_{i,j}=\left\{
\begin{array}
[c]{ll}%
i & \qquad\text{if }\mathrm{Z}_{i,j}\left(  \mathrm{RT}_{i,j}\right)
=\lambda\medskip\\
j & \qquad\text{if }\mathrm{Z}_{i,j}\left(  \mathrm{RT}_{i,j}\right)  =-\beta
\end{array}
\right.
\]

\end{itemize}

A different net evidence, $\lambda$ and $\beta$, accounts for the different
roles of alternatives as proposal and status quo. A DDM is pinned down by its
elements $\nu$, $\lambda$ and $\beta$. We thus write it as DDM $\left(
\nu,\lambda,\beta\right)  $. When $\lambda=\beta$ we say that the DDM is
\emph{symmetric}.

\begin{proposition}
\label{prop:ddm_bcm}The pair $\left(  \rho_{\mathrm{C}},\tau_{\mathrm{RT}%
}\right)  $ generated by a DDM $\left(  \nu,\lambda,\beta\right)  $ is a
chronometric tandem, with $\rho_{\mathrm{C}}$ positive (and transitive). It
has a binary value representation with%
\begin{equation}
v_{\mathrm{C}}=\lambda\nu\quad\text{;}\quad s_{\mathrm{C}}\left(  i,j\right)
=1+\dfrac{e^{\lambda\left\vert \nu\left(  i\right)  -\nu\left(  j\right)
\right\vert }-e^{\beta\left\vert \nu\left(  i\right)  -\nu\left(  j\right)
\right\vert }}{1-e^{\left(  \lambda+\beta\right)  \left\vert \nu\left(
i\right)  -\nu\left(  j\right)  \right\vert }}\quad\text{;}\quad
w_{\mathrm{C}}\text{ constant} \label{eq:bcp-uno}%
\end{equation}
and%
\begin{equation}
\varphi_{\mathrm{RT}}\left(  x\right)  =\frac{\lambda^{2}}{x}\left[
\frac{1-e^{\frac{\beta}{\lambda}x}}{e^{-x}-e^{\frac{\beta}{\lambda}x}}\left(
1+\frac{\beta}{\lambda}\right)  -\frac{\beta}{\lambda}\right]
\label{eq:bcp-due}%
\end{equation}
for all $x\in\mathbb{R}$.
\end{proposition}

Thus, in the DDM case the pair $\left(  \rho_{\mathrm{C}},\tau_{\mathrm{RT}%
}\right)  $ is a tandem with binary value representation%

\[
\rho_{\mathrm{C}}\left(  i\mid j\right)  =s_{\mathrm{C}}\left(  i,j\right)
\xi\left(  v_{\mathrm{C}}\left(  i\right)  -v_{\mathrm{C}}\left(  j\right)
\right)  =s_{\mathrm{C}}\left(  i,j\right)  \frac{e^{v_{\mathrm{C}}\left(
i\right)  }}{e^{v_{\mathrm{C}}\left(  i\right)  }+e^{v_{\mathrm{C}}\left(
j\right)  }}%
\]
and%

\[
\tau_{\mathrm{RT}}\left(  i\mid j\right)  =\varphi_{\mathrm{RT}}\left(
v_{\mathrm{C}}\left(  i\right)  -v_{\mathrm{C}}\left(  j\right)  \right)
=\lambda\frac{\lambda\rho_{\mathrm{C}}\left(  i\mid j\right)  -\beta\left(
1-\rho_{\mathrm{C}}\left(  i\mid j\right)  \right)  }{v_{\mathrm{C}}\left(
i\right)  -v_{\mathrm{C}}\left(  j\right)  }%
\]
In particular, we have a decomposition of the Fechnerian utility
$v_{\mathrm{C}}=\lambda\nu$ in terms of neural utility function $\nu$ and
acceptance threshold $\lambda$. Accordingly,%
\[
v_{\mathrm{C}}\left(  i\right)  -v_{\mathrm{C}}\left(  j\right)
=\lambda\left(  \nu\left(  i\right)  -\nu\left(  j\right)  \right)
\]
The Fechnerian utility difference is decomposed in the neural utility
difference $\nu\left(  i\right)  -\nu\left(  j\right)  $ weighted by the
coefficient $\lambda$. The higher the neural utility difference, the higher
can be viewed the intensity of the neural value for $i$ over $j$. The higher
$\lambda$, the higher the DM ability to perceive this value difference, so to
discriminate the alternatives' subjective values. In other words, $\lambda$
acts as a magnification lens for neural utility differences.

The next result gives a sharp empirical content to the DDM case. It is
convenient to state it using the log-odds%
\[
\ell_{ij}=\log\frac{\rho_{\mathrm{C}}\left(  i\mid j\right)  }{\rho
_{\mathrm{C}}\left(  j\mid i\right)  }\quad\text{and\quad}\bar{\ell}_{ij}%
=\log\frac{1-\rho_{\mathrm{C}}\left(  j\mid i\right)  }{1-\rho_{\mathrm{C}%
}\left(  i\mid j\right)  }%
\]

\begin{proposition}
\label{prop:ddm-bcm}The elements of a DDM $\left(  v,\lambda,\beta\right)  $
are uniquely identified by the tandem $\left(  \rho_{\mathrm{C}}%
,\tau_{\mathrm{RT}}\right)  $ that it generates. In particular, if $\ell
_{ij}\neq0$,%
\[
\lambda=\left\vert \ell_{ij}\right\vert \sqrt{\frac{\tau_{ij}}{\ell_{ij}%
\rho_{ij}+\bar{\ell}_{ij}\left(  \rho_{ij}-1\right)  }}%
\]
and%
\[
\beta=\lambda\frac{\bar{\ell}_{ij}}{\ell_{ij}}\quad;\quad\nu\left(  i\right)
=\frac{1}{\lambda}\log r_{\mathrm{C}}\left(  i,j^{\ast}\right)
\]
under the normalization $v\left(  j^{\ast}\right)  =0$ for some alternative
$j^{\ast}$.
\end{proposition}

Finally, we characterize symmetric DDMs by showing that symmetry is equivalent
to an unbiased $\rho_{\mathrm{C}}$ as well as to a symmetric $\tau
_{\mathrm{RT}}$.

\begin{proposition}
\label{prop:ddm-error}For a tandem $\left(  \rho_{\mathrm{C}},\tau
_{\mathrm{RT}}\right)  $ generated by a DDM $\left(  \nu,\lambda,\beta\right)
$, the following conditions are equivalent:

\begin{enumerate}
\item[(i)] the tandem is psychometric;

\item[(ii)] $\beta=\lambda$;

\item[(iii)] $\tau_{\mathrm{RT}}\left(  i\mid j\right)  =\tau_{\mathrm{RT}%
}\left(  j\mid i\right)  $ for some (all) $i\neq j$;

\item[(iv)] $\rho_{\mathrm{C}}\left(  i\mid j\right)  =1-\rho_{\mathrm{C}%
}\left(  j\mid i\right)  $ for some (all) $i\neq j$.
\end{enumerate}

In this case,%
\[
\varphi_{\mathrm{RT}}\left(  x\right)  =\frac{\lambda^{2}}{x}\tanh\frac{x}{2}%
\]
for all $x\in\mathbb{R}$.
\end{proposition}

We conclude this section by observing that a broad family of BCMs is given by
\emph{evidence threshold models}. They encompass integration models, like the
DDM just studied, as well as the extrema detection models discussed by Stine
et al. (2020). In Appendix \ref{sect:evid-mod} we discuss this family of BCMs
in some detail.

\subsection{Neural Metropolis Algorithm\label{sect:nma}}

The protagonist of our analysis is an algorithmic decision process that a
neural system might implement when facing a multialternative menu $A$. This
process consists of a sequence of pairwise comparisons conducted via a BCM,
whose contestants are selected by a Markovian mechanism in the sense of
Metropolis et al. (1953). This sequential structure is motivated, as discussed
in the Introduction, by the well-known limits of working memory and is
supported by classic and recent eye-tracking studies.\footnote{See Russo and
Rosen (1975), Krajbich and Rangel (2011) and Reutskaja et al. (2011) as well
as the discussion in Cerreia-Vioglio et al. (2022).}

In broad strokes, this algorithmic decision process:

\begin{enumerate}
\item starts from an arbitrary element $j$ of the menu, the\textbf{
}\emph{incumbent};

\item selects a candidate alternative $i$ in the menu, the \emph{proposal};

\item compares them via a BCM and makes the winner the new incumbent;

\item repeats steps 1-3 until deliberation time comes, with the last incumbent
being the chosen alternative in the menu.
\end{enumerate}

More in detail, the algorithm starts by selecting a first incumbent $j$
according to an initial distribution $\mu\in\Delta\left(  A\right)  $ that,
for example, may describe the \textquotedblleft first
fixation\textquotedblright\ of the decision unit. It proceeds through an
\emph{exploration }(\emph{stochastic})\emph{ matrix}
\[
Q=\left[  Q\left(  i\mid j\right)  :i,j\in A\right]
\]
of order $\left\vert A\right\vert $ that describes how the algorithm navigates
through alternatives. In particular, given the incumbent $j$, a proposal $i$
is selected with probability $Q\left(  i\mid j\right)  $. Incumbent and
proposal are then compared via a BCM $\left(  \mathrm{C},\mathrm{RT}\right)
$. After $\mathrm{RT}_{i,j}$ seconds, the new incumbent is $j^{\prime
}=\mathrm{C}_{i,j}$; a new proposal $i^{\prime}$ is then selected with
probability $Q\left(  i^{\prime}\mid j^{\prime}\right)  $, and so on so forth.

The algorithm terminates according to a posited random \emph{stopping}
\emph{number }$N$ that limits the number of allowed iterations because of
exogenously constrained computational resources (for instance, this number may
have a cost, say economic or physiological, for the decision unit). The last
incumbent is the algorithm output, so what the algorithm chooses from menu $A$.

After this preliminary discussion, next we formalize the \emph{Neural
Metropolis Algorithm}, our algorithmic decision process. Its constitutive
elements are a BCM $\left(  \mathrm{C},\mathrm{RT}\right)  $ and an
exploration strategy $\left(  \mu,Q\right)  $, summarized in the quartet%
\begin{equation}
\left(  \mathrm{C},\mathrm{RT},\mu,Q\right)  \label{eq:algo-elms}%
\end{equation}
For mathematical convenience we start the algorithm at time $-1$.

\begin{center}
\rule{18cm}{0.04cm}

\textbf{Neural Metropolis Algorithm}

\rule{18cm}{0.04cm}
\end{center}

\noindent\textbf{Input:}$\ $\emph{Given a stopping number }$N$\emph{.}\medskip

\noindent\textbf{Start: }\emph{Draw }$i\ $\emph{from}$\emph{\ }A$\emph{
according to }$\mu$ \emph{and}$\medskip$

$\bullet$ \emph{set }$t_{-1}=0$\emph{,\medskip}

$\bullet$\emph{ set }$j_{-1}=i$\emph{.\medskip}

\noindent\textbf{Repeat: }\emph{Draw }$i_{n}\ \emph{from\ }A$\emph{ according
to }$Q\left(  \cdot\mid j_{n-1}\right)  $\emph{ and compare it to }$j_{n-1}%
$\emph{:\medskip}

$\bullet$ \emph{set }$t_{n}=t_{n-1}+\mathrm{RT}_{i_{n},j_{n-1}}$%
\emph{,\medskip}

$\bullet$\emph{ set }$j_{n}=\mathrm{C}_{i_{n},j_{n-1}}$\emph{;\medskip}

\noindent\textbf{until }$n=N$\emph{.}\medskip

\noindent\textbf{Stop: }\emph{Set }$k=j_{n-1}$\emph{.\medskip}

\noindent\textbf{Output: }\emph{Choose }$k$\emph{ from }$A$\emph{.\vspace
{-20pt}}

\begin{center}
\rule{18cm}{0.04cm}
\end{center}

Along with stopping number $N$, the Neural Metropolis Algorithm
(\ref{eq:algo-elms}) selects alternative $j_{n-1}$ when $N=n$, where $n$ is
the iteration at which the decision process is interrupted by the stopping
number. The Neural Metropolis Algorithm generalizes the Metropolis-DDM
Algorithm of Cerreia-Vioglio et al. (2022), which is the special case when the
underlying BCM is generated by a DDM, the exploration matrix $Q\left(  i\mid
j\right)  $ is inversely proportional to the mean of $\mathrm{RT}_{i,j}$ and a
hard deadline is given.

\subsection{Algorithmic properties}

The Neural Metropolis Algorithm (\ref{eq:algo-elms}) generates a Markov chain
of incumbents%
\begin{equation}
J=\left\{  J_{-1},J_{0},J_{1},...\right\}  \label{eq:marko}%
\end{equation}
with $\mathbb{P}\left[  J_{-1}=j\right]  =\mu\left(  j\right)  $ for all
alternatives $j$ in $A$ and, for each $n\geq0$,
\[
\mathbb{P}\left[  J_{n}=i\mid J_{n-1}=j\right]  =\underset{\text{prob.
}i\text{ proposed}}{\underbrace{Q\left(  i\mid j\right)  }}\times
\underset{\text{prob. }i\text{ accepted}}{\underbrace{\rho_{\mathrm{C}}\left(
i\mid j\right)  }}=:M\left(  i\mid j\right)
\]
for all distinct alternatives $i$ and $j$ in $A$. The stochastic matrix $M$ is
the \emph{transition matrix} of the incumbents' Markov chain $J$. In
particular, we say that the Neural Metropolis Algorithm is \emph{reversible}
when its transition matrix is reversible and so the incumbents' Markov chain
(\ref{eq:marko}) is reversible.

The Neural Metropolis Algorithm induces, for each stopping number $N$, a:

\begin{enumerate}
\item[(i)] \emph{choice probability} $p_{N}\in\Delta\left(  A\right)
$,\footnote{Recall that $p_{N}$ is an $\left\vert A\right\vert $-dimensional
vector.} where $p_{N}\left(  i,A\right)  $ is the probability that alternative
$i$ is selected from menu $A$ by the algorithm;

\item[(ii)] \emph{mean response time} $\tau_{N}\in\left[  0,\infty\right)  $,
the average time that the algorithm takes to select an alternative from $A$.
\end{enumerate}

The possibility of computing these quantities in explicit form is what makes
the Neural Metropolis Algorithm empirically relevant. To this end, next we
introduce a class of stopping numbers amenable to computations.

\begin{definition}
A stopping number $N$ is \emph{simple} within a Neural Metropolis Algorithm
(\ref{eq:algo-elms}) if it is independent of the realizations of incumbents,
proposals and response times.
\end{definition}

Next we compute the choice probabilities and mean response times for a Neural
Metropolis Algorithm with a simple stopping number. A piece of notation: we
denote by%
\begin{equation}
\bar{\tau}_{j}=%
{\displaystyle\sum\limits_{i\in A}}
Q\left(  i\mid j\right)  \tau_{\mathrm{RT}}\left(  i\mid j\right)
\label{eq:tao-simple}%
\end{equation}
the average duration of an iteration when $j$ is the incumbent.

\begin{proposition}
\label{lem:comp}For a Neural Metropolis Algorithm (\ref{eq:algo-elms}) with a
simple stopping number $N$,\footnote{The r.h.s. of both formulas in
(\ref{eq:uella}) involve standard matrix-vector multiplications: $\mu$ and
$\bar{\tau}$ are $\left\vert A\right\vert $-dimensional vectors, while
$f_{N}\left(  M\right)  $ and $g_{N}\left(  M\right)  $ are the square
matrices of order $\left\vert A\right\vert $ defined by
(\ref{eq:matrix-gen-fct}).}%
\begin{equation}
p_{N}=f_{N}\left(  M\right)  \mu\quad\text{and\quad}\tau_{N}=\bar{\tau}\cdot
g_{N}\left(  M\right)  \mu\label{eq:uella}%
\end{equation}

\end{proposition}

Using the definitions of probability and survival generating functions, we can
rewrite the choice probabilities and mean response times (\ref{eq:uella}) as%
\begin{equation}
p_{N}=\left(
{\displaystyle\sum\limits_{n=0}^{\infty}}
\mathbb{P}\left[  N=n\right]  M^{n}\right)  \mu\quad\text{and\quad}\tau
_{N}=\bar{\tau}\cdot\left(
{\displaystyle\sum\limits_{n=0}^{\infty}}
\mathbb{P}\left[  N>n\right]  M^{n}\right)  \mu\label{eq:uella-bis}%
\end{equation}
An immediate consequence of this rewriting is that%
\[
N\geq N^{\prime}\Longrightarrow\tau_{N}\geq\tau_{N^{\prime}}%
\]
A less tight stopping number results, as natural, in a longer mean decision time.

In the following important case we can compute the choice probabilities and
mean response times in closed form.

\begin{definition}
Given two coefficients $\zeta\in\left(  0,1\right)  $ and $r\geq1$, the
\emph{negative binomial stopping number} $N_{r}\left(  \zeta\right)  $ is
defined by%
\[
\mathbb{P}\left[  N=n\right]  =\binom{n+r-1}{r-1}\zeta^{n}\left(
1-\zeta\right)  ^{r}\qquad\forall n\geq0
\]

\end{definition}

Under this distribution, the decision unit receives a \textquotedblleft
search\textquotedblright\ signal with probability $\zeta$ and a
\textquotedblleft stop\textquotedblright\ signal with probability $1-\zeta$;
it then proceeds to compare the alternatives when a search signal is received,
while it stops searching after $r$ stop signals.\footnote{For the first $r-1$
stop signals it just freezes, restarting in the next round.} When $r=1$, it
reduces to a \emph{geometric stopping number}%
\[
\mathbb{P}\left[  N_{1}\left(  \zeta\right)  =n\right]  =\zeta^{n}\left(
1-\zeta\right)  \qquad\forall n\geq0
\]
Now, the decision unit stops as soon as it receives the first stop signal.

\begin{proposition}
\label{prop:neg-bin}It holds%
\[
f_{N_{r}\left(  \zeta\right)  }\left(  M\right)  =\left(  1-\zeta\right)
^{r}\left(  1-\zeta M\right)  ^{-r}%
\]
and
\begin{equation}
g_{N_{r}\left(  \zeta\right)  }\left(  M\right)  =-\left(  \sum_{k=0}%
^{r}\binom{r}{k}\left(  -\zeta\right)  ^{k}\sum_{j=0}^{k-1}M^{j}\right)
\left(  1-\zeta M\right)  ^{-r} \label{eq:gnb}%
\end{equation}

\end{proposition}

By Proposition \ref{lem:comp}, for a simple negative binomial stopping
number\emph{ }we thus have%
\[
p_{N_{r}\left(  \zeta\right)  }=\left(  1-\zeta\right)  ^{r}\left(  1-\zeta
M\right)  ^{-r}\mu\quad\text{and\quad}\tau_{N_{r}\left(  \zeta\right)  }%
=-\bar{\tau}\cdot\left(  \sum_{k=0}^{r}\binom{r}{k}\left(  -\zeta\right)
^{k}\sum_{j=0}^{k-1}M^{j}\right)  \left(  1-\zeta M\right)  ^{-r}\mu
\]
In particular, in the geometric case $r=1$ we get%
\[
p_{N_{1}\left(  \zeta\right)  }=\left(  1-\zeta\right)  \left(  1-\zeta
M\right)  ^{-1}\mu\qquad\text{and}\qquad\tau_{N_{1}\left(  \zeta\right)
}=\bar{\tau}\cdot\zeta\left(  1-\zeta M\right)  ^{-1}\mu
\]
The formula for $p_{N_{1}\left(  \zeta\right)  }$ was first proved in
Valkanova (2020), all other formulas appear to be novel.

\subsection{Algorithmic value analysis\label{sect:rev}{}}

Earlier in the paper we discussed the value underpinning of binary choice
probabilities. Next we consider a similar concept for Neural Metropolis Algorithms.

\begin{definition}
A Neural Metropolis Algorithm (\ref{eq:algo-elms}) is \emph{value based} if
its binary choice probability $\rho_{\mathrm{C}}$ has a binary value
representation $\left(  s_{\mathrm{C}},v_{\mathrm{C}},w_{\mathrm{C}}\right)  $.
\end{definition}

This notion is the algorithmic counterpart of the binary value representation
of a binary choice probability. By Theorem \ref{thm:value}, value-based Neural
Metropolis Algorithms are characterized by transitive binary choice probabilities.

\begin{theorem}
\label{prop:value-bis}If a value-based Neural Metropolis Algorithm has a nice
exploration matrix, then%
\begin{equation}
\lim_{n\rightarrow\infty}\Pr\left[  J_{n}=i\right]  =\lim_{N_{k}%
\rightarrow\infty}p_{N_{k}}\left(  i,A\right)  =\left\{
\begin{array}
[c]{ll}%
\medskip\frac{e^{v_{\mathrm{C}}\left(  i\right)  }}{\sum_{j\in\arg\max
_{A}w_{\mathrm{C}}}e^{v_{\mathrm{C}}\left(  j\right)  }} & \qquad\text{if
}i\in\arg\max_{A}w_{\mathrm{C}}\\
0 & \qquad\text{else}%
\end{array}
\right.  \label{eq:algo-value}%
\end{equation}
for all sequences of divergent simple stopping rules $N_{k}$.
\end{theorem}

This result clarifies the nature of a value-based Neural Metropolis Algorithm.
To appreciate it, observe that $\Pr\left[  J_{n}=i\right]  $ is the
probability that, unstopped, the algorithm chooses alternative $i$ after $n$
iterations, while $\arg\max_{A}w_{\mathrm{C}}$ is the set of alternatives that
are maximal under $\succ^{\ast}$. Thus,%
\[
\lim_{n\rightarrow\infty}\Pr\left[  J_{n}=i\right]
\]
indicates the inherent tendency of the Neural Metropolis Algorithm to choose a
maximal alternative $i$, regardless of the exogenously posited stopping
number. As a result, it can be seen as representing the underlying value of
alternative $i$. When the algorithm satisfies (\ref{eq:algo-value}), we have,
for alternatives $i$ and $j$ that are maximal under $\succ^{\ast}$,
\[
v_{\mathrm{C}}\left(  i\right)  \geq v_{\mathrm{C}}\left(  j\right)
\Longleftrightarrow\lim_{n\rightarrow\infty}\Pr\left[  J_{n}=i\right]
\geq\lim_{n\rightarrow\infty}\Pr\left[  J_{n}=i\right]
\]
The inherent tendency of the algorithm is thus consistent with the Fechnerian
utility function $v_{\mathrm{C}}$, which in the limit governs the choices
between maximal alternatives (be they incumbents or proposals). The equality%
\[
\lim_{N_{k}\rightarrow\infty}p_{N_{k}}\left(  i,A\right)  =\lim_{n\rightarrow
\infty}\Pr\left[  J_{n}=i\right]
\]
shows that this limit behavior occurs when the stopping number is less and
less tight. This means, \emph{inter alia}, that the limit behavior is
unaffected by status quo biases, so $s_{\mathrm{C}}$ plays no role. Implicit
here is the view that these biases arise under external pressure, here
embodied by the posited stopping number, so they vanish when this pressure relaxes.

Finally, formula (\ref{eq:algo-value}) ensures that%
\[
\lim_{n\rightarrow\infty}\Pr\left[  J_{n}=i\right]  =\lim_{N_{k}%
\rightarrow\infty}p_{N_{k}}\left(  i,A\right)  =0
\]
for all alternatives $i$ in $A$ that are not maximal under $\succ^{\ast}$. In
other words, at the limit these alternatives have no chance to be selected --
as $\lim_{N_{k}\rightarrow\infty}p_{N_{k}}\left(  i,A\right)  =0$ -- and in
any event the algorithm has no tendency to select them -- as $\lim
_{n\rightarrow\infty}\Pr\left[  J_{n}=i\right]  =0$. This optimality property
ensures that, as stopping numbers are less and less tight, the algorithm
select alternatives that are maximal under $\succ^{\ast}$. Among them,
stochastic comparisons are then governed by the Fechnerian utility. In sum, at
the limit the Neural Metropolis Algorithm hard-maximizes $w_{\mathrm{C}}$ and
soft-maximizes $v_{\mathrm{C}}$.

A first important consequence of the previous theorem concerns the case in
which $\succ^{\ast}$ features a single maximal element.

\begin{corollary}
\label{prop:value-ter}A value-based Neural Metropolis Algorithm
(\ref{eq:algo-elms}), with nice exploration matrix $Q$, satisfies
\[
\lim_{n\rightarrow\infty}\Pr\left[  J_{n}=i\right]  =\lim_{N_{k}%
\rightarrow\infty}p_{N_{k}}\left(  i,A\right)  =\left\{
\begin{array}
[c]{ll}%
\medskip1 & \qquad\text{if }i\in\arg\max_{A}w_{\mathrm{C}}\\
0 & \qquad\text{else}%
\end{array}
\right.
\]
if and only if $\arg\max_{A}w_{\mathrm{C}}$ is a singleton.
\end{corollary}

For instance, in the deterministic case of a transitive Dirac binary choice
probability the Neural Metropolis Algorithm selects, at the limit, the best
alternative.\footnote{Recall that a Dirac and transitive $\rho_{\mathrm{C}}$
corresponds to a weakly complete and transitive $\succ^{\ast}$ (cf. Lemma
\ref{lm:dirac}). So. $\arg\max_{A}w_{\mathrm{C}}$ is a singleton consisting of
the best alternative under $\succ^{\ast}$.} This limit analysis is much in
line with the traditional assumption of unconstrained computational resources.
Traditional choice behavior is thus implemented computationally by the Neural
Metropolis Algorithm.

In the traditional case just considered, $\arg\max_{A}w_{\mathrm{C}}$ is a
singleton in $A$. In contrast, $\arg\max_{A}w_{\mathrm{C}}$ coincides with the
whole set $A$ when, like in the DDM case, the binary choice probability
$\rho_{\mathrm{C}}$ is positive. In this case, $w_{\mathrm{C}}$ is constant,
so all alternatives are maximal under $\succ^{\ast}$. We thus have a second
noteworthy special case of the last representation theorem.

\begin{corollary}
\label{prop:value-quater}A value-based Neural Metropolis Algorithm
(\ref{eq:algo-elms}), with nice exploration matrix $Q$, satisfies%
\begin{equation}
\lim_{n\rightarrow\infty}\Pr\left[  J_{n}=i\right]  =\lim_{N_{k}%
\rightarrow\infty}p_{N_{k}}\left(  i,A\right)  =\frac{e^{v_{\mathrm{C}}\left(
i\right)  }}{\sum_{j\in A}e^{v_{\mathrm{C}}\left(  j\right)  }}\qquad\forall
i\in A \label{eq:acqua}%
\end{equation}
if and only if its binary choice probability $\rho_{\mathrm{C}}$ is positive.
\end{corollary}

By Proposition \ref{prop:ddm_bcm}, in the DDM special case we have
$v_{\mathrm{C}}=\lambda\nu$ in (\ref{eq:acqua}) and so multinomial logit
behavior
\begin{equation}
\frac{e^{\lambda\nu\left(  i\right)  }}{\sum_{j\in A}e^{\lambda\nu\left(
j\right)  }} \label{eq:acqua-bis}%
\end{equation}
emerges at the limit, like in Baldassi et al. (2020) and Cerreia-Vioglio et
al. (2022), even though here the assumptions on the stopping numbers are different.

In the positive $\rho_{\mathrm{C}}$ case, value-based Neural Metropolis
Algorithms have a remarkable computational property, as the next theorem, our
last main result, shows.

\begin{theorem}
\label{thm:value-bis}A Neural Metropolis Algorithm (\ref{eq:algo-elms}) with
positive $\rho_{\mathrm{C}}$ and nice exploration matrix $Q$, is value based
if and only if its transition matrix $M$ is reversible.\footnote{If and only
if $\rho_{\mathrm{C}}$ is transitive.}
\end{theorem}

At a computational level, reversibility ensures that the transition matrix $M$
is diagonalizable with real eigenvalues. Therefore,
\[
M=U\operatorname*{diag}\left(  \lambda_{1},\lambda_{2},...,\lambda_{\left\vert
A\right\vert }\right)  U^{-1}%
\]
where $\operatorname*{diag}\left(  \cdot\right)  $ is the diagonal matrix of
the eigenvalues $\lambda_{i}$ of $M$, each repeated according to its
multiplicity, and the columns of $U$ form a basis of the respective
eigenvectors. In turn, this readily implies%
\begin{equation}
f_{N}\left(  M\right)  =U\operatorname*{diag}\left(  f_{N}\left(  \lambda
_{1}\right)  ,f_{N}\left(  \lambda_{2}\right)  ,...,f_{N}\left(
\lambda_{\left\vert A\right\vert }\right)  \right)  U^{-1} \label{eq:frev}%
\end{equation}
and
\begin{equation}
g_{N}\left(  M\right)  =U\operatorname*{diag}\left(  \dfrac{1-f_{N}\left(
\lambda_{1}\right)  }{1-\lambda_{1}},\dfrac{1-f\left(  \lambda_{2}\right)
}{1-\lambda_{2}},...,\dfrac{1-f\left(  \lambda_{\left\vert A\right\vert
}\right)  }{1-\lambda_{\left\vert A\right\vert }}\right)  U^{-1}
\label{eq:grev}%
\end{equation}
with the limit convention (\ref{eq:limit-bis}). These formulas permit to
compute choice probabilities and mean response times for simple stopping
numbers, as in formulas (\ref{eq:uella-bis}). This computational achievement
concludes the analysis of value-based Neural Metropolis Algorithms.

\section{Discussion: temporal constrains\label{sect:icemia}}

In our analysis we considered constrained resources as modelled by a stopping
number on iterations, which may have an economic or physiological cost for the
decision unit. For perspective, in this final section we consider a different
type of constraint, namely, a temporal constraint in the form of a hard time
constraint $t$. This deadline induces a stopping number $N_{t}$ with $N_{t}=n$
if $t\in\left[  t_{n-1},t_{n}\right)  $. In words, the decision unit cannot
conclude the $n$-th comparison when the duration $t_{n}$ of that comparison
exceeds the deadline $t$. To see how this stopping number affects the
analysis, observe that, if unstopped, a Neural Metropolis Algorithm realizes a
stochastic process%
\begin{equation}
\left(  J,I,T\right)  =\left(  J_{-1},I_{0},T_{0},J_{0},I_{1},T_{1}%
,...,J_{n-1},I_{n},T_{n},J_{n},I_{n+1},...\right)  \label{eq:story}%
\end{equation}
where the realization $j_{n-1}$ of $J_{n-1}$ is the incumbent at the end of
iteration $n-1$, the realization $i_{n}$ of $I_{n}$ is the proposal at
iteration $n$, and the realization $t_{n}$ of $T_{n}$ is the duration of
iteration $n$. The stopping number $N_{t}$ acts as follows:%
\[
N_{t}=n\iff T_{-1}+T_{0}+\cdot\cdot\cdot+T_{n-1}\leq t<T_{-1}+T_{0}+\cdot
\cdot\cdot+T_{n-1}+T_{n}%
\]
where $T_{-1}=0$. In this case, a closed form representation of $p_{N_{t}}$ is
not achievable in general and, by definition, $\tau_{N_{t}}=t$. Yet, we can
give a limit result, in the spirit of Theorem \ref{prop:value-bis}, under the
following assumption.

\bigskip

\noindent\textbf{Regularity} A binary choice model $\left(  \mathrm{C}%
,\mathrm{RT}\right)  $ is \emph{regular} if $\rho_{\mathrm{C}}$ is
positive\ and transitive, $\tau_{\mathrm{RT}}$ is positive and $\mathrm{RT}%
=\left[  \mathrm{RT}_{i,j}\right]  $ consists of random response times
$\mathrm{RT}_{i,j}$ with a continuous distribution at $0$ and with no singular
part.\footnote{These two conditions on the distributions of response times are
automatically satisfied when they all admit density.}

\bigskip

We can now state the limit result.\footnote{In this discussion section we
focus on the positive case, leaving the more general case to an in-depth
future analysis.}

\begin{proposition}
\label{thm:value-ter}If a Neural Metropolis Algorithm (\ref{eq:algo-elms})
with irreducible exploration matrix $Q$ is based on a regular BCM $\left(
\mathrm{C},\mathrm{RT}\right)  $, then%
\[
\lim_{t\rightarrow\infty}p_{N_{t}}\left(  i,A\right)  =\frac{e^{v_{\mathrm{C}%
}\left(  i\right)  }\bar{\tau}_{i}}{\sum_{j\in A}e^{v_{\mathrm{C}}\left(
j\right)  }\bar{\tau}_{j}}\qquad\forall i\in A
\]

\end{proposition}

As time pressure diminishes, the limit probability of choosing alternative $i$
becomes proportional to the limit probability with which $i$ is an incumbent
times the average duration of the comparisons in which $i$ is the incumbent.
The intuition is natural: the longer the time spent in comparing an
alternative with the other alternatives, the higher the probability of
choosing that alternative at the deadline $t$.

In the DDM special case, we get%
\begin{equation}
\lim_{t\rightarrow\infty}p_{N_{t}}\left(  i\right)  =\frac{e^{\lambda
\nu\left(  i\right)  }\bar{\tau}_{i}}{%
{\displaystyle\sum\limits_{j\in A}}
e^{\lambda\nu\left(  j\right)  }\bar{\tau}_{j}}=\frac{e^{\lambda\nu\left(
i\right)  +\alpha\left(  i\right)  }}{%
{\displaystyle\sum\limits_{j\in A}}
e^{\lambda\nu\left(  j\right)  +\alpha\left(  j\right)  }}\qquad\forall i\in A
\label{eq:alpha}%
\end{equation}
Thus, the limit probability is softmax with neural utility $\nu$ and
alternative specific bias
\[
\alpha\left(  i\right)  =\log\bar{\tau}_{i}\qquad\forall i\in A
\]
If, in addition, the DDM is symmetric and the off diagonal entries of the
exploration matrix $Q$ are inversely proportional to mean response times (as
in Cerreia-Vioglio et al. 2022, Section 2), then the $\bar{\tau}_{i}$'s are
approximately constant and multinomial logit behavior (\ref{eq:acqua-bis})
emerges.\footnote{See Appendix \ref{app:endicectomia} for details.}

\newpage

\appendix

\section{Appendix: Evidence threshold models\label{sect:evid-mod}}

As it will soon become clear, evidence threshold models are best introduced in
discrete time. For each pair $i,j$ of alternatives in $A$, let $\left\{
\mathrm{Z}_{i,j}\left(  t\right)  \right\}  _{t=0}^{\infty}$ be a
discrete-time stochastic process in which each variable $\mathrm{Z}%
_{i,j}\left(  t\right)  $ represents the net evidence -- accumulated or
instantaneous -- in favor of $i$ over $j$ that the neural system takes into
account at time $t$. Given two evidence thresholds $\lambda,\beta>0$, a
decision is taken when either the evidence in favor of $i$ reaches level
$\lambda$ or the evidence in favor of $j$ reaches level $\beta$. This happens
at (stochastic) time
\begin{equation}
\mathrm{RT}_{i,j}=\min\left\{  t:\mathrm{Z}_{i,j}\left(  t\right)  \geq
\lambda\text{ or }\mathrm{Z}_{i,j}\left(  t\right)  \leq-\beta\right\}
\label{BCM1}%
\end{equation}
With this, the choice variable is
\begin{equation}
\mathrm{C}_{i,j}=\left\{
\begin{array}
[c]{ll}%
i & \qquad\text{if }\mathrm{Z}_{i,j}\left(  \mathrm{RT}_{i,j}\right)
\geq\lambda\medskip\\
j & \qquad\text{if }\mathrm{Z}_{i,j}\left(  \mathrm{RT}_{i,j}\right)
\leq-\beta
\end{array}
\right.  \label{BCM2}%
\end{equation}

Evidence threshold models encompass integration models, like a discrete-time
version of the DDM, as well as the extrema detection models discussed by Stine
et al. (2020). To see why, consider the discrete-time Ornstein-Uhlenbeck
process%
\begin{equation}
\mathrm{Z}_{i,j}\left(  t\right)  =\underset{\text{past evidence}}%
{\underbrace{\left(  1-\eta\right)  \mathrm{Z}_{i,j}\left(  t-1\right)  }%
}+\underset{\text{new evidence}}{\underbrace{\zeta_{i,j}\left(  t\right)  }%
}\qquad\forall t\geq1\nonumber
\end{equation}
with initial condition $\mathrm{Z}_{i,j}\left(  0\right)  =0$. The scalar
$\eta\in\left[  0,1\right]  $ captures past evidence deterioration and the
variable
\begin{equation}
\zeta_{i,j}\left(  t\right)  =\left[  \nu\left(  i\right)  -\nu\left(
j\right)  \right]  \mu\left(  t-1\right)  +\sigma\varepsilon\left(  t\right)
\qquad\forall t\geq1 \label{eq:ou-ter}%
\end{equation}
is the instantaneous\ noisy evidence gathered at time $t$ in favor of either
alternative.\footnote{Evidence is in favor of $i$ over $j$ when $\zeta
_{i,j}\left(  t\right)  \geq0$ and in favor of $j$ over $i$ when $\zeta
_{i,j}\left(  t\right)  \leq0$. The possible dependence of $\mu$ on $t-1$
allows for urgency signals.} The shock $\varepsilon$ is a Gaussian white noise
process -- i.e., it consists of i.i.d. Gaussian random variables
$\varepsilon\left(  t\right)  \sim N\left(  0,1\right)  $; like in the DDM,
$\nu\left(  i\right)  $ is the value of alternative $i$. When
\begin{equation}
\eta=0\quad\text{,\quad}\mu=1\quad\text{and\quad}\sigma=\sqrt{2}
\label{eq:ddm-discrete}%
\end{equation}
process (\ref{eq:ou-ter}) reduces to the following discrete-time version of
the DDM%
\[
\mathrm{Z}_{i,j}\left(  t\right)  -\mathrm{Z}_{i,j}\left(  t-1\right)
=\left[  \nu\left(  i\right)  -\nu\left(  j\right)  \right]  +\sqrt
{2}\,\varepsilon\left(  t\right)  \qquad\forall t\geq1
\]
Through the discrete-time Wiener process $w\left(  t\right)  =\sum_{s=1}%
^{t}\varepsilon\left(  s\right)  $, it is immediate to see that $\mathrm{Z}%
_{i,j}\left(  t\right)  $ represents accumulated noisy evidence:%
\[
\mathrm{Z}_{i,j}\left(  t\right)  =\sum_{s=1}^{t}\zeta_{i,j}\left(  s\right)
=\left[  \nu\left(  i\right)  -\nu\left(  j\right)  \right]  t+\sqrt
{2}w\left(  t\right)
\]
In contrast, when $\eta=1$ the process (\ref{eq:ou-ter}) takes the
\emph{extrema detection} form%
\begin{equation}
\mathrm{Z}_{i,j}\left(  t\right)  =\zeta_{i,j}\left(  t\right)  =\left[
\nu\left(  i\right)  -\nu\left(  j\right)  \right]  \mu\left(  t-1\right)
+\sigma\varepsilon\left(  t\right)  \qquad\forall t\geq1
\label{eq:extrema-det}%
\end{equation}
Now $\mathrm{Z}_{i,j}\left(  t\right)  $ represents instantaneous noisy
evidence, as opposed to the DDM accumulated one.

In continuous time, the Ornstein-Uhlenbeck process becomes%
\[
\mathrm{dZ}_{i,j}\left(  t\right)  =-\eta\mathrm{Z}_{i,j}\left(  t\right)
\mathrm{d}t+\left[  \nu\left(  i\right)  -\nu\left(  j\right)  \right]
\mu\left(  t\right)  \mathrm{d}t+\sigma\mathrm{dW}%
\]
with solution%
\begin{equation}
\mathrm{Z}_{i,j}\left(  t\right)  =\left[  \nu\left(  i\right)  -\nu\left(
j\right)  \right]  \mu\left(  t\right)  \frac{1-e^{-\eta t}}{\eta}+\int
_{0}^{t}e^{-\eta\left(  t-s\right)  }\sigma\mathrm{dW}\left(  s\right)
\label{eq:ou-cts}%
\end{equation}
The DDM (\ref{eq:ddm}) is still the special case (\ref{eq:ddm-discrete}). More
difficult is to identify the continuous counterpart of the extrema detection
model (\ref{eq:extrema-det}) because of the technical issues that arise with
continuous time white noise. As these issues do not appear to have a
substantive neural underpinning, we introduced evidence threshold models in
discrete time.\footnote{The accumulated evidence used by integration models
like the DDM is properly formalized by Wiener processes. The instantaneous
evidence featured by extrema detection models would rely on a notion of
\textquotedblleft derivative\textquotedblright\ for Wiener processes, a
notoriously subtle issue as their paths are nowhere differentiable.}

Be that as it may, Bogacz et al. (2006) report formulas for the continuous
time Ornstein-Uhlenbeck process that generalize the DDM ones upon which
Proposition \ref{prop:ddm_bcm} is based.\ It is unclear, however, whether this
generalized formulas deliver a sharp Ornstein-Uhlenbeck extension of this
proposition. Nevertheless, (\ref{eq:ou-cts}) is a significant generalization
of the DDM that, via the obvious continuous time versions of (\ref{BCM1}) and
(\ref{BCM2}), can play the role of a BCM.\pagebreak

\section{Appendix: Proofs and related analysis}

\subsection{Section \ref{sect:bin-choice}}

In this section it is sometimes convenient to use the exponential
transformation $u=e^{v}$ of the Fenchel utility $v$. We call $u$ \emph{strict
utility}.

\subsubsection{Proof of Lemma \ref{lm:bcp-prop}}

(i) Asymmetry is easily checked. Assume \emph{per contra} that $\succ^{\ast}$
is not negatively transitive.$\ $Then, there exist $i$, $j$ and $k$ such that
$i\nsucc^{\ast}k\nsucc^{\ast}j$ but $i\succ^{\ast}j$. Alternatives $i$, $k$
and $j$ must be distinct: $i\succ^{\ast}j$ implies $i\neq j$, while $i=k$
would imply $i\nsucc^{\ast}j$ and so would $k=j$. Moreover,

\begin{enumerate}
\item[(a)] $i\nsucc^{\ast}k$ implies $\rho\left(  i\mid k\right)  <1$ and
$\rho\left(  k\mid i\right)  >0$,

\item[(b)] $k\nsucc^{\ast}j$ implies $\rho\left(  k\mid j\right)  <1$ and
$\rho\left(  j\mid k\right)  >0$,

\item[(c)] $i\succ^{\ast}j$ implies $\rho\left(  i\mid j\right)  =1$
and$\ \rho\left(  j\mid i\right)  =0$.
\end{enumerate}

Therefore,
\[
\rho\left(  j\mid i\right)  \rho\left(  k\mid j\right)  \rho\left(  i\mid
k\right)  =0\quad\text{and\quad}\rho\left(  k\mid i\right)  \rho\left(  j\mid
k\right)  \rho\left(  i\mid j\right)  \neq0
\]
which contradicts the transitivity of $\rho$. We conclude that $\succ^{\ast}$
is negatively transitive.

(ii) In view of (i), it follows from Fishburn (1970) p. 13.

(iii) Completeness is easily established. Assume \emph{per contra} that
$\succsim$ is not transitive.$\ $Then there exist $i$, $j$ and $k$ such that
$i\succsim k\succsim j$ but $i\not \succsim j$. Alternatives $i$, $k$ and $j$
must be distinct: $i\not \succsim j$ implies $i\neq j$, while $i=k$ would
imply $i\succsim j$, and so would $k=j$. Moreover,

\begin{enumerate}
\item[(a)] $i\succsim k$ implies $\rho\left(  i\mid k\right)  \geq\rho\left(
k\mid i\right)  $,

\item[(b)] $k\succsim j$ implies $\rho\left(  k\mid j\right)  \geq\rho\left(
j\mid k\right)  $,

\item[(c)] $i\not \succsim j$ implies $\rho\left(  j\mid i\right)
>\rho\left(  i\mid j\right)  $.
\end{enumerate}

It holds $\rho\left(  i\mid k\right)  >0$. Indeed, $\rho\left(  i\mid
k\right)  =0$ would imply $\rho\left(  k\mid i\right)  =1$, contradicting (a).
Similarly, $\rho\left(  k\mid j\right)  >0$. Then,%
\[
\rho\left(  i\mid k\right)  \rho\left(  k\mid j\right)  \geq\rho\left(  k\mid
i\right)  \rho\left(  j\mid k\right)  \quad\text{;\quad}\rho\left(  j\mid
i\right)  >\rho\left(  i\mid j\right)  \quad\text{;\quad}\rho\left(  i\mid
k\right)  \rho\left(  k\mid j\right)  >0
\]
If $\rho\left(  i\mid k\right)  \rho\left(  k\mid j\right)  =\rho\left(  k\mid
i\right)  \rho\left(  j\mid k\right)  $, then both terms are strictly
positive, and%
\[
\rho\left(  i\mid k\right)  \rho\left(  k\mid j\right)  \rho\left(  j\mid
i\right)  >\rho\left(  k\mid i\right)  \rho\left(  j\mid k\right)  \rho\left(
i\mid j\right)
\]
Else $\rho\left(  i\mid k\right)  \rho\left(  k\mid j\right)  >\rho\left(
k\mid i\right)  \rho\left(  j\mid k\right)  $, and then
\[
\rho\left(  i\mid k\right)  \rho\left(  k\mid j\right)  \rho\left(  j\mid
i\right)  >\rho\left(  k\mid i\right)  \rho\left(  j\mid k\right)  \rho\left(
j\mid i\right)  \geq\rho\left(  k\mid i\right)  \rho\left(  j\mid k\right)
\rho\left(  i\mid j\right)
\]
In both cases the transitivity of $\rho$ is contradicted. We conclude that
$\succsim$ is transitive.

(iv) Reflexivity of $\succsim^{\circ}$ is obvious. Let $i\succsim^{%
{{}^\circ}%
}j$ and $j\succsim^{%
{{}^\circ}%
}k$. By definition, $i\succsim j$ and $i\parallel^{\ast}j$ as well as
$j\succsim k$ and $j\parallel^{\ast}k$. As both $\succsim$ and $\parallel
^{\ast}$ are transitive, it follows that $i\succsim k$ and $i\parallel^{\ast
}k$, that is, $i\succsim^{%
{{}^\circ}%
}k$. We conclude that $\succsim^{%
{{}^\circ}%
}$ is transitive. Finally, assume $i\parallel^{\ast}j$, and note that also
$j\parallel^{\ast}i$. Since $\succsim$ is complete, then either $i\succsim j$
or $j\succsim i$, thus either $i\succsim^{%
{{}^\circ}%
}j$ or $j\succsim^{%
{{}^\circ}%
}i$.\hfill$\blacksquare$

\subsubsection{Proof of Lemma \ref{prop:vb}}

Point (i) is obvious.

(ii) Let $i$ and $j$ be any two alternatives with $w\left(  i\right)
=w\left(  j\right)  $, and let $\tilde{u}:A\rightarrow\left(  0,\infty\right)
$ be such that%
\[
\rho\left(  i\mid j\right)  =s\left(  i,j\right)  \dfrac{\tilde{u}\left(
i\right)  }{\tilde{u}\left(  i\right)  +\tilde{u}\left(  j\right)  }%
\]
We have%
\[
\frac{\rho\left(  i\mid j\right)  }{\rho\left(  j\mid i\right)  }%
=\frac{s\left(  i,j\right)  \dfrac{\tilde{u}\left(  i\right)  }{\tilde
{u}\left(  i\right)  +\tilde{u}\left(  j\right)  }}{s\left(  j,i\right)
\dfrac{\tilde{u}\left(  j\right)  }{\tilde{u}\left(  i\right)  +\tilde
{u}\left(  j\right)  }}=\dfrac{\tilde{u}\left(  i\right)  }{\tilde{u}\left(
j\right)  }%
\]
Similarly,%
\[
\frac{\rho\left(  i\mid j\right)  }{\rho\left(  j\mid i\right)  }%
=\frac{s\left(  i,j\right)  \dfrac{u\left(  i\right)  }{u\left(  i\right)
+u\left(  j\right)  }}{s\left(  j,i\right)  \dfrac{u\left(  j\right)
}{u\left(  i\right)  +u\left(  j\right)  }}=\dfrac{u\left(  i\right)
}{u\left(  j\right)  }%
\]
Therefore, for any $j^{\ast}\in A$,%
\[
\tilde{u}\left(  i\right)  =\frac{\tilde{u}\left(  j^{\ast}\right)  }{u\left(
j^{\ast}\right)  }u\left(  i\right)
\]
for all $i\in A$. We conclude that $u$ is unique up to a positive scalar multiple.

(iii) Let $i$ and $j$ be any two alternatives with $w\left(  i\right)
=w\left(  j\right)  $. By the symmetry of $s$,%
\begin{equation}
\rho\left(  i\mid j\right)  +\rho\left(  j\mid i\right)  =s\left(  i,j\right)
\dfrac{u\left(  i\right)  }{u\left(  i\right)  +u\left(  j\right)  }+s\left(
j,i\right)  \dfrac{u\left(  j\right)  }{u\left(  i\right)  +u\left(  j\right)
}=s\left(  i,j\right)  \label{eq:fig}%
\end{equation}
Then $s$ is unique on the level set of $w\left(  i\right)  $. The relations in
(\ref{eq:propvb}) follow. \hfill$\blacksquare$

\subsubsection{Proof of Lemma \ref{lm:dirac}}

Let $\rho$ be a binary choice probability. Suppose that $\rho$ is Dirac and
transitive. Let $i\neq j$. As $\rho$ is Dirac, $\rho\left(  i\mid j\right)
\in\left\{  0,1\right\}  $. If $\rho\left(  i\mid j\right)  =1$, then
$i\succ^{\ast}j$. If $\rho\left(  i\mid j\right)  =0$, then $\rho\left(  j\mid
i\right)  =1$ and so $j\succ^{\ast}i$. We conclude that $\succ^{\ast}$ is
weakly complete.

Let $i\succ^{\ast}j$ and $j\succ^{\ast}k$. Hence, we have that $j\not \succ
^{\ast}i$, $k\not \succ ^{\ast}j$, and $k\neq i$. As $\rho$ is transitive,
$\succ^{\ast}$ is negatively transitive by Lemma \ref{lm:bcp-prop}. Hence,
$k\not \succ ^{\ast}i$ thus $\rho\left(  i\mid k\right)  \neq0$. By the
definition of Dirac and since $i\neq k$, we have that $\rho\left(  i\mid
k\right)  =1$, thus $i\succ^{\ast}k$. We conclude that $\succ^{\ast}$ is transitive.

As to the converse, let $\succ^{\ast}$ be weakly complete and transitive. Let
$i\neq j$. As $\succ^{\ast}$ is weakly complete, either $i\succ^{\ast}j$ or
$j\succ^{\ast}i$. If $i\succ^{\ast}j$, then $\rho\left(  i\mid j\right)  =1$;
if $j\succ^{\ast}i$, then $\rho\left(  j\mid i\right)  =1$ and so $\rho\left(
i\mid j\right)  =0$. We conclude that $\rho\left(  i\mid j\right)  \in\left\{
0,1\right\}  $. This proves that $\rho$ is Dirac. Suppose, \emph{per contra},
that $\rho$ is not transitive. Then, there exist three distinct alternatives
$i$, $j$ and $k$ such that%
\[
\rho\left(  j\mid i\right)  \rho\left(  k\mid j\right)  \rho\left(  i\mid
k\right)  \neq\rho\left(  k\mid i\right)  \rho\left(  j\mid k\right)
\rho\left(  i\mid j\right)
\]
It is impossible that both sides contain a zero factor. Since $\rho$ is Dirac,
then either
\[
\rho\left(  j\mid i\right)  \rho\left(  k\mid j\right)  \rho\left(  i\mid
k\right)  =1
\]
or
\[
\rho\left(  k\mid i\right)  \rho\left(  j\mid k\right)  \rho\left(  i\mid
j\right)  =1
\]
In the former case, $i\succ^{\ast}k\succ^{\ast}j\succ^{\ast}i$ and so
$\succ^{\ast}$ is not transitive. In the latter case, $i\succ^{\ast}%
j\succ^{\ast}k\succ^{\ast}i$ and so, again, $\succ^{\ast}$ is not transitive.
We conclude that $\rho$ must be transitive.\hfill$\blacksquare$

\subsubsection{Theorem \ref{thm:value}}

We prove a more general result that provides a utility representation
$\bar{u}:A\rightarrow\mathbb{R}$ for the preference $\succsim$.

\begin{theorem}
Given a binary choice probability $\rho$, the following conditions are equivalent:

\begin{enumerate}
\item[(i)] $\rho$ is transitive;

\item[(ii)] there exist $w,u:A\rightarrow\left(  0,\infty\right)  $ and a
symmetric $s:A^{2}\rightarrow\left(  0,\infty\right)  $ such that%
\[
\rho\left(  i\mid j\right)  =\left\{
\begin{array}
[c]{ll}%
1\medskip & \qquad\text{if }w\left(  i\right)  >w\left(  j\right) \\
s\left(  i,j\right)  \dfrac{u\left(  i\right)  }{u\left(  i\right)  +u\left(
j\right)  }\medskip & \qquad\text{if }w\left(  i\right)  =w\left(  j\right) \\
0 & \qquad\text{if }w\left(  i\right)  <w\left(  j\right)
\end{array}
\right.
\]
for all $i,j\in A$;

\item[(iii)] there exist $\bar{u}:A\rightarrow\left(  0,\infty\right)  $,
$f:\operatorname{Im}\bar{u}\rightarrow\left(  0,\infty\right)  $ increasing,
and a symmetric $s:A^{2}\rightarrow\left(  0,\infty\right)  $ such that
\[
\rho\left(  i\mid j\right)  =\left\{
\begin{array}
[c]{ll}%
1\medskip & \qquad\text{if }f\left(  \bar{u}\left(  i\right)  \right)
>f\left(  \bar{u}\left(  j\right)  \right) \\
s\left(  i,j\right)  \dfrac{\bar{u}\left(  i\right)  }{\bar{u}\left(
i\right)  +\bar{u}\left(  j\right)  }\medskip & \qquad\text{if }f\left(
\bar{u}\left(  i\right)  \right)  =f\left(  \bar{u}\left(  j\right)  \right)
\\
0 & \qquad\text{if }f\left(  \bar{u}\left(  i\right)  \right)  <f\left(
\bar{u}\left(  j\right)  \right)
\end{array}
\right.
\]
for all $i,j\in A$.
\end{enumerate}
\end{theorem}

By setting $v=\log u$ we recover Theorem \ref{thm:value} (note that since $w$
is ordinally unique we can always assume it to be strictly positive).

\bigskip

\noindent\textbf{Proof} (i) implies (iii). Since $A$ is finite there exists
$w:A\rightarrow\left(  0,\infty\right)  $ that represents $\succ^{\ast}$ in
the sense of (\ref{eq:stoch-ut-pre}). Then,%
\begin{align*}
w\left(  i\right)   &  >w\left(  j\right)  \iff i\succ^{\ast}j\iff\rho\left(
i\mid j\right)  =1\\
w\left(  i\right)   &  <w\left(  j\right)  \iff j\succ^{\ast}i\iff\rho\left(
i\mid j\right)  =0\\
w\left(  i\right)   &  =w\left(  j\right)  \iff i\parallel^{\ast}j\iff
\rho\left(  i\mid j\right)  \in\left(  0,1\right)
\end{align*}
By Lemma \ref{lm:bcp-prop}, $\parallel^{\ast}$ is an equivalence relation on
$A$. Since $w$ is unique up to a strictly increasing transformation, if
$\left\vert \operatorname{Im}w\right\vert =m$ we can assume $\operatorname{Im}%
w=\left\{  1,2,...,m\right\}  $. For all $h=1,2,...,m$ we can choose
$i_{h}^{\ast}\in w^{-1}\left(  h\right)  $. With this, $\left[  i_{1}^{\ast
}\right]  ,...,\left[  i_{m}^{\ast}\right]  $ is the partition of $A$ induced
by $\parallel^{\ast}$. For each $h=1,...,m$, set
\[
u_{h}^{\ast}\left(  j\right)  =\dfrac{\rho\left(  j\mid i_{h}^{\ast}\right)
}{\rho\left(  i_{h}^{\ast}\mid j\right)  }\qquad\forall j\in\left[
i_{h}^{\ast}\right]
\]
The ratio is well defined because $w\left(  j\right)  =w\left(  i_{h}^{\ast
}\right)  $ implies $\rho\left(  j\mid i_{h}^{\ast}\right)  ,\rho\left(
i_{h}^{\ast}\mid j\right)  \in\left(  0,1\right)  $. With this,%
\[
\frac{u_{h}^{\ast}\left(  j\right)  }{u_{h}^{\ast}\left(  k\right)  }%
=\frac{\dfrac{\rho\left(  j\mid i_{h}^{\ast}\right)  }{\rho\left(  i_{h}%
^{\ast}\mid j\right)  }}{\dfrac{\rho\left(  k\mid i_{h}^{\ast}\right)  }%
{\rho\left(  i_{h}^{\ast}\mid k\right)  }}=\dfrac{\rho\left(  j\mid
i_{h}^{\ast}\right)  }{\rho\left(  i_{h}^{\ast}\mid j\right)  }\dfrac
{\rho\left(  i_{h}^{\ast}\mid k\right)  }{\rho\left(  k\mid i_{h}^{\ast
}\right)  }\qquad\forall j,k\in\left[  i_{h}^{\ast}\right]
\]
By transitivity, we have that%
\[
\rho\left(  j\mid i\right)  \rho\left(  k\mid j\right)  \rho\left(  i\mid
k\right)  =\rho\left(  k\mid i\right)  \rho\left(  j\mid k\right)  \rho\left(
i\mid j\right)
\]
for all $i,$ $j,$ $k\in A$,\footnote{As previously observed, transitivity
implies the above \textquotedblleft product rule\textquotedblright\ for all
triplets of alternatives in $A$ and not only triplets of distinct ones.} and%
\[
\dfrac{\rho\left(  j\mid i_{h}^{\ast}\right)  }{\rho\left(  i_{h}^{\ast}\mid
j\right)  }\dfrac{\rho\left(  i_{h}^{\ast}\mid k\right)  }{\rho\left(  k\mid
i_{h}^{\ast}\right)  }=\frac{\rho\left(  j\mid k\right)  }{\rho\left(  k\mid
j\right)  }%
\]
for all $j,k\in\left[  i_{h}^{\ast}\right]  $. Therefore,%
\[
\frac{u_{h}^{\ast}\left(  j\right)  }{u_{h}^{\ast}\left(  k\right)  }%
=\dfrac{\rho\left(  j\mid i_{h}^{\ast}\right)  }{\rho\left(  i_{h}^{\ast}\mid
j\right)  }\dfrac{\rho\left(  i_{h}^{\ast}\mid k\right)  }{\rho\left(  k\mid
i_{h}^{\ast}\right)  }=\frac{\rho\left(  j\mid k\right)  }{\rho\left(  k\mid
j\right)  }\qquad\forall j,k\in\left[  i_{h}^{\ast}\right]
\]
for all $h=1,...,m$.

Set $\sigma_{1}=1$ and for each $h=2,...,m$, choose a strictly positive
constant $\sigma_{h}$ such that
\[
\max_{j\in\left[  i_{h-1}^{\ast}\right]  }\sigma_{h-1}u_{h-1}^{\ast}\left(
j\right)  <\min_{j\in\left[  i_{h}^{\ast}\right]  }\sigma_{h}u_{h}^{\ast
}\left(  j\right)
\]
that is,
\[
\sigma_{h}>\sigma_{h-1}\frac{\max_{j\in\left[  i_{h-1}^{\ast}\right]  }%
u_{h-1}^{\ast}\left(  j\right)  }{\min_{j\in\left[  i_{h}^{\ast}\right]
}u_{h}^{\ast}\left(  j\right)  }%
\]
Define%
\[
\bar{u}\left(  j\right)  =\sigma_{h}u_{h}^{\ast}\left(  j\right)
\qquad\forall j\in\left[  i_{h}^{\ast}\right]  ,\forall h=1,...,m
\]
Note that for all $h=2,...,m$, all $j_{h-1}\in\left[  i_{h-1}^{\ast}\right]  $
and all $j_{h}\in\left[  i_{h}^{\ast}\right]  $%
\[
\bar{u}\left(  j_{h-1}\right)  <\bar{u}\left(  j_{h}\right)
\]
that is,
\[
\bar{u}\left(  j_{1}\right)  <\bar{u}\left(  j_{2}\right)  <\cdot\cdot
\cdot<\bar{u}\left(  j_{m}\right)
\]
whenever $j_{h}\in\left[  i_{h}^{\ast}\right]  $ for all $h=1,2,...,m$. Then,
if $\bar{u}\left(  k\right)  \geq\bar{u}\left(  j\right)  $, with
$k\in\lbrack i_{h_{k}}^{\ast}]$ and $j\in\lbrack i_{h_{j}}^{\ast}]$, it cannot
be the case that
\[
h_{j}>h_{k}%
\]
Thus, $h_{k}\geq h_{j}$, $w\left(  i_{h_{k}}^{\ast}\right)  \geq w\left(
i_{h_{j}}^{\ast}\right)  $, and $w\left(  k\right)  \geq w\left(  j\right)  $.
Therefore there exists $f:\bar{u}\left(  A\right)  \rightarrow\left(
0,\infty\right)  $ increasing and such that
\[
f\circ\bar{u}=w
\]
Thus,%
\begin{align*}
f\left(  \bar{u}\left(  i\right)  \right)   &  >f\left(  \bar{u}\left(
j\right)  \right)  \iff\rho\left(  i\mid j\right)  =1\quad\text{;\quad
}f\left(  \bar{u}\left(  i\right)  \right)  =f\left(  \bar{u}\left(
j\right)  \right)  \iff\rho\left(  i\mid j\right)  \in\left(  0,1\right) \\
f\left(  \bar{u}\left(  i\right)  \right)   &  <f\left(  \bar{u}\left(
j\right)  \right)  \iff\rho\left(  i\mid j\right)  =0
\end{align*}
For all $j\neq k$ in $A$ such that $f\left(  \bar{u}\left(  j\right)
\right)  =f\left(  \bar{u}\left(  k\right)  \right)  $, we have $\rho\left(
j\mid k\right)  \in\left(  0,1\right)  $, and so $j\parallel^{\ast}k$, then
there exists $h=1,...,m$ such that, for each $j,k\in\left[  i_{h}^{\ast
}\right]  $,%
\[
\frac{\bar{u}\left(  j\right)  }{\bar{u}\left(  j\right)  +\bar{u}\left(
k\right)  }=\frac{1}{1+\frac{\bar{u}\left(  k\right)  }{\bar{u}\left(
j\right)  }}=\frac{1}{1+\frac{\sigma_{h}u^{\ast}\left(  k\right)  }{\sigma
_{h}u^{\ast}\left(  j\right)  }}=\frac{1}{1+\frac{\rho\left(  k\mid j\right)
}{\rho\left(  j\mid k\right)  }}=\frac{\rho\left(  j\mid k\right)  }%
{\rho\left(  j\mid k\right)  +\rho\left(  k\mid j\right)  }%
\]
and%
\[
\rho\left(  j\mid k\right)  =\,\underset{=s\left(  j,k\right)  }%
{\underbrace{(\rho\left(  j\mid k\right)  +\rho\left(  k\mid j\right)  )}%
}\frac{\bar{u}\left(  j\right)  }{\bar{u}\left(  j\right)  +\bar{u}\left(
k\right)  }%
\]
By setting $s\left(  j,k\right)  =1$ if $f\left(  \bar{u}\left(  i\right)
\right)  \neq f\left(  \bar{u}\left(  j\right)  \right)  $ we conclude the argument.

Since (iii) trivially implies (ii), it remains to prove that (ii) implies (i).
Let $u$, $w$ and $s$ represent $\rho$ as in (ii). We have already observed
that $w$ represents $\succ^{\ast}$.

For any triplet $i,j,k$ of distinct elements of $A$, consider the two products%
\[
\rho\left(  j\mid i\right)  \rho\left(  k\mid j\right)  \rho\left(  i\mid
k\right)  \qquad\text{and}\qquad\rho\left(  k\mid i\right)  \rho\left(  j\mid
k\right)  \rho\left(  i\mid j\right)
\]

Suppose first that $i,j,k$ do not belong to the same level set of $w$. Without
loss of generality, we can then set $\rho\left(  j\mid i\right)  =0$. Hence,
$\rho\left(  i\mid j\right)  =1$ and so $i\succ^{\ast}j$, that is, $w\left(
i\right)  >w\left(  j\right)  $. There are two cases to consider.

\begin{enumerate}
\item[(1)] If $w\left(  k\right)  \geq w\left(  i\right)  $, then $w\left(
k\right)  >w\left(  j\right)  $ and so $\rho\left(  j\mid k\right)  =0$.

\item[(2)] Else $w\left(  i\right)  >w\left(  k\right)  $, then $\rho\left(
k\mid i\right)  =0$.
\end{enumerate}

In both cases, the two products are null, so equal. Next suppose that $i,j,k$
belong to the same level set of $w$. Then,%
\begin{align*}
\rho\left(  j\mid i\right)  \rho\left(  k\mid j\right)  \rho\left(  i\mid
k\right)   &  =s\left(  j,i\right)  \dfrac{u\left(  j\right)  }{u\left(
j\right)  +u\left(  i\right)  }s\left(  k,j\right)  \dfrac{u\left(  k\right)
}{u\left(  k\right)  +u\left(  j\right)  }s\left(  i,k\right)  \dfrac{u\left(
i\right)  }{u\left(  i\right)  +u\left(  k\right)  }\\
&  =s\left(  k,i\right)  \dfrac{u\left(  i\right)  }{u\left(  k\right)
+u\left(  i\right)  }s\left(  j,k\right)  \dfrac{u\left(  k\right)  }{u\left(
j\right)  +u\left(  k\right)  }s\left(  i,j\right)  \dfrac{u\left(  j\right)
}{u\left(  i\right)  +u\left(  j\right)  }\\
&  =s\left(  k,i\right)  \dfrac{u\left(  k\right)  }{u\left(  k\right)
+u\left(  i\right)  }s\left(  j,k\right)  \dfrac{u\left(  j\right)  }{u\left(
j\right)  +u\left(  k\right)  }s\left(  i,j\right)  \dfrac{u\left(  i\right)
}{u\left(  i\right)  +u\left(  j\right)  }\\
&  =\rho\left(  k\mid i\right)  \rho\left(  j\mid k\right)  \rho\left(  i\mid
j\right)
\end{align*}
We conclude that $\rho$ is transitive.\hfill$\blacksquare$

\subsubsection{Theorem \ref{prop:chrono}}

\textquotedblleft Only if.\textquotedblright\ If a tandem has a binary value
representation $\left(  v,w,s,\varphi\right)  $, then $\rho$ is transitive
(see Theorem \ref{thm:value}). Consider the set
\begin{align*}
\mathbb{D}  &  =\left\{  \left(  i,j\right)  :\tau\left(  i\mid j\right)
\neq0\right\}  =\left\{  \left(  i,j\right)  :\rho\left(  i\mid j\right)
\in\left(  0,1\right)  \right\} \\
&  =\left\{  \left(  i,j\right)  :i\parallel^{\ast}j\right\}  =\left\{
\left(  i,j\right)  :w\left(  i\right)  =w\left(  j\right)  \right\}
\end{align*}
Note that for all $\left(  i,j\right)  \in\mathbb{D}$,
\[
\ell_{ij}=\ln\frac{\rho\left(  i\mid j\right)  }{\rho\left(  j\mid i\right)
}=v\left(  i\right)  -v\left(  j\right)
\]
thus (\ref{eq:response-time-bis}) delivers%
\[
\ell_{ij}=\ell_{hk}\implies v\left(  i\right)  -v\left(  j\right)  =v\left(
h\right)  -v\left(  k\right)  \implies\tau\left(  i\mid j\right)  =\tau\left(
h\mid k\right)
\]
for all $\left(  i,j\right)  ,\left(  h,k\right)  \in\mathbb{D}$. Thus
(\ref{eq:qe}) is satisfied.

Since $\varphi$ is strictly quasiconcave and unimodal, then there exists a
unique $l\in\mathbb{R}$ such that $\varphi$ is strictly increasing on $\left(
-\infty,l\right]  $ and strictly decreasing on $\left[  l,\infty\right)  $,
then
\begin{align*}
l  &  \leq\ell_{ij}<\ell_{hk}\implies l\leq v\left(  i\right)  -v\left(
j\right)  <v\left(  h\right)  -v\left(  k\right) \\
&  \implies\varphi\left(  v\left(  i\right)  -v\left(  j\right)  \right)
>\varphi\left(  v\left(  h\right)  -v\left(  k\right)  \right)  \implies
\tau\left(  i\mid j\right)  >\tau\left(  h\mid k\right)
\end{align*}
and
\begin{align*}
\ell_{ij}  &  <\ell_{hk}\leq l\implies v\left(  i\right)  -v\left(  j\right)
<v\left(  h\right)  -v\left(  k\right)  \leq l\\
&  \implies\varphi\left(  v\left(  i\right)  -v\left(  j\right)  \right)
<\varphi\left(  v\left(  h\right)  -v\left(  k\right)  \right)  \implies
\tau\left(  i\mid j\right)  <\tau\left(  h\mid k\right)
\end{align*}
for all $\left(  i,j\right)  ,\left(  h,k\right)  \in\mathbb{D}$. Thus
(\ref{eq:qi}) and (\ref{eq:qo}) are satisfied, as desired.

\textquotedblleft If.\textquotedblright\ If a tandem is chronometric then
$\rho$ is transitive and there exist $v,w:A\rightarrow\mathbb{R}$ and a
symmetric $s:A^{2}\rightarrow\left(  0,\infty\right)  $ such that
(\ref{eq:stoch-utt}) holds. Consider the set
\begin{align*}
\mathbb{D}  &  =\left\{  \left(  i,j\right)  :\tau\left(  i\mid j\right)
\neq0\right\}  =\left\{  \left(  i,j\right)  :\rho\left(  i\mid j\right)
\in\left(  0,1\right)  \right\} \\
&  =\left\{  \left(  i,j\right)  :i\parallel^{\ast}j\right\}  =\left\{
\left(  i,j\right)  :w\left(  i\right)  =w\left(  j\right)  \right\}
\end{align*}
Note that for all $\left(  i,j\right)  \in\mathbb{D}$,
\[
\ell_{ij}=\ln\frac{\rho\left(  i\mid j\right)  }{\rho\left(  j\mid i\right)
}=v\left(  i\right)  -v\left(  j\right)
\]
and set $L=\left\{  \ell_{ij}:\left(  i,j\right)  \in\mathbb{D}\right\}
=\left\{  v\left(  i\right)  -v\left(  j\right)  :\left(  i,j\right)
\in\mathbb{D}\right\}  $. With this, (\ref{eq:qe}) implies that there exists
$\psi:L\rightarrow\left(  0,\infty\right)  $ such that%
\[
\tau\left(  i\mid j\right)  =\psi\left(  \ell_{ij}\right)  =\psi\left(
v\left(  i\right)  -v\left(  j\right)  \right)
\]
for all $\left(  i,j\right)  \in\mathbb{D}$. Moreover, by (\ref{eq:qi}), if
$x,y\in L$ are such that $l\leq x<y$, taking $\left(  i,j\right)  $ and
$\left(  h,k\right)  \ $in $\mathbb{D}$ such that $x=\ell_{ij}$ and
$y=\ell_{hk}$, it follows that
\[
l\leq\ell_{ij}<\ell_{hk}\implies\tau\left(  i\mid j\right)  >\tau\left(  h\mid
k\right)  \implies\psi\left(  \ell_{ij}\right)  >\psi\left(  \ell_{hk}\right)
\implies\psi\left(  x\right)  >\psi\left(  y\right)
\]
Analogously, by (\ref{eq:qo}), if $x,y\in L$ are such that $x<y\leq l$, taking
$\left(  i,j\right)  $ and $\left(  h,k\right)  \ $in $\mathbb{D}$ such that
$x=\ell_{ij}$ and $y=\ell_{hk}$, it follows that
\[
\ell_{ij}<\ell_{hk}\leq l\implies\tau\left(  i\mid j\right)  <\tau\left(
h\mid k\right)  \implies\psi\left(  \ell_{ij}\right)  <\psi\left(  \ell
_{hk}\right)  \implies\psi\left(  x\right)  <\psi\left(  y\right)
\]
Summing up, $L$ is a finite subset of $\mathbb{R}$ and $\psi:L\rightarrow
\left(  0,\infty\right)  $ is such that there exists $l\in\mathbb{R}$ for
which%
\begin{align*}
l  &  \leq x<y\implies\psi\left(  x\right)  >\psi\left(  y\right) \\
x  &  <y\leq l\implies\psi\left(  x\right)  <\psi\left(  y\right)
\end{align*}
for all $x,y\in L$.

This allows to extend $\psi$ to a function $\varphi:\mathbb{R}\rightarrow
\left(  0,\infty\right)  $ such that $\varphi$ is strictly increasing on
$\left(  -\infty,l\right]  $ and strictly decreasing on $\left[
l,\infty\right)  $. Thus there exists a strictly quasiconcave and unimodal
$\varphi:\mathbb{R}\rightarrow\left(  0,\infty\right)  $ such that
\[
\tau\left(  i\mid j\right)  =\varphi\left(  v\left(  i\right)  -v\left(
j\right)  \right)
\]
if $w\left(  i\right)  =w\left(  j\right)  $.

Finally, if $w\left(  i\right)  \neq w\left(  j\right)  $, by
(\ref{eq:stoch-utt}), $\rho\left(  i\mid j\right)  \in\left\{  0,1\right\}  $
and by definition of tandem $\tau\left(  i\mid j\right)  =0$.\hfill
$\blacksquare$

\subsubsection{Theorem \ref{prop:psycho}}

Note that if either $\left(  \rho,\tau\right)  $ has a binary value
representation or it is psychometric, then $\rho$ is transitive and there
exist $v,w:A\rightarrow\mathbb{R}$ and a symmetric $s:A^{2}\rightarrow\left(
0,\infty\right)  $ such that (\ref{eq:stoch-utt}) holds. Therefore the set of
pairs of alternatives with nonzero response time is
\begin{align*}
\mathbb{D}  &  =\left\{  \left(  i,j\right)  :\tau\left(  i\mid j\right)
\neq0\right\}  =\left\{  \left(  i,j\right)  :\rho\left(  i\mid j\right)
\in\left(  0,1\right)  \right\} \\
&  =\left\{  \left(  i,j\right)  :i\parallel^{\ast}j\right\}  =\left\{
\left(  i,j\right)  :w\left(  i\right)  =w\left(  j\right)  \right\}
\end{align*}
Arbitrarily choose $\left(  i,j\right)  \in\mathbb{D}$, since $i\parallel
^{\ast}j$, by Lemma \ref{lm:bcp-prop}, we can assume, without loss of
generality, that $i\succsim^{%
{{}^\circ}%
}j$, thus%
\[
\rho\left(  i\mid j\right)  \geq\rho\left(  j\mid i\right)  \quad
\text{and\quad}1-\rho\left(  j\mid i\right)  \geq1-\rho\left(  i\mid j\right)
\]
A second-type error is the probability of accepting an inferior proposal, that
is,%
\[
\mathrm{ER}_{i,j}^{\mathrm{II}}=\rho\left(  j\mid i\right)  =\min\left\{
\rho\left(  i\mid j\right)  ,\rho\left(  j\mid i\right)  \right\}
\]
A first-type error is the probability of rejecting a superior proposal, that
is,%
\[
\mathrm{ER}_{i,j}^{\mathrm{I}}=1-\rho\left(  i\mid j\right)  =\min\left\{
1-\rho\left(  i\mid j\right)  ,1-\rho\left(  j\mid i\right)  \right\}
\]

Since (\ref{eq:stoch-utt}) holds, $i\succsim^{%
{{}^\circ}%
}j$ if and only if $v\left(  i\right)  \geq v\left(  j\right)  $. Therefore:%
\begin{align*}
\mathrm{ER}_{i,j}^{\mathrm{II}}  &  =\rho\left(  j\mid i\right)  =s\left(
j,i\right)  \frac{1}{1+e^{-\left(  v\left(  j\right)  -v\left(  i\right)
\right)  }}=s\left(  j,i\right)  \frac{1}{1+e^{\left\vert v\left(  j\right)
-v\left(  i\right)  \right\vert }}\\
&  =s\left(  i,j\right)  \frac{1}{1+e^{\left\vert v\left(  i\right)  -v\left(
j\right)  \right\vert }}\\
\mathrm{ER}_{i,j}^{\mathrm{I}}  &  =1-\rho\left(  i\mid j\right)  =1-s\left(
i,j\right)  \frac{1}{1+e^{-\left(  v\left(  i\right)  -v\left(  j\right)
\right)  }}=1-s\left(  i,j\right)  \frac{1}{1+e^{-\left\vert v\left(
i\right)  -v\left(  j\right)  \right\vert }}\\
&  =1-\mathrm{ER}_{i,j}^{\mathrm{II}}\frac{1+e^{\left\vert v\left(  i\right)
-v\left(  j\right)  \right\vert }}{1+e^{-\left\vert v\left(  i\right)
-v\left(  j\right)  \right\vert }}%
\end{align*}
Summing up, for each $\left(  i,j\right)  \in\mathbb{D}$,%
\begin{align*}
\mathrm{ER}_{i,j}^{\mathrm{II}}  &  =s\left(  i,j\right)  \frac{1}%
{1+e^{\left\vert v\left(  i\right)  -v\left(  j\right)  \right\vert }}\\
\mathrm{ER}_{i,j}^{\mathrm{I}}  &  =1-\mathrm{ER}_{i,j}^{\mathrm{II}}%
\frac{1+e^{\left\vert v\left(  i\right)  -v\left(  j\right)  \right\vert }%
}{1+e^{-\left\vert v\left(  i\right)  -v\left(  j\right)  \right\vert }}%
\end{align*}
But then, for all $\left(  i,j\right)  $ and $\left(  h,k\right)  $ in
$\mathbb{D}$ such that $\mathrm{ER}_{i,j}^{\mathrm{I}}<\mathrm{ER}%
_{h,k}^{\mathrm{I}}\ $and\ $\mathrm{ER}_{i,j}^{\mathrm{II}}<\mathrm{ER}%
_{h,k}^{\mathrm{II}}$, it follows that%
\begin{gather*}
1-\mathrm{ER}_{i,j}^{\mathrm{II}}\frac{1+e^{\left\vert v\left(  i\right)
-v\left(  j\right)  \right\vert }}{1+e^{-\left\vert v\left(  i\right)
-v\left(  j\right)  \right\vert }}<1-\mathrm{ER}_{h,k}^{\mathrm{II}}%
\frac{1+e^{\left\vert v\left(  h\right)  -v\left(  k\right)  \right\vert }%
}{1+e^{-\left\vert v\left(  h\right)  -v\left(  k\right)  \right\vert }}\\
\mathrm{ER}_{h,k}^{\mathrm{II}}\frac{1+e^{\left\vert v\left(  h\right)
-v\left(  k\right)  \right\vert }}{1+e^{-\left\vert v\left(  h\right)
-v\left(  k\right)  \right\vert }}<\mathrm{ER}_{i,j}^{\mathrm{II}}%
\frac{1+e^{\left\vert v\left(  i\right)  -v\left(  j\right)  \right\vert }%
}{1+e^{-\left\vert v\left(  i\right)  -v\left(  j\right)  \right\vert }}\\
\frac{\frac{1+e^{\left\vert v\left(  h\right)  -v\left(  k\right)  \right\vert
}}{1+e^{-\left\vert v\left(  h\right)  -v\left(  k\right)  \right\vert }}%
}{\frac{1+e^{\left\vert v\left(  i\right)  -v\left(  j\right)  \right\vert }%
}{1+e^{-\left\vert v\left(  i\right)  -v\left(  j\right)  \right\vert }}%
}<\frac{\mathrm{ER}_{i,j}^{\mathrm{II}}}{\mathrm{ER}_{h,k}^{\mathrm{II}}}<1\\
\frac{1+e^{\left\vert v\left(  h\right)  -v\left(  k\right)  \right\vert }%
}{1+e^{-\left\vert v\left(  h\right)  -v\left(  k\right)  \right\vert }}%
<\frac{1+e^{\left\vert v\left(  i\right)  -v\left(  j\right)  \right\vert }%
}{1+e^{-\left\vert v\left(  i\right)  -v\left(  j\right)  \right\vert }}\\
\left\vert v\left(  h\right)  -v\left(  k\right)  \right\vert <\left\vert
v\left(  i\right)  -v\left(  j\right)  \right\vert
\end{gather*}
in other words%
\begin{equation}
\mathrm{ER}_{i,j}^{\mathrm{I}}<\mathrm{ER}_{h,k}^{\mathrm{I}}\quad
\text{and\quad}\mathrm{ER}_{i,j}^{\mathrm{II}}<\mathrm{ER}_{h,k}^{\mathrm{II}%
}\Longrightarrow\left\vert v\left(  i\right)  -v\left(  j\right)  \right\vert
>\left\vert v\left(  h\right)  -v\left(  k\right)  \right\vert \label{eq:due}%
\end{equation}
A similar argument shows that
\begin{equation}
\mathrm{ER}_{i,j}^{\mathrm{I}}\leq\mathrm{ER}_{h,k}^{\mathrm{I}}%
\quad\text{and\quad}\mathrm{ER}_{i,j}^{\mathrm{II}}\leq\mathrm{ER}%
_{h,k}^{\mathrm{II}}\Longrightarrow\left\vert v\left(  i\right)  -v\left(
j\right)  \right\vert \geq\left\vert v\left(  h\right)  -v\left(  k\right)
\right\vert \label{eq:due-bis}%
\end{equation}

\textquotedblleft If.\textquotedblright\ If $\left(  \rho,\tau\right)  $ is
psychometric, then (\ref{eq:due}) and (\ref{eq:due-bis}) imply%
\begin{equation}
\tau\left(  i\mid j\right)  <\tau\left(  h\mid k\right)  \Longrightarrow
\left\vert v\left(  i\right)  -v\left(  j\right)  \right\vert >\left\vert
v\left(  h\right)  -v\left(  k\right)  \right\vert \label{eq:tre-bis}%
\end{equation}
and%
\begin{equation}
\tau\left(  i\mid j\right)  \leq\tau\left(  h\mid k\right)  \Longrightarrow
\left\vert v\left(  i\right)  -v\left(  j\right)  \right\vert \geq\left\vert
v\left(  h\right)  -v\left(  k\right)  \right\vert \label{eq:quattro-bis}%
\end{equation}
for all $\left(  i,j\right)  $ and $\left(  h,k\right)  $ in $\mathbb{D}$. As
a result%
\[
\left\vert v\left(  i\right)  -v\left(  j\right)  \right\vert \geq\left\vert
v\left(  h\right)  -v\left(  k\right)  \right\vert \Longleftrightarrow
\tau\left(  i\mid j\right)  \leq\tau\left(  h\mid k\right)
\]
Therefore, setting $M=\left\{  \left\vert v\left(  i\right)  -v\left(
j\right)  \right\vert :\left(  i,j\right)  \in\mathbb{D}\right\}  $, there is
a strictly decreasing function $\psi:M\rightarrow\left(  0,\infty\right)  $
such that $\tau\left(  i\mid j\right)  =\psi\left(  \left\vert v\left(
i\right)  -v\left(  j\right)  \right\vert \right)  $. We can first extend
$\psi$ from $M$ to $\left[  0,\infty\right)  $ as a strictly decreasing
function and then set $\varphi\left(  x\right)  =\psi\left(  \left\vert
x\right\vert \right)  $ for all $x\in\mathbb{R}$. With this, there exists a
strictly quasiconcave, unimodal, and even $\varphi:\mathbb{R}\rightarrow
\left(  0,\infty\right)  $ such that
\[
\tau\left(  i\mid j\right)  =\varphi\left(  v\left(  i\right)  -v\left(
j\right)  \right)
\]
if $w\left(  i\right)  =w\left(  j\right)  $.

Finally, if $w\left(  i\right)  \neq w\left(  j\right)  $, by
(\ref{eq:stoch-utt}), $\rho\left(  i\mid j\right)  \in\left\{  0,1\right\}  $
and by definition of tandem $\tau\left(  i\mid j\right)  =0$.

\textquotedblleft Only if.\textquotedblright\ If a tandem has a binary value
representation $\left(  v,w,s,\varphi\right)  $, then $\rho$ is transitive
(see Theorem \ref{thm:value}). Now $\varphi$ is strictly quasiconcave,
unimodal, and even $\varphi:\mathbb{R}\rightarrow\left(  0,\infty\right)  $,
with strong maximum at $0$ and strictly decreasing on $\left[  0,\infty
\right)  $. In particular,%
\[
\tau\left(  i\mid j\right)  =\tau\left(  j\mid i\right)
\]
for all alternatives $i$ and $j$. But then $\rho$ is unbiased, and so
$s\left(  i,j\right)  =1$ for all $\left(  i,j\right)  \in\mathbb{D}$, and so
\[
\mathrm{ER}_{i,j}^{\mathrm{I}}=\mathrm{ER}_{i,j}^{\mathrm{II}}=\frac
{1}{1+e^{\left\vert v\left(  i\right)  -v\left(  j\right)  \right\vert }}%
\]
Moreover, for all $\left(  i,j\right)  $ and $\left(  h,k\right)  $ in
$\mathbb{D}$,
\begin{align*}
\left.  \tau\left(  i\mid j\right)  <\tau\left(  h\mid k\right)  \right.   &
\iff\varphi\left(  \left\vert v\left(  i\right)  -v\left(  j\right)
\right\vert \right)  <\varphi\left(  \left\vert v\left(  h\right)  -v\left(
k\right)  \right\vert \right) \\
&  \iff\left\vert v\left(  i\right)  -v\left(  j\right)  \right\vert
>\left\vert v\left(  h\right)  -v\left(  k\right)  \right\vert \\
&  \iff\mathrm{ER}_{i,j}<\mathrm{ER}_{h,k}%
\end{align*}
This proves that $\left(  \rho,\tau\right)  $ is psychometric.\hfill
$\blacksquare$

\subsection{Section \ref{sect:algo}}

\subsubsection{Subsection \ref{sect:bin}}

To ease notation, we set $\Lambda_{ij}=\nu\left(  i\right)  -\nu\left(
j\right)  $ as well as%
\[
\rho_{\mathrm{C}}\left(  i\mid j\right)  =\rho_{ij}\quad\text{;}\quad
\rho_{\mathrm{C}}\left(  j\mid i\right)  =\rho_{ji}\quad\text{;}\quad
\tau_{\mathrm{RT}}\left(  i\mid j\right)  =\tau_{ij}\quad\text{;}\quad
\tau_{\mathrm{RT}}\left(  j\mid i\right)  =\tau_{ji}\quad\text{;}\quad
s_{\mathrm{C}}\left(  i,j\right)  =s_{ij}%
\]
By Theorems 8.1 and 8.2 of Pinsky and Karlin (2011),%
\begin{equation}
\rho_{ij}=\frac{1-e^{\beta\Lambda_{ij}}}{e^{-\lambda\Lambda_{ij}}%
-e^{\beta\Lambda_{ij}}}\quad\text{and\quad}\tau_{ij}=\frac{1}{\Lambda_{ij}%
}\left[  \rho_{ij}\left(  \lambda+\beta\right)  -\beta\right]
\label{eq:zero-ddm}%
\end{equation}
also note that
\[
\rho_{ij}=\frac{1-e^{-\beta\Lambda_{ij}}}{1-e^{-\left(  \lambda+\beta\right)
\Lambda_{ij}}}%
\]

We begin with a few preliminary lemmas.

\begin{lemma}
\label{lm:uno-ddm}For each $\left(  i,j\right)  \in A^{2}$, it holds
\begin{equation}
\lambda\Lambda_{ij}=\ln\frac{\rho_{ij}}{\rho_{ji}}\quad\text{;}\quad
\beta\Lambda_{ij}=\ln\frac{1-\rho_{ji}}{1-\rho_{ij}} \label{eq:uno-ddm}%
\end{equation}

\end{lemma}

\noindent\textbf{Proof} Let $\left(  i,j\right)  \in A^{2}$. We have%
\[
\frac{\rho_{ij}}{\rho_{ji}}=\frac{\frac{1-e^{\beta\Lambda_{ij}}}%
{e^{-\lambda\Lambda_{ij}}-e^{\beta\Lambda_{ij}}}}{\frac{1-e^{-\beta
\Lambda_{ij}}}{e^{\lambda\Lambda_{ij}}-e^{-\beta\Lambda_{ij}}}}=e^{\lambda
\Lambda_{ij}}%
\]
and so $\lambda\Lambda_{i,j}=\ln\rho_{ij}/\rho_{ji}$. We also have%
\[
\frac{1-\rho_{ji}}{1-\rho_{ij}}=\frac{1-\frac{1-e^{-\beta\Lambda_{ij}}%
}{e^{\lambda\Lambda_{ij}}-e^{-\beta\Lambda_{ij}}}}{1-\frac{1-e^{\beta
\Lambda_{ij}}}{e^{-\lambda\Lambda_{ij}}-e^{\beta\Lambda_{ij}}}}=\frac
{\frac{e^{\lambda\Lambda_{ij}}-1}{e^{\lambda\Lambda_{ij}}-e^{-\beta
\Lambda_{ij}}}}{\frac{1-e^{-\lambda\Lambda_{ij}}}{e^{\beta\Lambda_{ij}%
}-e^{-\lambda\Lambda_{ij}}}}=e^{\beta\Lambda_{ij}}%
\]
and so $\beta\Lambda_{i,j}=\ln\left(  1-\rho_{ji}\right)  /\left(  1-\rho
_{ij}\right)  $.\hfill$\blacksquare$

\begin{lemma}
Let $\left(  i,j\right)  \in A^{2}$ with $\Lambda_{ij}\neq0$. It holds%
\begin{equation}
\tau_{ij}=\frac{\lambda^{2}}{\ln\rho_{ij}-\ln\rho_{ji}}\left[  \rho_{ij}%
+\frac{\ln\left(  1-\rho_{ji}\right)  -\ln\left(  1-\rho_{ij}\right)  }%
{\ln\rho_{ij}-\ln\rho_{ji}}\left(  \rho_{ij}-1\right)  \right]
\label{eq:due-ddm}%
\end{equation}

\end{lemma}

\noindent\textbf{Proof} By (\ref{eq:uno-ddm}),%
\begin{equation}
\frac{\beta}{\lambda}=\frac{\beta\Lambda_{ij}}{\lambda\Lambda_{ij}}=\frac
{\ln\frac{1-\rho_{ji}}{1-\rho_{ij}}}{\ln\frac{\rho_{ij}}{\rho_{ji}}}=\frac
{\ln\left(  1-\rho_{ji}\right)  -\ln\left(  1-\rho_{ij}\right)  }{\ln\rho
_{ij}-\ln\rho_{ji}} \label{eq:tre-ddm}%
\end{equation}
Hence,%
\begin{align*}
\tau_{ij}  &  =\frac{1}{\Lambda_{ij}}\left[  \rho_{ij}\left(  \lambda
+\beta\right)  -\beta\right]  =\frac{\lambda^{2}}{\lambda\Lambda_{ij}}\left[
\rho_{ij}\left(  1+\frac{\beta}{\lambda}\right)  -\frac{\beta}{\lambda
}\right]  =\frac{\lambda^{2}}{\lambda\Lambda_{ij}}\left[  \rho_{ij}%
+\frac{\beta}{\lambda}\left(  \rho_{ij}-1\right)  \right] \\
&  =\frac{\lambda^{2}}{\ln\rho_{ij}-\ln\rho_{ji}}\left[  \rho_{ij}+\frac
{\ln\left(  1-\rho_{ji}\right)  -\ln\left(  1-\rho_{ij}\right)  }{\ln\rho
_{ij}-\ln\rho_{ji}}\left(  \rho_{ij}-1\right)  \right]
\end{align*}
as desired.\hfill$\blacksquare$

\begin{lemma}
\label{lm:tre-ddm}Let $\left(  i,j\right)  \in A^{2}$ with $\Lambda_{ij}\neq
0$. If $\tau_{ij}=\tau_{ji}$, then $\beta=\lambda$.
\end{lemma}

\noindent\textbf{Proof} To further ease notation, set $x=\rho_{ij}$ and
$y=\rho_{ji}$. Since $\Lambda_{ij}\neq0$, by (\ref{eq:uno-ddm}) we have $x\neq
y$. By (\ref{eq:due-ddm}), we have%

\begin{align*}
\tau_{ij}  &  =\tau_{ji}\\
&  \Longleftrightarrow\frac{\lambda^{2}}{\ln\frac{x}{y}}\left[  x+\frac
{\ln\frac{1-y}{1-x}}{\ln\frac{x}{y}}\left(  x-1\right)  \right]
=\frac{\lambda^{2}}{\ln\frac{y}{x}}\left[  y+\frac{\ln\frac{1-x}{1-y}}%
{\ln\frac{y}{x}}\left(  y-1\right)  \right] \\
&  \Longleftrightarrow\frac{\ln\frac{y}{x}}{\ln\frac{x}{y}}\left[  x+\frac
{\ln\frac{1-y}{1-x}}{\ln\frac{x}{y}}\left(  x-1\right)  \right]  =y+\frac
{\ln\frac{1-x}{1-y}}{\ln\frac{y}{x}}\left(  y-1\right) \\
&  \Longleftrightarrow-x-\frac{\ln\frac{1-y}{1-x}}{\ln\frac{x}{y}}\left(
x-1\right)  =y+\frac{\ln\frac{1-x}{1-y}}{\ln\frac{y}{x}}\left(  y-1\right) \\
&  \Longleftrightarrow\frac{\ln\frac{1-y}{1-x}}{\ln\frac{y}{x}}\left(
y-1\right)  +\frac{\ln\frac{1-y}{1-x}}{\ln\frac{y}{x}}\left(  x-1\right)
=y+x\Longleftrightarrow\frac{2}{x+y}+\frac{\ln\frac{y}{x}}{\ln\frac{1-y}{1-x}%
}=1
\end{align*}
The locus of pairs $\left(  x,y\right)  \in\left(  0,1\right)  \times\left(
0,1\right)  $, with $x\neq y$, that solve this equation is%
\[
\left\{  \left(  x,y\right)  \in\left(  0,1\right)  _{\neq}^{2}:x=1-y\right\}
\]
Thus, $\rho_{ij}=1-\rho_{ji}$. By (\ref{eq:uno-ddm}),
\[
\frac{\beta}{\lambda}=\frac{\beta\Lambda_{ij}}{\lambda\Lambda_{ij}}=\frac
{\ln\frac{1-\rho_{ji}}{1-\rho_{ij}}}{\ln\frac{\rho_{ij}}{\rho_{ji}}}=\frac
{\ln\frac{\rho_{ij}}{\rho_{ji}}}{\ln\frac{\rho_{ij}}{\rho_{ji}}}=1
\]
We conclude that $\beta=\lambda$.\hfill$\blacksquare$

\paragraph{Proof of Proposition \ref{prop:ddm_bcm}}

Positivity of $\rho=\rho_{\mathrm{C}}$ follows immediately from
(\ref{eq:zero-ddm}) and hence $\rho$ is a binary choice probability. So, to
establish whether $\left(  \rho,\tau\right)  $ is a tandem we need to check
only condition (\ref{eq:tao-rho2}). When $\beta=\lambda$, this condition
trivially holds because $\rho$ is unbiased and $\tau$ is symmetric. Since
$\nu$ is injective, we have $\Lambda_{ij}\neq0$ for all distinct $i$ and $j$
in $A$. By Lemma \ref{lm:tre-ddm}, it must then be the case that $\tau
_{ij}\neq\tau_{ji}$; so, condition (\ref{eq:tao-rho2}) now vacuously holds. We
conclude that $\left(  \rho,\tau\right)  $ is a tandem. Finally, the
transitivity of $\rho$ follows by Theorem \ref{thm:value-bis} because Baldassi
et al. (2020) show that, given any nice exploration matrix $Q$, the transition
matrix $M$ is reversible.\footnote{Of course, it can also be verified by brute
force from (\ref{eq:zero-ddm}).}

We now turn to the binary value representation. By positivity, $w_{\mathrm{C}%
}$ is constant. For all $i\ $and $j$ in $A$,%
\[
s_{ij}=\rho_{ij}+\rho_{ji}=\dfrac{1-e^{-\beta\Lambda_{ij}}}{1-e^{-\left(
\lambda+\beta\right)  \Lambda_{ij}}}+\dfrac{1-e^{\beta\Lambda_{ij}}%
}{1-e^{\left(  \lambda+\beta\right)  \Lambda_{ij}}}=1+\dfrac{e^{\lambda
\Lambda_{ij}}-e^{\beta\Lambda_{ij}}}{1-e^{\left(  \lambda+\beta\right)
\Lambda_{ij}}}%
\]
by symmetry%
\[
s_{ij}=s_{ji}=1+\dfrac{e^{\lambda\left(  -\Lambda_{ij}\right)  }%
-e^{\beta\left(  -\Lambda_{ij}\right)  }}{1-e^{\left(  \lambda+\beta\right)
\left(  -\Lambda_{ij}\right)  }}%
\]
and so%
\[
s_{\mathrm{C}}\left(  i,j\right)  =s_{ij}=1+\dfrac{e^{\lambda\Lambda_{ij}%
}-e^{\beta\Lambda_{ij}}}{1-e^{\left(  \lambda+\beta\right)  \Lambda_{ij}}%
}=1+\dfrac{e^{\lambda\left\vert \Lambda_{ij}\right\vert }-e^{\beta\left\vert
\Lambda_{ij}\right\vert }}{1-e^{\left(  \lambda+\beta\right)  \left\vert
\Lambda_{ij}\right\vert }}%
\]
Moreover,%
\[
s_{ij}\frac{1}{1+e^{-\lambda\Lambda_{ij}}}=\left(  1+\dfrac{e^{\lambda
\Lambda_{ij}}-e^{\beta\Lambda_{ij}}}{1-e^{\left(  \lambda+\beta\right)
\Lambda_{ij}}}\right)  \frac{1}{1+e^{-\lambda\Lambda_{ij}}}=\dfrac
{1-e^{-\beta\Lambda_{ij}}}{1-e^{-\left(  \lambda+\beta\right)  \Lambda_{ij}}%
}=\rho_{ij}%
\]
This proves that $v_{\mathrm{C}}=\lambda\nu$, thus completing the proof of
(\ref{eq:bcp-uno}).

As to (\ref{eq:bcp-due}),
\begin{align*}
\tau_{ij}  &  =\frac{\lambda^{2}}{\lambda\Lambda_{ij}}\left[  \rho_{ij}\left(
1+\frac{\beta}{\lambda}\right)  -\frac{\beta}{\lambda}\right]  =\frac
{\lambda^{2}}{\lambda\Lambda_{ij}}\left[  \frac{1-e^{\beta\Lambda_{ij}}%
}{e^{-\lambda\Lambda_{ij}}-e^{\beta\Lambda_{ij}}}\left(  1+\frac{\beta
}{\lambda}\right)  -\frac{\beta}{\lambda}\right] \\
&  =\frac{\lambda^{2}}{\lambda\Lambda_{ij}}\left[  \frac{1-e^{\frac{\beta
}{\lambda}\lambda\Lambda_{ij}}}{e^{-\lambda\Lambda_{ij}}-e^{\frac{\beta
}{\lambda}\lambda\Lambda_{ij}}}\left(  1+\frac{\beta}{\lambda}\right)
-\frac{\beta}{\lambda}\right]
\end{align*}
but $\lambda\Lambda_{ij}=v_{\mathrm{C}}\left(  i\right)  -v_{\mathrm{C}%
}\left(  j\right)  $. We can then define $\varphi_{\mathrm{RT}}:\mathbb{R}%
\rightarrow\mathbb{R}$ by%
\[
\varphi_{\mathrm{RT}}\left(  x\right)  =\frac{\lambda^{2}}{x}\left[
\frac{1-e^{\frac{\beta}{\lambda}x}}{e^{-x}-e^{\frac{\beta}{\lambda}x}}\left(
1+\frac{\beta}{\lambda}\right)  -\frac{\beta}{\lambda}\right]
\]
and obtain $\tau_{ij}=\varphi_{\mathrm{RT}}\left(  v_{\mathrm{C}}\left(
i\right)  -v_{\mathrm{C}}\left(  j\right)  \right)  $.\hfill$\blacksquare$

\paragraph{\textbf{Proof of Proposition \ref{prop:ddm-bcm}}}

By (\ref{eq:due-ddm}),%
\[
\tau_{ij}=\frac{\lambda^{2}}{\ell_{ij}}\left[  \rho_{ij}+\frac{\bar{\ell}%
_{ij}}{\ell_{ij}}\left(  \rho_{ij}-1\right)  \right]  =\frac{\lambda^{2}}%
{\ell_{ij}^{2}}\left[  \ell_{ij}\rho_{ij}+\bar{\ell}_{ij}\left(  \rho
_{ij}-1\right)  \right]
\]
thus%
\[
\lambda=\left\vert \ell_{ij}\right\vert \sqrt{\frac{\tau_{ij}}{\ell_{ij}%
\rho_{ij}+\bar{\ell}_{ij}\left(  \rho_{ij}-1\right)  }}%
\]
as desired. By (\ref{eq:tre-ddm}), $\beta=\lambda\bar{\ell}_{ij}/\ell_{ij}$.
Finally, set $\nu\left(  j^{\ast}\right)  =0$ for some alternative $j^{\ast}$.
By (\ref{eq:uno-ddm}), for each $i$ we have
\[
\nu\left(  i\right)  =\nu\left(  i\right)  -\nu\left(  j^{\ast}\right)
=\Lambda_{ij^{\ast}}=\frac{1}{\lambda}\ell_{ij^{\ast}}%
\]
concluding the proof.\hfill$\blacksquare$

\paragraph{\textbf{Proof of Proposition \ref{prop:ddm-error}}}

By Proposition \ref{prop:ddm_bcm}, the tandem $\left(  \rho_{\mathrm{C}}%
,\tau_{\mathrm{RT}}\right)  $ is chronometric.

(i) implies (iii) If $\varphi_{\mathrm{RT}}$ is even, then
\[
\tau_{\mathrm{RT}}\left(  i\mid j\right)  =\varphi_{\mathrm{RT}}\left(
v_{\mathrm{C}}\left(  i\right)  -v_{\mathrm{C}}\left(  j\right)  \right)
=\varphi_{\mathrm{RT}}\left(  v_{\mathrm{C}}\left(  j\right)  -v_{\mathrm{C}%
}\left(  i\right)  \right)  =\tau_{\mathrm{RT}}\left(  j\mid i\right)
\]
for all $i\neq j$.

(iii) implies (ii). If $\tau_{\mathrm{RT}}\left(  i\mid j\right)
=\tau_{\mathrm{RT}}\left(  j\mid i\right)  $ for some $i\neq j$, since $\nu$
is injective, then $\Lambda_{ij}\neq0$ and so, by Lemma \ref{lm:tre-ddm},
$\beta=\lambda$.

(ii) implies (iv). Indeed if $\beta=\lambda$, then
\[
s_{\mathrm{C}}\left(  i,j\right)  =1+\dfrac{e^{\lambda\left\vert \nu\left(
i\right)  -\nu\left(  j\right)  \right\vert }-e^{\beta\left\vert \nu\left(
i\right)  -\nu\left(  j\right)  \right\vert }}{1-e^{\left(  \lambda
+\beta\right)  \left\vert \nu\left(  i\right)  -\nu\left(  j\right)
\right\vert }}=1
\]
for all $i\ $and $j$, and so $\rho_{\mathrm{C}}$ is unbiased.

(iv) implies (i) If $\rho_{\mathrm{C}}\left(  i\mid j\right)  =1-\rho
_{\mathrm{C}}\left(  j\mid i\right)  $ for some $i\neq j$, then $\rho
_{\mathrm{C}}\left(  j\mid i\right)  =1-\rho_{\mathrm{C}}\left(  i\mid
j\right)  $, and, by Lemma \ref{lm:uno-ddm},%
\[
\lambda\Lambda_{ij}=\beta\Lambda_{ij}%
\]
Since $\nu$ is injective, then $\Lambda_{ij}\neq0$ and $\beta=\lambda$. In
particular,%
\[
\varphi_{\mathrm{RT}}\left(  x\right)  =\frac{\lambda^{2}}{x}\left(
2\frac{1-e^{x}}{e^{-x}-e^{x}}-1\right)  =\frac{\lambda^{2}}{x}\tanh\left(
\frac{x}{2}\right)
\]
is even.\hfill$\blacksquare$

\subsubsection{Stochastic matrices\label{sect:more-stoch-matrices}}

A sequence $a=\left\{  a_{n}\right\}  $ of non-negative scalars is
\emph{summable} if $%
{\displaystyle\sum\limits_{n=0}^{\infty}}
a_{n}<\infty$. Its \emph{generating function} given by%
\begin{equation}
f_{a}\left(  z\right)  =%
{\displaystyle\sum\limits_{n=0}^{\infty}}
a_{n}z^{n} \label{eq:Mirsky}%
\end{equation}
is defined where the power series on the right hand side converges.
Summability of $\left\{  a_{n}\right\}  $ guarantees that the radius of
convergence $R$ satisfies $R\geq1$ and that $f_{a}\left(  z\right)  $ is
defined and continuous on the unit disk$\mathbb{\ }\left\{  z\in
\mathbb{C}:\left\vert z\right\vert \leq1\right\}  $.

\begin{lemma}
\label{prop:hw}If $a=\left\{  a_{n}\right\}  $ is a non-negative and summable
sequence, then the matrix power series $%
{\displaystyle\sum\limits_{n=0}^{\infty}}
a_{n}B^{n}$ converges (entry by entry) for all stochastic matrices $B$.
\end{lemma}

\noindent\textbf{Proof} The $\left(  i,j\right)  $-th entry $b_{ij}^{(n)}$ of
the matrix $B^{n}$ belongs to $\left[  0,1\right]  $ because $B^{n}$ is a
stochastic matrix too. Then $\sum_{n=0}^{\infty}a_{n}b_{ij}^{(n)}$ is a
non-negative series such that $0\leq a_{n}b_{ij}^{(n)}\leq a_{n}$ and it
converges because $\sum_{n=0}^{\infty}a_{n}$ does.\hfill$\blacksquare\bigskip$

As a consequence the function%
\[
f_{a}\left(  B\right)  =%
{\displaystyle\sum\limits_{n=0}^{\infty}}
a_{n}B^{n}%
\]
is well defined in the strong sense of Weyr (see e.g. Rinehart, 1955), for all
stochastic matrices $B$.\bigskip

Denote by $Q$ and $M$ the exploration and transition matrices defined in the
main text, which, as observed, are stochastic.

\begin{lemma}
\label{lem:ma}Let $\rho_{\mathrm{C}}$ be positive. If $Q$ is irreducible
(quasi-positive), then $M$ is primitive (positive).
\end{lemma}

\noindent\textbf{Proof} To ease notation we write $\rho$ in place of
$\rho_{\mathrm{C}}$. Let $Q$ be irreducible. Recall that%
\begin{equation}
M\left(  i\mid j\right)  =Q\left(  i\mid j\right)  \rho\left(  i\mid j\right)
\qquad\forall i\neq j \label{eq:emme}%
\end{equation}
and $M\left(  j\mid j\right)  =1-%
{\displaystyle\sum\limits_{k\neq j}}
Q\left(  k\mid j\right)  \rho\left(  k\mid j\right)  $, for all $j\in A$.
Given any $j\in A$, since $Q$ is irreducible, then it cannot be the case that
$Q\left(  k\mid j\right)  =0$ for all $k\neq j$. Positivity of the BCM implies
that
\[
M\left(  j\mid j\right)  =1-%
{\displaystyle\sum\limits_{k\neq j}}
Q\left(  k\mid j\right)  \rho\left(  k\mid j\right)  >1-%
{\displaystyle\sum\limits_{k\neq j}}
Q\left(  k\mid j\right)  \geq1-%
{\displaystyle\sum\limits_{k\in A}}
Q\left(  k\mid j\right)  =0
\]
and so $M\left(  j\mid j\right)  >0$ for all $j\in A$.

Moreover, if $i\neq j$, then there exist $n\geq1$ and $k_{0},...,k_{n}$ in
$A$, with $k_{0}=i$, $k_{n}=j$, and $k_{h}\neq k_{h-1}$ for all $h=1,...,n$,
such that
\[
Q\left(  k_{1}\mid k_{0}\right)  Q\left(  k_{2}\mid k_{1}\right)  \cdot
\cdot\cdot Q\left(  k_{n}\mid k_{n-1}\right)  >0
\]
and positivity of the BCM implies that%
\[
M\left(  k_{1}\mid k_{0}\right)  M\left(  k_{2}\mid k_{1}\right)  \cdot
\cdot\cdot M\left(  k_{n}\mid k_{n-1}\right)  >0
\]
Together with positivity of $M$ on the diagonal, this yields primitivity of
$M$ itself.\footnote{Because $M$ is then irreducible and non-traceless.}

Finally, if $Q$ is quasi-positive, the argument above shows that $M$ is
positive on the diagonal, and (\ref{eq:emme}) shows that $M$ is positive also
off the diagonal.\hfill$\blacksquare$

\subsubsection{\textbf{Proof of Proposition \ref{lem:comp}}}

By Lemma \ref{prop:hw} the two matrix power series
\[
\sum_{n=0}^{\infty}\mathbb{P}\left[  N=n\right]  M^{n}\qquad\text{and\qquad
}\sum_{n=0}^{\infty}\mathbb{P}\left[  N>n\right]  M^{n}%
\]
converge (note that $\sum_{n=0}^{\infty}\mathbb{P}\left[  N>n\right]
=\mathbb{E}\left[  N\right]  <\infty$). Recall that, if the algorithm stops at
iteration $n\in\mathbb{N}=\left\{  0,1,...\right\}  $, it chooses the
incumbent $j_{n-1}$. By independence, the joint probability of stopping at
iteration $n$ and choosing $j\in A$ is
\[
\mathbb{P}\left[  N=n,J_{n-1}=j\right]  =\mathbb{P}\left[  N=n\right]
\mathbb{P}\left[  J_{n-1}=j\right]
\]
Now, for $n=0$ we have%
\[
\mathbb{P}\left[  J_{-1}=j\right]  =\mu_{j}=\left(  M^{0}\mu\right)  _{j}%
\]
Assume that for $n=m$ we have
\[
\mathbb{P}\left[  J_{m-1}=j\right]  =\left(  M^{m}\mu\right)  _{j}%
\]
Then, for $n=m+1$ we have
\begin{align*}
\mathbb{P}\left[  J_{\left(  m+1\right)  -1}=j\right]   &  =\mathbb{P}\left[
J_{m}=j\right]  =\sum_{i\in A}\mathbb{P}\left[  J_{m}=j,J_{m-1}=i\right]
=\sum_{i\in A}\mathbb{P}\left[  J_{m}=j\mid J_{m-1}=i\right]  \mathbb{P}%
\left[  J_{m-1}=i\right] \\
&  =\sum_{i\in A}m_{ji}\left(  M^{m}\mu\right)  _{i}=\left(  M\left(  M^{m}%
\mu\right)  \right)  _{j}=\left(  M^{m+1}\mu\right)  _{j}%
\end{align*}
We have proved by induction that%
\[
\mathbb{P}\left[  J_{n-1}=j\right]  =\left(  M^{n}\mu\right)  _{j}%
\qquad\forall n\in\mathbb{N}%
\]
It follows that the probability of choosing $j$ is%
\[%
{\displaystyle\sum\limits_{n=0}^{\infty}}
\mathbb{P}\left[  N=n,J_{n-1}=j\right]  =%
{\displaystyle\sum\limits_{n=0}^{\infty}}
\mathbb{P}\left[  N=n\right]  \mathbb{P}\left[  J_{n-1}=j\right]  =%
{\displaystyle\sum\limits_{n=0}^{\infty}}
\mathbb{P}\left[  N=n\right]  \left(  M^{n}\mu\right)  _{j}%
\]
Then
\begin{align*}
p_{N}  &  =%
{\displaystyle\sum\limits_{n=0}^{\infty}}
\mathbb{P}\left[  N=n\right]  \left(  M^{n}\mu\right)  =\lim_{k\rightarrow
\infty}%
{\displaystyle\sum\limits_{n=0}^{k}}
\mathbb{P}\left[  N=n\right]  \left(  M^{n}\mu\right)  =\lim_{k\rightarrow
\infty}\left(  \left[
{\displaystyle\sum\limits_{n=0}^{k}}
\mathbb{P}\left[  N=n\right]  M^{n}\right]  \mu\right) \\
&  =\left(  \lim_{k\rightarrow\infty}\left[
{\displaystyle\sum\limits_{n=0}^{k}}
\mathbb{P}\left[  N=n\right]  M^{n}\right]  \right)  \mu
\end{align*}
and so $p_{N}=f_{N}\left(  M\right)  \mu$ holds. The average duration of an
iteration starting with incumbent $j$ is%
\[
\tau_{j}=%
{\displaystyle\sum\limits_{i\in A}}
Q\left(  i\mid j\right)  \tau_{\mathrm{RT}}\left(  i\mid j\right)
\]
where to ease notation we write $\tau$ in place of $\bar{\tau}$. Since
$\mathbb{P}\left[  J_{n-1}=j\right]  =\left(  M^{n}\mu\right)  _{j}$, the
average duration of iteration $k$ (if it takes place, i.e., if $N>k$) is%
\[%
{\displaystyle\sum\limits_{j\in A}}
\tau_{j}\mathbb{P}\left[  J_{k-1}=j\right]  =%
{\displaystyle\sum\limits_{j\in A}}
\tau_{j}\left(  M^{k}\mu\right)  _{j}=\tau\cdot M^{k}\mu
\]
The average duration if $N=n$ is then%
\[%
{\displaystyle\sum\limits_{k=0}^{n-1}}
\tau\cdot M^{k}\mu=\tau\cdot\left(
{\displaystyle\sum\limits_{k=0}^{n-1}}
M^{k}\right)  \mu
\]
with the convention $%
{\displaystyle\sum\limits_{k=0}^{-1}}
M^{k}=0$ (the zero matrix). Since the probability of stopping at $n$ is
$\mathbb{P}\left[  N=n\right]  $, it follows that
\begin{align}
\tau_{N}  &  =%
{\displaystyle\sum\limits_{n=0}^{\infty}}
\mathbb{P}\left[  N=n\right]  \tau\cdot\left(
{\displaystyle\sum\limits_{k=0}^{n-1}}
M^{k}\right)  \mu=\tau\cdot\left(
{\displaystyle\sum\limits_{n=0}^{\infty}}
\mathbb{P}\left[  N=n\right]  \left(
{\displaystyle\sum\limits_{k=0}^{n-1}}
M^{k}\right)  \right)  \mu\label{eq:tn-pk}\\
&  =\tau\cdot\left(
{\displaystyle\sum\limits_{n=1}^{\infty}}
\mathbb{P}\left[  N=n\right]  \left(
{\displaystyle\sum\limits_{k=0}^{n-1}}
M^{k}\right)  \right)  \mu\nonumber
\end{align}
because $%
{\displaystyle\sum\limits_{k=0}^{-1}}
M^{k}=0$. Now%
\begin{align*}%
{\displaystyle\sum\limits_{n=1}^{\infty}}
\mathbb{P}\left[  N=n\right]  \left(
{\displaystyle\sum\limits_{k=0}^{n-1}}
M^{k}\right)   &  =%
{\displaystyle\sum\limits_{n=1}^{\infty}}
\mathbb{P}\left[  N=n\right]  \left(
{\displaystyle\sum\limits_{k=1}^{n}}
M^{k-1}\right) \\
&  =%
{\displaystyle\sum\limits_{n=1}^{\infty}}
{\displaystyle\sum\limits_{k=1}^{\infty}}
1_{\left\{  k\leq n\right\}  }\mathbb{P}\left[  N=n\right]  M^{k-1}=%
{\displaystyle\sum\limits_{k=1}^{\infty}}
{\displaystyle\sum\limits_{n=1}^{\infty}}
1_{\left\{  k\leq n\right\}  }\mathbb{P}\left[  N=n\right]  M^{k-1}\\
&  =%
{\displaystyle\sum\limits_{k=1}^{\infty}}
M^{k-1}%
{\displaystyle\sum\limits_{n=1}^{\infty}}
1_{\left\{  k\leq n\right\}  }\mathbb{P}\left[  N=n\right]  =%
{\displaystyle\sum\limits_{k=1}^{\infty}}
M^{k-1}%
{\displaystyle\sum\limits_{n=k}^{\infty}}
\mathbb{P}\left[  N=n\right] \\
&  =%
{\displaystyle\sum\limits_{k=1}^{\infty}}
\mathbb{P}\left[  N\geq k\right]  M^{k-1}=%
{\displaystyle\sum\limits_{n=0}^{\infty}}
\mathbb{P}\left[  N\geq n+1\right]  M^{n}=%
{\displaystyle\sum\limits_{n=0}^{\infty}}
\mathbb{P}\left[  N>n\right]  M^{n}%
\end{align*}
This proves that $\tau_{N}=\tau\cdot g_{N}\left(  M\right)  \mu$
holds.\hfill$\blacksquare$

\subsubsection{Proof of Proposition \ref{prop:neg-bin}}

If $N$ is negative binomial, then
\begin{align*}
f_{N}\left(  z\right)   &  =\sum_{n=0}^{\infty}\binom{n+r-1}{r-1}\zeta
^{n}\left(  1-\zeta\right)  ^{r}z^{n}=\left(  1-\zeta\right)  ^{r}\sum
_{n=0}^{\infty}\binom{n+r-1}{r-1}\left(  \zeta z\right)  ^{n}=\frac{\left(
1-\zeta\right)  ^{r}}{\left(  1-\zeta z\right)  ^{r}}\\
g_{N}\left(  z\right)   &  =\frac{1-f_{N}\left(  z\right)  }{1-z}%
=\frac{1-\frac{\left(  1-\zeta\right)  ^{r}}{\left(  1-\zeta z\right)  ^{r}}%
}{1-z}=\frac{\frac{\left(  1-\zeta z\right)  ^{r}}{\left(  1-\zeta z\right)
^{r}}-\frac{\left(  1-\zeta\right)  ^{r}}{\left(  1-\zeta z\right)  ^{r}}%
}{1-z}=\frac{\left(  1-\zeta z\right)  ^{r}-\left(  1-\zeta\right)  ^{r}%
}{\left(  1-z\right)  \left(  1-\zeta z\right)  ^{r}}%
\end{align*}
For $r=1$ it yields%
\begin{align*}
f_{N}\left(  z\right)   &  =\left(  1-\zeta\right)  \left(  1-\zeta z\right)
^{-1}\\
g_{N}\left(  z\right)   &  =\frac{1-\zeta z-1+\zeta}{\left(  1-z\right)
\left(  1-\zeta z\right)  }=\frac{\zeta-\zeta z}{\left(  1-z\right)  \left(
1-\zeta z\right)  }=\frac{\zeta\left(  1-z\right)  }{\left(  1-z\right)
\left(  1-\zeta z\right)  }=\zeta\left(  1-\zeta z\right)  ^{-1}%
\end{align*}
In general, note that $z=1$ is a root of $\left(  1-\zeta z\right)
^{r}-\left(  1-\zeta\right)  ^{r}$. Thus, the ratio%
\[
\frac{\left(  1-\zeta z\right)  ^{r}-\left(  1-\zeta\right)  ^{r}}{1-z}%
\]
appearing above is a polynomial of degree $r-1$ in $z$. Next we compute it. It
holds%
\begin{align*}
\left(  1-\zeta z\right)  ^{r}-\left(  1-\zeta\right)  ^{r}  &  =\sum
_{k=0}^{r}\binom{r}{k}\left(  -\zeta z\right)  ^{k}-\sum_{k=0}^{r}\binom{r}%
{k}\left(  -\zeta\right)  ^{k}\\
&  =\sum_{k=0}^{r}\left(  -1\right)  ^{k}\binom{r}{k}\zeta^{k}z^{k}-\sum
_{k=0}^{r}\left(  -1\right)  ^{k}\binom{r}{k}\zeta^{k}\\
&  =\sum_{k=0}^{r}\left(  -1\right)  ^{k}\binom{r}{k}\zeta^{k}\left(
z^{k}-1\right)  =\sum_{k=0}^{r}\binom{r}{k}\left(  -\zeta\right)  ^{k}\left(
z-1\right)  \sum_{j=0}^{k-1}z^{j}%
\end{align*}
because%
\[
z^{k}-1=\left(  z-1\right)  \left(  1+z+\cdot\cdot\cdot+z^{k-1}\right)
=\left(  z-1\right)  \sum_{j=0}^{k-1}z^{j}%
\]
with the convention $\sum_{j=0}^{-1}z^{j}=0$. Then,%
\begin{align*}
g_{N}\left(  z\right)   &  =\frac{\left(  1-\zeta z\right)  ^{r}-\left(
1-\zeta\right)  ^{r}}{\left(  1-z\right)  \left(  1-\zeta z\right)  ^{r}%
}=\left(  \sum_{k=0}^{r}\binom{r}{k}\left(  -\zeta\right)  ^{k}\left(
z-1\right)  \sum_{j=0}^{k-1}z^{j}\right)  \frac{1}{\left(  1-z\right)  \left(
1-\zeta z\right)  ^{r}}\\
&  =-\left(  \sum_{k=0}^{r}\binom{r}{k}\left(  -\zeta\right)  ^{k}\sum
_{j=0}^{k-1}z^{j}\right)  \left(  1-\zeta z\right)  ^{-r}%
\end{align*}
showing that (\ref{eq:gnb}) holds.

\subsubsection{Equations (\ref{eq:frev}) and (\ref{eq:grev})}

\paragraph{Preamble}

A reversible matrix $B$ is diagonalizable with real eigenvalues. Indeed, from
the detailed balance condition (\ref{eq:balance-pre}) it readily follows that
the matrix $B^{\ast}$ with off diagonal entries $b_{ij}^{\ast}=b_{ij}%
\sqrt{p_{j}/p_{i}}$ is symmetric and has the same eigenvalues as $B$. A
stochastic reversible matrix $B$ has then a largest eigenvalue $\lambda_{1}$
equal to $1$ and all its other eigenvalues have absolute values $\leq
\lambda_{1}$, i.e., they belong to $\left[  -1,1\right]  $. If, in addition,
$B$ is primitive, by Perron's Theorem their absolute values are actually
$<\lambda_{1}$, so they belong to $\left(  -1,1\right)  $.

\paragraph{Equations}

Let the transition matrix $M$ be diagonalizable (e.g., because it is
reversible) and let $\Lambda=\operatorname*{diag}\left(  \lambda_{1}%
,\lambda_{2},...,\lambda_{m}\right)  $ be the diagonal matrix of its
eigenvalues, each repeated according to its multiplicity. For any summable
sequence $a=\left\{  a_{n}\right\}  $ of non-negative scalars we then have
\begin{align*}
f_{a}\left(  M\right)   &  =%
{\displaystyle\sum\limits_{n=0}^{\infty}}
a_{n}M^{n}=\lim_{l\rightarrow\infty}%
{\displaystyle\sum\limits_{n=0}^{l}}
a_{n}U\Lambda^{n}U^{-1}=\lim_{l\rightarrow\infty}\left[  U\left(
{\displaystyle\sum\limits_{n=0}^{l}}
a_{n}\Lambda^{n}\right)  U^{-1}\right] \\
&  =U\left[  \lim_{l\rightarrow\infty}\left(
{\displaystyle\sum\limits_{n=0}^{l}}
a_{n}\Lambda^{n}\right)  \right]  U^{-1}=U\left[  \operatorname*{diag}\left(
{\displaystyle\sum\limits_{n=0}^{\infty}}
a_{n}\lambda_{1}^{n},%
{\displaystyle\sum\limits_{n=0}^{\infty}}
a_{n}\lambda_{2}^{n},...,%
{\displaystyle\sum\limits_{n=0}^{\infty}}
a_{n}\lambda_{m}^{n}\right)  \right]  U^{-1}\\
&  =U\left[  \operatorname*{diag}\left(  f_{a}\left(  \lambda_{1}\right)
,f_{a}\left(  \lambda_{2}\right)  ,...,f_{a}\left(  \lambda_{\left\vert
A\right\vert }\right)  \right)  \right]  U^{-1}%
\end{align*}
This immediately yields (\ref{eq:frev}) and (\ref{eq:grev}).

\subsection{Section \ref{sect:rev}}

\subsubsection{Proof of Theorem \ref{prop:value-bis}, Corollary
\ref{prop:value-ter}, and Corollary \ref{prop:value-quater}}

\begin{proposition}
If $\rho_{\mathrm{C}}$ is has a binary value representation and the
exploration matrix $Q$ is nice, then the probability distribution%
\[
\pi\left(  i\right)  =\left\{
\begin{array}
[c]{ll}%
\dfrac{u\left(  i\right)  }{\sum_{j\in\left.  \arg\max\right.  _{A}w}u\left(
j\right)  }\medskip & \qquad\text{if }i\in\left.  \arg\max\right.  _{A}w\\
0 & \qquad\text{else}%
\end{array}
\right.
\]
is the only stationary distribution for $M$, and there exists $\varepsilon
\in\left(  0,1\right)  $ such that, for all $n\in\mathbb{N}$ and all $\mu
\in\Delta\left(  A\right)  $,%
\begin{equation}
\left\Vert M^{n}\mu-\pi\right\Vert _{1}\leq2\left(  1-\varepsilon\right)  ^{n}
\label{eq:lim_ergo}%
\end{equation}
Moreover, if $\left\{  N_{k}\right\}  _{k=0}^{\infty}$ is a sequence of
stopping numbers that diverges, then, for all $\mu\in\Delta\left(  A\right)  $%
\begin{equation}
\left(
{\displaystyle\sum\limits_{n=0}^{\infty}}
\mathbb{P}\left[  N_{k}=n\right]  M^{n}\right)  \mu\rightarrow\pi
\qquad\text{as }k\rightarrow\infty\label{eq:lim_pk}%
\end{equation}

\end{proposition}

In particular, (\ref{eq:lim_ergo}) implies that
\[
\lim_{n\rightarrow\infty}\Pr\left[  J_{n}=i\right]  =\pi\left(  i\right)
\]
and (\ref{eq:lim_pk}) implies that
\begin{equation}
\lim_{N_{k}\rightarrow\infty}p_{N_{k}}=\pi\label{eq:asintotico}%
\end{equation}
when the stopping numbers are simple. Hence,
\[
\lim_{N_{k}\rightarrow\infty}p_{N_{k}}\left(  i,A\right)  =\lim_{n\rightarrow
\infty}\Pr\left[  J_{n}=i\right]  =\left\{
\begin{array}
[c]{ll}%
\dfrac{u\left(  i\right)  }{\sum_{j\in\left.  \arg\max\right.  _{A}w}u\left(
j\right)  }\medskip & \qquad\text{if }i\in\left.  \arg\max\right.  _{A}w\\
0 & \qquad\text{else}%
\end{array}
\right.
\]
This proves Theorem \ref{prop:value-bis}. Corollaries \ref{prop:value-ter} and
\ref{prop:value-quater} follow immediately.

\bigskip

\noindent\textbf{Proof} To ease notation we write $\rho$ in place of
$\rho_{\mathrm{C}}$. We first show that
\begin{equation}
M\left(  k\mid j\right)  \pi\left(  j\right)  =M\left(  j\mid k\right)
\pi\left(  k\right)  \label{eq:db}%
\end{equation}
for all $j$ and $k$ in $A$. Denote, for brevity, $W\left(  A\right)  =\left.
\arg\max\right.  _{A}w$. If $j=k$, the equality is trivial. Let $j\neq k$ in
$A$.

\begin{itemize}
\item If $j,k\in W\left(  A\right)  $, then
\begin{align*}
M\left(  k\mid j\right)  \pi\left(  j\right)   &  =Q\left(  k\mid j\right)
\rho\left(  k\mid j\right)  \dfrac{u\left(  j\right)  }{\sum_{i\in W\left(
A\right)  }u\left(  i\right)  }\\
&  =Q\left(  k\mid j\right)  s\left(  k,j\right)  \dfrac{u\left(  k\right)
}{u\left(  k\right)  +u\left(  j\right)  }\dfrac{u\left(  j\right)  }%
{\sum_{i\in W\left(  A\right)  }u\left(  i\right)  }\\
&  =Q\left(  j\mid k\right)  s\left(  j,k\right)  \dfrac{u\left(  j\right)
}{u\left(  j\right)  +u\left(  k\right)  }\dfrac{u\left(  k\right)  }%
{\sum_{i\in W\left(  A\right)  }u\left(  i\right)  }=Q\left(  j\mid k\right)
\rho\left(  j\mid k\right)  \pi\left(  k\right) \\
&  =M\left(  j\mid k\right)  \pi\left(  k\right)
\end{align*}

\item If $j,k\notin W\left(  A\right)  $, then $\pi\left(  j\right)
=\pi\left(  k\right)  =0$ and
\[
M\left(  k\mid j\right)  \pi\left(  j\right)  =M\left(  j\mid k\right)
\pi\left(  k\right)
\]

\item If $j\in W\left(  A\right)  $ and $k\notin W\left(  A\right)  $, then
$w\left(  j\right)  >w\left(  k\right)  $ and so $\rho\left(  k\mid j\right)
=0=\pi\left(  k\right)  $, thus
\[
M\left(  k\mid j\right)  \pi\left(  j\right)  =Q\left(  k\mid j\right)
\rho\left(  k\mid j\right)  \dfrac{u\left(  j\right)  }{\sum_{i\in W\left(
A\right)  }u\left(  i\right)  }=0=M\left(  j\mid k\right)  \pi\left(
k\right)
\]

\item If $j\notin W\left(  A\right)  $ and $k\in W\left(  A\right)  $, then
$w\left(  k\right)  >w\left(  j\right)  $ and so $\rho\left(  j\mid k\right)
=0=\pi\left(  j\right)  $, thus
\[
M\left(  k\mid j\right)  \pi\left(  j\right)  =0=Q\left(  j\mid k\right)
\rho\left(  j\mid k\right)  \pi\left(  k\right)  =M\left(  j\mid k\right)
\pi\left(  k\right)
\]

\end{itemize}

The \textquotedblleft detailed balance\textquotedblright\ condition
(\ref{eq:db}) implies that
\[%
{\displaystyle\sum\limits_{j\in A}}
M\left(  k\mid j\right)  \pi\left(  j\right)  =%
{\displaystyle\sum\limits_{j\in A}}
M\left(  j\mid k\right)  \pi\left(  k\right)  =\pi\left(  k\right)
{\displaystyle\sum\limits_{j\in A}}
M\left(  j\mid k\right)  =\pi\left(  k\right)
\]
for all $k\in A$, then $M\pi=\pi$. Thus $\pi$ is a stationary distribution for
$M$.

Take $j_{0}\in W\left(  A\right)  $. Then, $w\left(  j_{0}\right)  \geq
w\left(  i\right)  $ for all $i\neq j_{0}$, and so%
\[
M\left(  j_{0}\mid i\right)  =Q\left(  j_{0}\mid i\right)  \rho\left(
j_{0}\mid i\right)  =\left\{
\begin{array}
[c]{ll}%
Q\left(  j_{0}\mid i\right)  \medskip & \qquad\text{if }w\left(  j_{0}\right)
>w\left(  i\right) \\
Q\left(  j_{0}\mid i\right)  s\left(  j_{0},i\right)  \dfrac{u\left(
j_{0}\right)  }{u\left(  j_{0}\right)  +u\left(  i\right)  } & \qquad\text{if
}w\left(  j_{0}\right)  =w\left(  i\right)
\end{array}
\right.
\]
For $i=j_{0}$, we have that $\rho\left(  k\mid j_{0}\right)  =0$ if $k\notin
W\left(  A\right)  $ and $\rho\left(  k\mid j_{0}\right)  \in\left(
0,1\right)  $ if $k\in W\left(  A\right)  $ (provided $k\neq j_{0}$),%
\[
M\left(  j_{0}\mid j_{0}\right)  =1-%
{\displaystyle\sum\limits_{k\neq j_{0}}}
Q\left(  k\mid j_{0}\right)  \rho\left(  k\mid j_{0}\right)  >1-%
{\displaystyle\sum\limits_{k\neq j_{0}}}
Q\left(  k\mid j_{0}\right)  \geq1-%
{\displaystyle\sum\limits_{k\in A}}
Q\left(  k\mid j_{0}\right)  =0
\]
By Doeblin's Theorem, $\pi$ is the only stationary distribution for $M$ and
there exists $\varepsilon\in\left(  0,1\right)  $ such that
\[
\left\Vert M^{n}\mu-\pi\right\Vert _{1}\leq2\left(  1-\varepsilon\right)  ^{n}%
\]
for all $n\in\mathbb{N}$ and all $\mu\in\Delta\left(  A\right)  $.

Given any $\mu\in\Delta\left(  A\right)  $, set, for each $k\in\mathbb{N}$,%
\[
P_{k}\left(  n\right)  =\mathbb{P}\left[  N_{k}=n\right]  \qquad\forall
n\in\mathbb{N}%
\]
and
\[
p_{k}=\left(
{\displaystyle\sum\limits_{n=0}^{\infty}}
P_{k}\left(  n\right)  M^{n}\right)  \mu
\]
Then%
\[
p_{k}=\lim_{m\rightarrow\infty}%
{\displaystyle\sum\limits_{n=0}^{m}}
P_{k}\left(  n\right)  M^{n}\mu\text{\quad and\quad}\pi=\lim_{m\rightarrow
\infty}%
{\displaystyle\sum\limits_{n=0}^{m}}
P_{k}\left(  n\right)  \pi
\]
and so
\[
p_{k}-\pi=\lim_{m\rightarrow\infty}%
{\displaystyle\sum\limits_{n=0}^{m}}
P_{k}\left(  n\right)  \left(  M^{n}\mu-\pi\right)
\]
Thus,%
\begin{align*}
\left\Vert p_{k}-\pi\right\Vert _{1}  &  =\lim_{m\rightarrow\infty}\left\Vert
{\displaystyle\sum\limits_{n=0}^{m}}
P_{k}\left(  n\right)  \left(  M^{n}\mu-\pi\right)  \right\Vert _{1}\leq
\lim_{m\rightarrow\infty}%
{\displaystyle\sum\limits_{n=0}^{m}}
P_{k}\left(  n\right)  \left\Vert M^{n}\mu-\pi\right\Vert _{1}\\
&  \leq\lim_{m\rightarrow\infty}%
{\displaystyle\sum\limits_{n=0}^{m}}
P_{k}\left(  n\right)  2\left(  1-\varepsilon\right)  ^{n}=%
{\displaystyle\sum\limits_{n=0}^{\infty}}
P_{k}\left(  n\right)  2\left(  1-\varepsilon\right)  ^{n}%
\end{align*}
The sequence $\left\{  a_{k}\right\}  _{k\in\mathbb{N}}$ of functions $a_{k}%
$\thinspace$:\mathbb{N\rightarrow}\left[  0,\infty\right)  \ $given by%
\[
a_{k}\left(  n\right)  =P_{k}\left(  n\right)  2\left(  1-\varepsilon\right)
^{n}%
\]
is bounded above by the function $a:\mathbb{N\rightarrow}\left[
0,\infty\right)  $ given by%
\[
a\left(  n\right)  =2\left(  1-\varepsilon\right)  ^{n}%
\]
The latter is summable with respect to the counting measure $\gamma$ on
$\mathbb{N}$. In addition, $\lim_{k\rightarrow\infty}a_{k}\left(  n\right)
=0$ for all $n\in\mathbb{N}$. By the Lebesgue Dominated Convergence Theorem,%
\[
\lim_{k\rightarrow\infty}%
{\displaystyle\sum\limits_{n=0}^{\infty}}
P_{k}\left(  n\right)  2\left(  1-\varepsilon\right)  ^{n}=\lim_{k\rightarrow
\infty}\int_{\mathbb{N}}a_{k}\left(  n\right)  \mathrm{d}\gamma\left(
n\right)  =0
\]
Therefore, $\lim_{k\rightarrow\infty}\left\Vert p_{k}-\pi\right\Vert _{1}%
=0$.\hfill$\blacksquare$

\subsubsection{Proof of Theorem \ref{thm:value-bis}}

Given a menu $A$, with typical elements $i$, $j$ and $k$, we denote by
$P=\left[  P\left(  i\mid j\right)  \right]  _{i,j\in A}$ a $\left\vert
A\right\vert \times\left\vert A\right\vert $ a\emph{ }stochastic matrix\emph{
}such that $P\left(  i\mid j\right)  $ is interpreted as the probability with
which a system moves from state $j$ to state $i$. Clearly, $P\left(  \cdot\mid
j\right)  \in\Delta\left(  A\right)  $ for all $j\in A$.

\begin{definition}
A stochastic matrix $P$ is \emph{transitive} if
\begin{equation}
P\left(  j\mid i\right)  P\left(  k\mid j\right)  P\left(  i\mid k\right)
=P\left(  k\mid i\right)  P\left(  j\mid k\right)  P\left(  i\mid j\right)
\qquad\forall i,j,k\in A \label{eq:kolmo}%
\end{equation}

\end{definition}

Transitivity is known as the \emph{Kolmogorov criterion} in the Markov chains
literature (see, e.g., Kelly, 1979, p. 24) and as the \emph{product rule} in
the stochastic choice literature (Luce and Suppes, 1965, p. 341).

Transitivity is automatically satisfied if at least two of the three states
$i$, $j$, and $k$ in $A$ coincide. In fact,

\begin{itemize}
\item if $i=j$, then
\begin{align*}
P\left(  j\mid i\right)  P\left(  k\mid j\right)  P\left(  i\mid k\right)   &
=P\left(  i\mid i\right)  P\left(  k\mid i\right)  P\left(  i\mid k\right) \\
P\left(  k\mid i\right)  P\left(  j\mid k\right)  P\left(  i\mid j\right)   &
=P\left(  k\mid i\right)  P\left(  i\mid k\right)  P\left(  i\mid i\right)
\end{align*}

\item if $i=k$, then
\begin{align*}
P\left(  j\mid i\right)  P\left(  k\mid j\right)  P\left(  i\mid k\right)   &
=P\left(  j\mid i\right)  P\left(  i\mid j\right)  P\left(  i\mid i\right) \\
P\left(  k\mid i\right)  P\left(  j\mid k\right)  P\left(  i\mid j\right)   &
=P\left(  i\mid i\right)  P\left(  j\mid i\right)  P\left(  i\mid j\right)
\end{align*}

\item if $j=k$, then
\begin{align*}
P\left(  j\mid i\right)  P\left(  k\mid j\right)  P\left(  i\mid k\right)   &
=P\left(  j\mid i\right)  P\left(  j\mid j\right)  P\left(  i\mid j\right) \\
P\left(  k\mid i\right)  P\left(  j\mid k\right)  P\left(  i\mid j\right)   &
=P\left(  j\mid i\right)  P\left(  j\mid j\right)  P\left(  i\mid j\right)
\end{align*}

\end{itemize}

\noindent Therefore, transitivity can be restated as
\[
P\left(  j\mid i\right)  P\left(  k\mid j\right)  P\left(  i\mid k\right)
=P\left(  k\mid i\right)  P\left(  j\mid k\right)  P\left(  i\mid j\right)
\]
for all distinct$\ i$, $j$ and $k$ in $A$.\footnote{This argument applies to
any function $P:A\times A\rightarrow\mathbb{R}$ and is independent of its
\textquotedblleft diagonal\textquotedblright\ values $P\left(  i\mid i\right)
$.}

The next result, which relates reversibility and transitivity, builds upon
Kolmogorov (1936) and Luce and Suppes (1965).

\begin{proposition}
\label{prop:kolmo}Let $P$ be a positive stochastic matrix. The following
conditions are equivalent:

\begin{enumerate}
\item[(i)] $P$ is reversible under some $\pi\in\Delta\left(  A\right)  $;

\item[(ii)] $P$ is transitive.
\end{enumerate}

In this case, given any $i\in A$, it holds%
\[
\pi\left(  j\right)  =\frac{r\left(  j\mid i\right)  }{%
{\displaystyle\sum\limits_{k\in A}}
r\left(  k\mid i\right)  }\qquad\forall j\in A
\]
where $r\left(  j\mid i\right)  =P\left(  j\mid i\right)  /P\left(  i\mid
j\right)  $. In particular, $\pi$ is unique.
\end{proposition}

\noindent\textbf{Proof} Assume that there exists $\pi\in\Delta\left(
A\right)  $ such that $P\left(  i\mid j\right)  \pi\left(  j\right)  =P\left(
j\mid i\right)  \pi\left(  i\right)  $ for all distinct $i,j\in A$ (note that
this is weaker that reversibility in that $\pi$ is not assumed to be
positive), then
\begin{equation}
P\left(  i\mid j\right)  \pi\left(  j\right)  =P\left(  j\mid i\right)
\pi\left(  i\right)  \qquad\forall i,j\in A \label{eq:heart}%
\end{equation}
If $\pi\left(  i^{\ast}\right)  =0$ for some $i^{\ast}\in A$, then (being $P$
positive)%
\begin{equation}
\pi\left(  j\right)  =\frac{P\left(  j\mid i^{\ast}\right)  }{P\left(
i^{\ast}\mid j\right)  }\pi\left(  i^{\ast}\right)  =0\qquad\forall j\in A
\label{eq:star}%
\end{equation}
But, this is impossible since $%
{\displaystyle\sum\limits_{j\in A}}
\pi\left(  j\right)  =1$. Hence, $\pi$ is positive. Moreover, by
(\ref{eq:star}) we have%
\[
\frac{\dfrac{P\left(  j\mid i^{\ast}\right)  }{P\left(  i^{\ast}\mid j\right)
}}{%
{\displaystyle\sum\limits_{k\in A}}
\dfrac{P\left(  k\mid i^{\ast}\right)  }{P\left(  i^{\ast}\mid k\right)  }%
}=\frac{\dfrac{P\left(  j\mid i^{\ast}\right)  }{P\left(  i^{\ast}\mid
j\right)  }\pi\left(  i^{\ast}\right)  }{%
{\displaystyle\sum\limits_{k\in A}}
\dfrac{P\left(  k\mid i^{\ast}\right)  }{P\left(  i^{\ast}\mid k\right)  }%
\pi\left(  i^{\ast}\right)  }=\frac{\pi\left(  j\right)  }{%
{\displaystyle\sum\limits_{k\in A}}
\pi\left(  k\right)  }=\pi\left(  j\right)  \qquad\forall j\in A
\]
irrespective of the choice of $i^{\ast}\in A$. Hence, $\pi$ is unique.
Finally, given any $i,j,k\in A$, by (\ref{eq:heart}) we have:%
\begin{align*}
\frac{\pi\left(  j\right)  }{\pi\left(  i\right)  }\frac{\pi\left(  k\right)
}{\pi\left(  j\right)  }\frac{\pi\left(  i\right)  }{\pi\left(  k\right)  }
&  =1\implies\frac{P\left(  j\mid i\right)  }{P\left(  i\mid j\right)  }%
\frac{P\left(  k\mid j\right)  }{P\left(  j\mid k\right)  }\frac{P\left(
i\mid k\right)  }{P\left(  k\mid i\right)  }=1\\
&  \implies\frac{P\left(  j\mid i\right)  P\left(  k\mid j\right)  P\left(
i\mid k\right)  }{P\left(  k\mid i\right)  P\left(  j\mid k\right)  P\left(
i\mid j\right)  }=1\\
&  \implies P\left(  j\mid i\right)  P\left(  k\mid j\right)  P\left(  i\mid
k\right)  =P\left(  k\mid i\right)  P\left(  j\mid k\right)  P\left(  i\mid
j\right)
\end{align*}
So, transitivity holds.

Conversely, if transitivity holds, choose arbitrarily $i^{\ast}\in A$ and set%
\begin{equation}
\pi^{\ast}\left(  j\right)  :=\frac{\dfrac{P\left(  j\mid i^{\ast}\right)
}{P\left(  i^{\ast}\mid j\right)  }}{%
{\displaystyle\sum\limits_{k\in A}}
\dfrac{P\left(  k\mid i^{\ast}\right)  }{P\left(  i^{\ast}\mid k\right)  }%
}=\dfrac{P\left(  j\mid i^{\ast}\right)  }{P\left(  i^{\ast}\mid j\right)
}\zeta\qquad\forall j\in A \label{eq:tr1}%
\end{equation}
where $1/\zeta=%
{\displaystyle\sum\limits_{k\in A}}
P\left(  k\mid i^{\ast}\right)  /P\left(  i^{\ast}\mid k\right)  >0$. By
transitivity, for all $i,j\in A$,%
\[
P\left(  j\mid i\right)  P\left(  i^{\ast}\mid j\right)  P\left(  i\mid
i^{\ast}\right)  =P\left(  i^{\ast}\mid i\right)  P\left(  j\mid i^{\ast
}\right)  P\left(  i\mid j\right)
\]
and, since $P$ is positive,%
\[
P\left(  j\mid i\right)  \frac{P\left(  i\mid i^{\ast}\right)  }{P\left(
i^{\ast}\mid i\right)  }=P\left(  i\mid j\right)  \frac{P\left(  j\mid
i^{\ast}\right)  }{P\left(  i^{\ast}\mid j\right)  }%
\]
Thus, for all $i,j\in A$,%
\[
P\left(  j\mid i\right)  \frac{P\left(  i\mid i^{\ast}\right)  }{P\left(
i^{\ast}\mid i\right)  }\zeta=P\left(  i\mid j\right)  \frac{P\left(  j\mid
i^{\ast}\right)  }{P\left(  i^{\ast}\mid j\right)  }\zeta
\]
In view of (\ref{eq:tr1}), reversibility with respect to $\pi^{\ast}$ holds
(note that $\pi^{\ast}$ is strictly positive).\hfill$\blacksquare\bigskip$

\noindent\textbf{Proof of Theorem \ref{thm:value-bis}} To ease notation we
write $\rho$ in place of $\rho_{\mathrm{C}}$.

\textquotedblleft If.\textquotedblright\ By Lemma \ref{lem:ma}, since $Q$ is
quasi-positive, then $M$ is positive. By assumption $M$ is reversible. But
then, by Proposition \ref{prop:kolmo}, $M$ is transitive, thus
\[
M\left(  j\mid i\right)  M\left(  k\mid j\right)  M\left(  i\mid k\right)
=M\left(  k\mid i\right)  M\left(  j\mid k\right)  M\left(  i\mid j\right)
\]
for all distinct$\ i$, $j$ and $k$ in $A$. By definition of $M$,%
\begin{align*}
&  Q\left(  j\mid i\right)  \rho\left(  j\mid i\right)  Q\left(  k\mid
j\right)  \rho\left(  k\mid j\right)  Q\left(  i\mid k\right)  \rho\left(
i\mid k\right) \\
&  =Q\left(  k\mid i\right)  \rho\left(  k\mid i\right)  Q\left(  j\mid
k\right)  \rho\left(  j\mid k\right)  Q\left(  i\mid j\right)  \rho\left(
i\mid j\right)
\end{align*}
for all distinct$\ i$, $j$ and $k$ in $A$. By symmetry and quasi-positivity of
$Q$, this implies
\[
\rho\left(  j\mid i\right)  \rho\left(  k\mid j\right)  \rho\left(  i\mid
k\right)  =\rho\left(  k\mid i\right)  \rho\left(  j\mid k\right)  \rho\left(
i\mid j\right)
\]
for all distinct$\ i$, $j$ and $k$ in $A$. Therefore, $\rho$ is transitive,
and by Theorem \ref{thm:value} it admits a binary value representation, so
that the Neural Metropolis Algorithm is value based.

\textquotedblleft Only if.\textquotedblright\ If the Neural Metropolis
Algorithm is value based, by definition, $\rho$ admits a binary value
representation, and by Theorem \ref{thm:value} it is transitive. Thus
\[
\rho\left(  j\mid i\right)  \rho\left(  k\mid j\right)  \rho\left(  i\mid
k\right)  =\rho\left(  k\mid i\right)  \rho\left(  j\mid k\right)  \rho\left(
i\mid j\right)
\]
for all distinct$\ i$, $j$ and $k$ in $A$. Since $Q$ is symmetric and
quasi-positivite, then%
\begin{align*}
&  Q\left(  j\mid i\right)  \rho\left(  j\mid i\right)  Q\left(  k\mid
j\right)  \rho\left(  k\mid j\right)  Q\left(  i\mid k\right)  \rho\left(
i\mid k\right) \\
&  =Q\left(  k\mid i\right)  \rho\left(  k\mid i\right)  Q\left(  j\mid
k\right)  \rho\left(  j\mid k\right)  Q\left(  i\mid j\right)  \rho\left(
i\mid j\right)
\end{align*}
for all distinct$\ i$, $j$ and $k$ in $A$. By definition of $M$,%
\[
M\left(  j\mid i\right)  M\left(  k\mid j\right)  M\left(  i\mid k\right)
=M\left(  k\mid i\right)  M\left(  j\mid k\right)  M\left(  i\mid j\right)
\]
for all distinct$\ i$, $j$ and $k$ in $A$. By Lemma \ref{lem:ma}, since $Q$ is
quasi-positive, then $M$ is positive. But then, by Proposition
\ref{prop:kolmo}, $M$ is reversible.\hfill$\blacksquare$

\subsection{Section \ref{sect:icemia}\label{app:endicectomia}}

Proposition \ref{thm:value-ter} is a consequence of the following result.

\begin{proposition}
\label{prop:amul}Given any positive $\rho_{\mathrm{C}}$ and any irreducible
exploration matrix $Q$, the transition matrix $M$ is primitive. Moreover,
denoting by $\pi$ the stationary distribution of $M$, it follows that%
\[
\lim_{t\rightarrow\infty}p_{N_{t}}\left(  j\right)  =\frac{\pi\left(
j\right)  \bar{\tau}_{j}}{%
{\displaystyle\sum\limits_{k\in A}}
\pi\left(  k\right)  \bar{\tau}_{k}}\qquad\forall j\in A
\]
provided the distribution of $\mathrm{RT}_{i,j}$ has (strictly) positive
expectation, is continuous at $0$, and has no singular part for all $\left(
i,j\right)  \in A^{2}$.
\end{proposition}

\noindent\textbf{Proof of Proposition \ref{prop:amul}} To ease notation we
write $\rho$ in place of $\rho_{\mathrm{C}}$. The stochastic process $\left(
I,J,T\right)  $ produces sequences%
\[
\left(  \underset{\text{state }x_{0}}{\underbrace{j_{-1},i_{0}}}%
,t_{0},\underset{\text{state }x_{1}}{\underbrace{j_{0},i_{1}}},t_{1}%
,...,\underset{\text{state }x_{n}}{\underbrace{j_{n-1},i_{n}}},t_{n}%
,\underset{\text{state }x_{n+1}}{\underbrace{j_{n},i_{n+1}}},...\right)
\]
it then can be seen as a semi Markov chain with state space
\[
\mathcal{X}=\left\{  \left(  j,i\right)  \in A^{2}:Q\left(  i\mid j\right)
>0\right\}
\]
where state $x=\left(  j,i\right)  \in\mathcal{X}$ represents the comparison
between incumbent $j$ and proposal $i$. Since the comparison between $j$ and
$i$ produces incumbent $k=i$ with probability $\rho\left(  i\mid j\right)  $,
$k=j$ with probability $1-\rho\left(  i\mid j\right)  $, and all other
incumbents with probability $0$, then the probability of switching from
comparison $\left(  j,i\right)  $ to comparison $\left(  k,h\right)  $ is
given by%
\[
\mathbb{P}\left[  X_{n+1}=\left(  k,h\right)  \mid X_{n}=\left(  j,i\right)
\right]  =\left(  \delta_{i}\left(  k\right)  \rho\left(  i\mid j\right)
+\delta_{j}\left(  k\right)  \left(  1-\rho\left(  i\mid j\right)  \right)
\right)  Q\left(  h\mid k\right)
\]
In fact,

\begin{itemize}
\item if $i=j$, then the comparison between $j$ and $i$ produces new incumbent
$i$ for sure and

\begin{itemize}
\item[$\circ$] if $k=i$, then%
\begin{align*}
\mathbb{P}\left[  X_{n+1}=\left(  k,h\right)  \mid X_{n}=\left(  i,i\right)
\right]   &  =Q\left(  h\mid i\right) \\
&  =\left(  \underset{=1}{\underbrace{\delta_{i}\left(  k\right)  }}%
\rho\left(  i\mid j\right)  +~\underset{=1}{\underbrace{\delta_{j}\left(
k\right)  }}\left(  1-\rho\left(  i\mid j\right)  \right)  \right)  Q\left(
h\mid k\right)
\end{align*}

\item[$\circ$] else $k\neq i$, and
\begin{align*}
\mathbb{P}\left[  X_{n+1}=\left(  k,h\right)  \mid X_{n}=\left(  i,i\right)
\right]   &  =0\\
&  =\left(  \underset{=0}{\underbrace{\delta_{i}\left(  k\right)  }}%
\rho\left(  i\mid j\right)  +~\underset{=0}{\underbrace{\delta_{j}\left(
k\right)  }}\left(  1-\rho\left(  i\mid j\right)  \right)  \right)  Q\left(
h\mid k\right)
\end{align*}

\end{itemize}

\item if $i\neq j$, then the comparison between $j$ and $i$ produces new
incumbent $k=i$ with probability $\rho\left(  i\mid j\right)  $ and $k=j$ with
probability $1-\rho\left(  i\mid j\right)  $ and

\begin{itemize}
\item[$\circ$] if $k=i$, then%
\begin{align*}
\mathbb{P}\left[  X_{n+1}=\left(  k,h\right)  \mid X_{n}=\left(  j,i\right)
\right]   &  =\rho\left(  i\mid j\right)  Q\left(  h\mid i\right) \\
&  =\left(  \underset{=1}{\underbrace{\delta_{i}\left(  k\right)  }}%
\rho\left(  i\mid j\right)  +~\underset{=0}{\underbrace{\delta_{j}\left(
k\right)  }}\left(  1-\rho\left(  i\mid j\right)  \right)  \right)  Q\left(
h\mid k\right)
\end{align*}

\item[$\circ$] if $k=j$, then, we have%
\begin{align*}
\mathbb{P}\left[  X_{n+1}=\left(  k,h\right)  \mid X_{n}=\left(  j,i\right)
\right]   &  =\left(  1-\rho\left(  i\mid j\right)  \right)  Q\left(  h\mid
j\right) \\
&  =\left(  \underset{=0}{\underbrace{\delta_{i}\left(  k\right)  }}%
\rho\left(  i\mid j\right)  +~\underset{=1}{\underbrace{\delta_{j}\left(
k\right)  }}\left(  1-\rho\left(  i\mid j\right)  \right)  \right)  Q\left(
h\mid k\right)
\end{align*}

\item[$\circ$] else $k\neq i$ and $k\neq j$, thus
\[
\mathbb{P}\left[  X_{n+1}=\left(  k,h\right)  \mid X_{n}=\left(  j,i\right)
\right]  =0=\left(  \underset{=0}{\underbrace{\delta_{i}\left(  k\right)  }%
}\rho\left(  i\mid j\right)  +~\underset{=0}{\underbrace{\delta_{j}\left(
k\right)  }}\left(  1-\rho\left(  i\mid j\right)  \right)  \right)  Q\left(
h\mid k\right)
\]

\end{itemize}
\end{itemize}

Set
\[
\hat{M}\left(  \left(  k,h\right)  \mid\left(  j,i\right)  \right)  =\left(
\delta_{i}\left(  k\right)  \rho\left(  i\mid j\right)  +\delta_{j}\left(
k\right)  \left(  1-\rho\left(  i\mid j\right)  \right)  \right)  Q\left(
h\mid k\right)  \qquad\forall\left(  k,h\right)  ,\left(  j,i\right)
\in\mathcal{X}%
\]
Next we show that $\hat{M}$ is a \emph{bona fide} stochastic matrix. Given any
$\left(  j,i\right)  \in\mathcal{X}$,%
\begin{align*}%
{\displaystyle\sum_{\left(  k,h\right)  \in\mathcal{X}}}
\hat{M}\left(  \left(  k,h\right)  \mid\left(  j,i\right)  \right)   &  =%
{\displaystyle\sum_{\left(  k,h\right)  \in\mathcal{X}}}
\left(  \delta_{i}\left(  k\right)  \rho\left(  i\mid j\right)  +\delta
_{j}\left(  k\right)  \left(  1-\rho\left(  i\mid j\right)  \right)  \right)
Q\left(  h\mid k\right) \\
&  =%
{\displaystyle\sum_{\left(  k,h\right)  \in A^{2}}}
\left(  \delta_{i}\left(  k\right)  \rho\left(  i\mid j\right)  +\delta
_{j}\left(  k\right)  \left(  1-\rho\left(  i\mid j\right)  \right)  \right)
Q\left(  h\mid k\right) \\
&  =%
{\displaystyle\sum_{h\in A}}
\left(
{\displaystyle\sum_{k\in A}}
\left(  \delta_{i}\left(  k\right)  \rho\left(  i\mid j\right)  +\delta
_{j}\left(  k\right)  \left(  1-\rho\left(  i\mid j\right)  \right)  \right)
Q\left(  h\mid k\right)  \right)
\end{align*}
where equality in the second line follows from the fact that if $\left(
k,h\right)  \in A^{2}\setminus\mathcal{X}$ then $Q\left(  h\mid k\right)  =0$.
We will use this fact repeatedly. Now,

\begin{itemize}
\item if $i=j$, then%
\[%
{\displaystyle\sum_{k\in A}}
\left(  \delta_{i}\left(  k\right)  \rho\left(  i\mid j\right)  +\delta
_{j}\left(  k\right)  \left(  1-\rho\left(  i\mid j\right)  \right)  \right)
Q\left(  h\mid k\right)  =Q\left(  h\mid i\right)
\]
hence%
\[%
{\displaystyle\sum_{\left(  k,h\right)  \in\mathcal{X}}}
\hat{M}\left(  \left(  k,h\right)  \mid\left(  j,i\right)  \right)  =%
{\displaystyle\sum_{h\in A}}
Q\left(  h\mid i\right)  =1
\]

\item else
\[%
{\displaystyle\sum_{k\in A}}
\left(  \delta_{i}\left(  k\right)  \rho\left(  i\mid j\right)  +\delta
_{j}\left(  k\right)  \left(  1-\rho\left(  i\mid j\right)  \right)  \right)
Q\left(  h\mid k\right)  =\rho\left(  i\mid j\right)  Q\left(  h\mid i\right)
+\left(  1-\rho\left(  i\mid j\right)  \right)  Q\left(  h\mid j\right)
\]
hence%
\begin{align*}%
{\displaystyle\sum_{\left(  k,h\right)  \in\mathcal{X}}}
\hat{M}\left(  \left(  k,h\right)  \mid\left(  j,i\right)  \right)   &  =%
{\displaystyle\sum_{h\in A}}
\left(  \rho\left(  i\mid j\right)  Q\left(  h\mid i\right)  +\left(
1-\rho\left(  i\mid j\right)  \right)  Q\left(  h\mid j\right)  \right) \\
&  =\rho\left(  i\mid j\right)
{\displaystyle\sum_{h\in A}}
Q\left(  h\mid i\right)  +\left(  1-\rho\left(  i\mid j\right)  \right)
{\displaystyle\sum_{h\in A}}
Q\left(  h\mid j\right)  =1
\end{align*}

\end{itemize}

Next we show that if $\pi\in\Delta\left(  A\right)  $ and $M\pi=\pi$, then
setting
\[
\hat{\pi}\left(  j,i\right)  =Q\left(  i\mid j\right)  \pi\left(  j\right)
\qquad\forall\left(  j,i\right)  \in\mathcal{X}%
\]
defines an element of $\Delta\left(  \mathcal{X}\right)  $ such that $\hat
{M}\hat{\pi}=\hat{\pi}$. Clearly
\[%
{\displaystyle\sum_{\left(  j,i\right)  \in\mathcal{X}}}
\hat{\pi}\left(  j,i\right)  =%
{\displaystyle\sum_{\left(  j,i\right)  \in\mathcal{X}}}
Q\left(  i\mid j\right)  \pi\left(  j\right)  =%
{\displaystyle\sum_{\left(  j,i\right)  \in A^{2}}}
Q\left(  i\mid j\right)  \pi\left(  j\right)  =%
{\displaystyle\sum_{j\in A}}
\left(
{\displaystyle\sum_{i\in A}}
Q\left(  i\mid j\right)  \pi\left(  j\right)  \right)  =1
\]
Moreover, for all $\forall\left(  k,h\right)  \in\mathcal{X}$,%
\begin{align*}
\left(  \hat{M}\hat{\pi}\right)  _{\left(  k,h\right)  }  &  =%
{\displaystyle\sum_{\left(  j,i\right)  \in\mathcal{X}}}
\hat{M}\left(  \left(  k,h\right)  \mid\left(  j,i\right)  \right)  \hat{\pi
}\left(  j,i\right)  =\\
&  =%
{\displaystyle\sum_{\left(  j,i\right)  \in\mathcal{X}}}
\left(  \delta_{i}\left(  k\right)  \rho\left(  i\mid j\right)  +\delta
_{j}\left(  k\right)  \left(  1-\rho\left(  i\mid j\right)  \right)  \right)
Q\left(  h\mid k\right)  Q\left(  i\mid j\right)  \pi\left(  j\right) \\
&  =%
{\displaystyle\sum_{\left(  j,i\right)  \in A^{2}}}
\left(  \delta_{i}\left(  k\right)  \rho\left(  i\mid j\right)  +\delta
_{j}\left(  k\right)  \left(  1-\rho\left(  i\mid j\right)  \right)  \right)
Q\left(  h\mid k\right)  Q\left(  i\mid j\right)  \pi\left(  j\right) \\
&  =Q\left(  h\mid k\right)
{\displaystyle\sum_{\left(  j,i\right)  \in A^{2}}}
\left(  \delta_{i}\left(  k\right)  \rho\left(  i\mid j\right)  +\delta
_{j}\left(  k\right)  \left(  1-\rho\left(  i\mid j\right)  \right)  \right)
Q\left(  i\mid j\right)  \pi\left(  j\right)
\end{align*}
Next we show that, for all $k\in A$,%
\[%
{\displaystyle\sum_{\left(  j,i\right)  \in A^{2}}}
\left(  \delta_{i}\left(  k\right)  \rho\left(  i\mid j\right)  +\delta
_{j}\left(  k\right)  \left(  1-\rho\left(  i\mid j\right)  \right)  \right)
Q\left(  i\mid j\right)  \pi\left(  j\right)  =\pi\left(  k\right)
\]
obtaining $\left(  \hat{M}\hat{\pi}\right)  _{\left(  k,h\right)  }=Q\left(
h\mid k\right)  \pi\left(  k\right)  =\hat{\pi}_{\left(  k,h\right)  }$.
Indeed%
\begin{align*}
&
{\displaystyle\sum_{\left(  j,i\right)  \in A^{2}}}
\left(  \delta_{i}\left(  k\right)  \rho\left(  i\mid j\right)  +\delta
_{j}\left(  k\right)  \left(  1-\rho\left(  i\mid j\right)  \right)  \right)
Q\left(  i\mid j\right)  \pi\left(  j\right) \\
&  =%
{\displaystyle\sum_{\left(  j,i\right)  \in A^{2}}}
\delta_{i}\left(  k\right)  \rho\left(  i\mid j\right)  Q\left(  i\mid
j\right)  \pi\left(  j\right)  +%
{\displaystyle\sum_{\left(  j,i\right)  \in A^{2}}}
\delta_{j}\left(  k\right)  Q\left(  i\mid j\right)  \pi\left(  j\right)  -%
{\displaystyle\sum_{\left(  j,i\right)  \in A^{2}}}
\delta_{j}\left(  k\right)  \rho\left(  i\mid j\right)  Q\left(  i\mid
j\right)  \pi\left(  j\right) \\
&  =%
{\displaystyle\sum_{j\in A}}
\rho\left(  k\mid j\right)  Q\left(  k\mid j\right)  \pi\left(  j\right)  +%
{\displaystyle\sum_{i\in A}}
Q\left(  i\mid k\right)  \pi\left(  k\right)  -%
{\displaystyle\sum_{i\in A}}
\rho\left(  i\mid k\right)  Q\left(  i\mid k\right)  \pi\left(  k\right)
\end{align*}
now the central summand $%
{\displaystyle\sum_{i\in A}}
Q\left(  i\mid k\right)  \pi\left(  k\right)  $ is $\pi\left(  k\right)  $, we
conclude by showing that $M\pi=\pi$ implies that
\[%
{\displaystyle\sum_{j\in A}}
\rho\left(  k\mid j\right)  Q\left(  k\mid j\right)  \pi\left(  j\right)  =%
{\displaystyle\sum_{i\in A}}
\rho\left(  i\mid k\right)  Q\left(  i\mid k\right)  \pi\left(  k\right)
\]
in fact%
\begin{align*}%
{\displaystyle\sum_{j\in A}}
\rho\left(  k\mid j\right)  Q\left(  k\mid j\right)  \pi\left(  j\right)   &
=%
{\displaystyle\sum_{j\neq k}}
\rho\left(  k\mid j\right)  Q\left(  k\mid j\right)  \pi\left(  j\right)
+\rho\left(  k\mid k\right)  Q\left(  k\mid k\right)  \pi\left(  k\right) \\
&  =%
{\displaystyle\sum_{j\neq k}}
M\left(  k\mid j\right)  \pi\left(  j\right)  +\rho\left(  k\mid k\right)
Q\left(  k\mid k\right)  \pi\left(  k\right) \\
&  =-M\left(  k\mid k\right)  \pi\left(  k\right)  +%
{\displaystyle\sum_{j\in A}}
M\left(  k\mid j\right)  \pi\left(  j\right)  +\rho\left(  k\mid k\right)
Q\left(  k\mid k\right)  \pi\left(  k\right) \\
&  =-M\left(  k\mid k\right)  \pi\left(  k\right)  +\pi\left(  k\right)
+\rho\left(  k\mid k\right)  Q\left(  k\mid k\right)  \pi\left(  k\right) \\
&  =\left(  1-M\left(  k\mid k\right)  \right)  \pi\left(  k\right)
+\rho\left(  k\mid k\right)  Q\left(  k\mid k\right)  \pi\left(  k\right) \\
&  =\left(
{\displaystyle\sum_{i\neq k}}
\rho\left(  i\mid k\right)  Q\left(  i\mid k\right)  \right)  \pi\left(
k\right)  +\rho\left(  k\mid k\right)  Q\left(  k\mid k\right)  \pi\left(
k\right) \\
&  =%
{\displaystyle\sum_{i\in A}}
\rho\left(  i\mid k\right)  Q\left(  i\mid k\right)  \pi\left(  k\right)
\end{align*}

So far, we did not use the fact that $M$ is primitive, this is used next to
show that $\hat{M}$ is primitive too. By irreducibility of $M$, for all
$i,k\in A$, there exists of a finite sequence
\begin{equation}
i=i_{0},i_{1},i_{2},...,i_{n}=k\label{eq:basechain}%
\end{equation}
satisfying $i_{a+1}\neq i_{a}$ and $M\left(  i_{a+1}\mid i_{a}\right)  >0$,
for all $a=0,...,n-1$, by the definition of $M$, it follows that%
\begin{equation}
Q\left(  i_{a+1}\mid i_{a}\right)  >0\label{eq:chain}%
\end{equation}
for all $a=0,...,n-1$.

Consider any $\left(  j,i\right)  ,\left(  k,h\right)  \in\mathcal{X}$, and a
chain $i_{0},i_{1},i_{2},...,i_{n}$ satisfying (\ref{eq:basechain}) and
(\ref{eq:chain}). The derived chain%
\[
\left(  j,i\right)  =\left(  j,i_{0}\right)  ,\left(  i_{0},i_{1}\right)
,\left(  i_{1},i_{2}\right)  ...,\left(  i_{n-1},i_{n}\right)  ,\left(
i_{n},h\right)  =\left(  k,h\right)
\]
belongs to $\mathcal{X}$ because $\left(  j,i_{0}\right)  =\left(  j,i\right)
\in\mathcal{X}$, $\left(  i_{n},h\right)  =\left(  k,h\right)  \in\mathcal{X}%
$, and also $\left(  i_{a},i_{a+1}\right)  \in\mathcal{X}$ because of
(\ref{eq:chain}). Now
\[
\hat{M}\left(  \left(  i_{0},i_{1}\right)  \mid\left(  j,i_{0}\right)
\right)  =\left(  \delta_{i_{0}}\left(  i_{0}\right)  \rho\left(  i_{0}\mid
j\right)  +\delta_{j}\left(  i_{0}\right)  \left(  1-\rho\left(  i_{0}\mid
j\right)  \right)  \right)  Q\left(  i_{1}\mid i_{0}\right)
\]
which is strictly positive because, $\rho\left(  i_{0}\mid j\right)
=\rho\left(  i\mid j\right)  >0$ (since $\rho$ is positive) and $Q\left(
i_{1}\mid i_{0}\right)  >0$. Moreover,%
\[
\hat{M}\left(  \left(  i_{a+1},i_{a+2}\right)  \mid\left(  i_{a}%
,i_{a+1}\right)  \right)  =\left(  \delta_{i_{a+1}}\left(  i_{a+1}\right)
\rho\left(  i_{a+1}\mid i_{a}\right)  +\delta_{i_{a}}\left(  i_{a+1}\right)
\left(  1-\rho\left(  i_{a+1}\mid i_{a}\right)  \right)  \right)  Q\left(
i_{a+2}\mid i_{a+1}\right)
\]
which is strictly positive for $a=0,...,n-2$, because $\rho\left(  i_{a+1}\mid
i_{a}\right)  >0$ and $Q\left(  i_{a+2}\mid i_{a+1}\right)  >0$. Finally,%
\[
\hat{M}\left(  \left(  i_{n},h\right)  \mid\left(  i_{n-1},i_{n}\right)
\right)  =\left(  \delta_{i_{n}}\left(  i_{n}\right)  \rho\left(  i_{n}\mid
i_{n-1}\right)  +\delta_{i_{n-1}}\left(  i_{n}\right)  \left(  1-\rho\left(
i_{n}\mid i_{n-1}\right)  \right)  \right)  Q\left(  h\mid i_{n}\right)
\]
which is strictly positive, because $\rho\left(  i_{n}\mid i_{n-1}\right)  >0$
and $Q\left(  h\mid i_{n}\right)  =Q\left(  h\mid k\right)  >0$.

This shows irreducibility of $\hat{M}$. Having proved irreducibility,
primitivity can be established by exhibiting a non-zero diagonal element in
the transition matrix $\hat{M}$. By definition, for all $\left(  j,i\right)
\in\mathcal{X}$%
\[
\hat{M}\left(  \left(  j,i\right)  \mid\left(  j,i\right)  \right)  =\left(
\delta_{i}\left(  j\right)  \rho\left(  i\mid j\right)  +\delta_{j}\left(
j\right)  \left(  1-\rho\left(  i\mid j\right)  \right)  \right)  Q\left(
i\mid j\right)
\]
Positivity of $\rho$ guarantees that $\delta_{j}\left(  j\right)  \left(
1-\rho\left(  i\mid j\right)  \right)  >0$, while $Q\left(  i\mid j\right)
>0$ by definition of $\mathcal{X}$.

All the assumptions of Howard (1971) p. 713 are then satisfied by the
semi-Markov chain with embedded Markov chain $\hat{M}\left(  \left(
k,h\right)  \mid\left(  j,i\right)  \right)  $ and holding times $\hat
{T}\left(  \left(  k,h\right)  \mid\left(  j,i\right)  \right)  =\mathrm{RT}%
_{i,j}$.\footnote{Observe that holding times are independent of the
\textquotedblleft next state\textquotedblright\ $\left(  k,h\right)  $
therefore \textquotedblleft average waiting times\textquotedblright\ are just
average holding times (see Howard, 1971, p. 691).} Hence the limit as
$t\rightarrow\infty$ of the probability $\phi_{\left(  j,i\right)  }\left(
t\right)  $ with which comparison $\left(  j,i\right)  $ is taking place at
time $t$ is given by%
\begin{align*}
\phi_{\left(  j,i\right)  }  &  =\frac{\hat{\pi}\left(  j,i\right)
\tau_{\mathrm{RT}}\left(  i\mid j\right)  }{%
{\displaystyle\sum_{\left(  k,h\right)  \in\mathcal{X}}}
\hat{\pi}\left(  k,h\right)  \tau_{\mathrm{RT}}\left(  h\mid k\right)  }%
=\frac{Q\left(  i\mid j\right)  \pi\left(  j\right)  \tau_{\mathrm{RT}}\left(
i\mid j\right)  }{%
{\displaystyle\sum_{\left(  k,h\right)  \in\mathcal{X}}}
Q\left(  h\mid k\right)  \pi\left(  k\right)  \tau_{\mathrm{RT}}\left(  h\mid
k\right)  }\\
&  =\frac{Q\left(  i\mid j\right)  \pi\left(  j\right)  \tau_{\mathrm{RT}%
}\left(  i\mid j\right)  }{%
{\displaystyle\sum_{\left(  k,h\right)  \in A^{2}}}
Q\left(  h\mid k\right)  \pi\left(  k\right)  \tau_{\mathrm{RT}}\left(  h\mid
k\right)  }\qquad\forall\left(  j,i\right)  \in\mathcal{X}%
\end{align*}
The same is true if $\left(  j,i\right)  \notin\mathcal{X}$, because in that
case $\phi_{\left(  j,i\right)  }\left(  t\right)  =0$ for all $t$ and
$Q\left(  i\mid j\right)  =0$. Thus%
\begin{align*}
\lim_{t\rightarrow\infty}p_{N_{t}}\left(  j\right)   &  =\lim_{t\rightarrow
\infty}%
{\displaystyle\sum_{i\in A}}
\phi_{\left(  j,i\right)  }\left(  t\right)  =%
{\displaystyle\sum_{i\in A}}
\lim_{t\rightarrow\infty}\phi_{\left(  j,i\right)  }\left(  t\right)  =%
{\displaystyle\sum_{i\in A}}
\phi_{\left(  j,i\right)  }\\
&  =\frac{%
{\displaystyle\sum_{i\in A}}
Q\left(  i\mid j\right)  \pi\left(  j\right)  \tau_{\mathrm{RT}}\left(  i\mid
j\right)  }{%
{\displaystyle\sum_{\left(  k,h\right)  \in A^{2}}}
Q\left(  h\mid k\right)  \pi\left(  k\right)  \tau_{\mathrm{RT}}\left(  h\mid
k\right)  }=\frac{\pi\left(  j\right)
{\displaystyle\sum_{i\in A}}
Q\left(  i\mid j\right)  \tau_{\mathrm{RT}}\left(  i\mid j\right)  }{%
{\displaystyle\sum_{k\in A}}
\pi\left(  k\right)
{\displaystyle\sum_{h\in A}}
Q\left(  h\mid k\right)  \tau_{\mathrm{RT}}\left(  h\mid k\right)  }\\
&  =\frac{\pi\left(  j\right)  \tau_{j}}{%
{\displaystyle\sum_{k\in A}}
\pi\left(  k\right)  \tau_{k}}%
\end{align*}
as desired.\hfill$\blacksquare$

\bigskip

Consider, like in Cerreia-Vioglio et al. (2022, Section 2), a symmetric DDM
and an exploration matrix with off diagonal entries that are inversely
proportional to mean response times. This means that there exists $w>0$ such
that%
\[
Q\left(  i\mid j\right)  =\frac{w}{\tau_{\mathrm{RT}}\left(  i\mid j\right)
}\text{\qquad}\forall i\neq j
\]
so that
\[
Q\left(  j\mid j\right)  =1-%
{\displaystyle\sum\limits_{i\neq j}}
\frac{w}{\tau_{\mathrm{RT}}\left(  i\mid j\right)  }\text{\qquad}\forall j
\]
Also assume that $\tau_{\mathrm{RT}}\left(  j\mid j\right)  <\delta$ (small)
for all $j\in A$.\footnote{The Drift Diffusion Model makes no predictions on
the response time for comparisons of an alternative with itself. But, it makes
sense to assume that the decision unit takes almost no time in realizing that
no actual comparison needs to be made.} With this, for all $j\in A$,%
\begin{align*}
\bar{\tau}_{j}  &  =%
{\displaystyle\sum\limits_{i\in A}}
Q\left(  i\mid j\right)  \tau_{\mathrm{RT}}\left(  i\mid j\right) \\
&  =%
{\displaystyle\sum\limits_{i\neq j}}
\frac{w}{\tau_{\mathrm{RT}}\left(  i\mid j\right)  }\tau_{\mathrm{RT}}\left(
i\mid j\right)  +\left(  1-%
{\displaystyle\sum\limits_{i\neq j}}
\frac{w}{\tau_{\mathrm{RT}}\left(  i\mid j\right)  }\right)  \tau
_{\mathrm{RT}}\left(  j\mid j\right)  =\left(  \left\vert A\right\vert
-1\right)  w+\delta_{j}%
\end{align*}
with $0\leq\delta_{j}<\delta$. Since $\delta$ is small, then%
\[
\bar{\tau}_{j}\approx\left(  \left\vert A\right\vert -1\right)  w
\]
irrespective of $j$, so that $\alpha\left(  j\right)  $ is approximately
constant in (\ref{eq:alpha}) and
\[
\lim_{t\rightarrow\infty}p_{N_{t}}\left(  j\right)  \approx\frac{e^{\lambda
v\left(  j\right)  }}{%
{\displaystyle\sum\limits_{k\in A}}
e^{\lambda v\left(  k\right)  }}\qquad\forall j\in A
\]
that is, the limit probability is approximately of the multinomial logit type.

\newpage

\begin{Hidden}
\noindent\textsc{Powers of a matrix}\medskip

\noindent$N=\left\{  1,2,...,n\right\}  $, $T$ an $n\times n$ square matrix,
$T^{0}=I$, $T^{m}=TT^{m-1}$ for all $m\geq1$. Denoting by $t_{ij}^{\left(
m\right)  }$ the $\left(  i,j\right)  $-th element of $T^{m}$, for $m=2$ we
have%
\[
t_{ij}^{\left(  2\right)  }=\sum\limits_{k=1}^{n}t_{ik}t_{kj}=\sum
\limits_{k_{1}\in N^{2-1}}t_{ik_{1}}t_{k_{1}j}\qquad\forall i,j\in N
\]

\begin{proposition}
By induction if $m\geq3$,%
\[
t_{ij}^{\left(  m\right)  }=\sum\limits_{\left(  k_{1},k_{2},...,k_{m-1}%
\right)  \in N^{m-1}}t_{ik_{1}}t_{k_{1}k_{2}}\cdot\cdot\cdot t_{k_{m-1}%
j}\qquad\forall i,j\in N
\]

\end{proposition}

\noindent\textbf{Proof} The result is true for $m=3$, in fact,%
\begin{align*}
t_{ij}^{\left(  3\right)  }  &  =\sum\limits_{h=1}^{n}t_{ih}t_{hj}^{\left(
2\right)  }=\sum\limits_{h\in N}t_{ih}\sum\limits_{k\in N}t_{hk}t_{kj}%
=\sum\limits_{h\in N}\sum\limits_{k\in N}t_{ih}t_{hk}t_{kj}\\
&  =\sum\limits_{\left(  h,k\right)  \in N^{2}}t_{ih}t_{hk}t_{kj}%
=\sum\limits_{\left(  k_{1},k_{2}\right)  \in N^{2}}t_{ik_{1}}t_{k_{1}k_{2}%
}t_{k_{2}j}%
\end{align*}
Let $m>3$ and assume true for $m-1$, then%
\begin{align*}
t_{ij}^{\left(  m\right)  }  &  =\sum\limits_{h=1}^{n}t_{ih}t_{hj}^{\left(
m-1\right)  }=\sum\limits_{h=1}^{n}t_{ih}\sum\limits_{\left(  k_{1}%
,k_{2},...,k_{m-2}\right)  \in N^{m-2}}t_{hk_{1}}t_{k_{1}k_{2}}\cdot\cdot\cdot
t_{k_{m-2}j}\\
&  =\sum\limits_{h\in N}t_{ih}\sum\limits_{\left(  h_{2},...,h_{m-1}\right)
\in N^{m-2}}t_{hh_{2}}t_{h_{2}h_{3}}\cdot\cdot\cdot t_{h_{m-1}j}%
=\sum\limits_{h\in N}\sum\limits_{\left(  h_{2},...,h_{m-1}\right)  \in
N^{m-2}}t_{ih_{1}}t_{h_{1}h_{2}}\cdot\cdot\cdot t_{h_{m-1}j}\\
&  =\sum\limits_{\left(  h_{1},h_{2},...,h_{m-1}\right)  \in N^{m-1}}%
t_{ih_{1}}t_{h_{1}h_{2}}\cdot\cdot\cdot t_{h_{m-1}j}%
\end{align*}
as wanted.\hfill$\blacksquare$
\end{Hidden}

\bigskip

\begin{Hidden}
\noindent\textsc{Quasipositivity and primitivity}\medskip

\noindent Let%
\[
T=\left[
\begin{array}
[c]{cc}%
0 & 1\\
1 & 0
\end{array}
\right]
\]
then%
\[
T^{2}=\left[
\begin{array}
[c]{cc}%
0 & 1\\
1 & 0
\end{array}
\right]  \left[
\begin{array}
[c]{cc}%
0 & 1\\
1 & 0
\end{array}
\right]  =\left[
\begin{array}
[c]{cc}%
1 & 0\\
0 & 1
\end{array}
\right]  =I
\]
Therefore $T^{2m}=I$ and $T^{2m+1}=T$ for all $m\geq0$. $T$ is quasi-positive
but not primitive.

\begin{proposition}
If $n\geq3$, then any quasi-positive matrix $T$ is primitive, indeed $T^{2}$
is positive.
\end{proposition}

\noindent\textbf{Proof} For all $i,j\in N$,%
\[
t_{ij}^{\left(  2\right)  }=\sum\limits_{k\in N}t_{ik}t_{kj}%
\]
where, by quasipositivity, all summands are non-negative and $t_{ik}t_{kj}>0$
for all $k\notin\left\{  i,j\right\}  $.\hfill$\blacksquare$
\end{Hidden}

\bigskip

\begin{Hidden}
\noindent\textsc{Reversible stochastic matrices are diagonalizable}\medskip

\begin{fact}
If $M$ is any square matrix and $D=\mathrm{diag}\left\{  d_{i}%
:i=1,...,n\right\}  $, then%
\begin{align*}
\left(  DM\right)  _{ij}  &  =d_{i}m_{ij}\\
\left(  MD\right)  _{ij}  &  =m_{ij}d_{j}%
\end{align*}
and in particular, if $d_{i}\neq0$ for all $i\in N$, $D^{-1}=\mathrm{diag}%
\left\{  d_{i}^{-1}:i=1,...,n\right\}  $.
\end{fact}

Now let $B$ be reversible, then
\[
b_{ij}p_{j}=b_{ji}p_{i}%
\]
and since $p>0$, then
\[
b_{ij}=p_{i}b_{ji}p_{j}^{-1}%
\]
Consider the diagonal matrix $D=\mathrm{diag}\left\{  p_{i}^{1/2}%
:i=1,...,n\right\}  $, so that $D^{-1}=\mathrm{diag}\left\{  p_{i}%
^{-1/2}:i=1,...,n\right\}  $. Then%
\[
\left(  D^{-1}B\right)  _{i,j}=p_{i}^{-1/2}b_{ij}%
\]
and
\[
\left(  D^{-1}BD\right)  _{i,j}=p_{i}^{-1/2}b_{ij}p_{j}^{1/2}%
\]
by reversibility%
\[
\left(  D^{-1}BD\right)  _{i,j}=p_{i}^{-1/2}p_{i}b_{ji}p_{j}^{-1}p_{j}%
^{1/2}=p_{i}^{1/2}b_{ji}p_{j}^{-1/2}=p_{j}^{-1/2}b_{ji}p_{i}^{1/2}=\left(
D^{-1}BD\right)  _{j,i}%
\]
Therefore, a reversible matrix is similar to a symmetric matrix which is
similar to a diagonal matrix. But similarity is an equivalence relation.
\end{Hidden}

\bigskip

\begin{Hidden}
\noindent\textsc{Transitivities}\medskip

An unbiased, $0$-$1$ valued stochastic choice kernel defines a strict
preference%
\[
j\succ_{\rho}i\iff\rho\left(  j\mid i\right)  =1
\]
note that, by unbiasedness,
\[
j\succ_{\rho}i\iff1-\rho\left(  i\mid j\right)  =1\iff\rho\left(  i\mid
j\right)  =0
\]
thus, $\succ_{\rho}$ is \emph{total}: for all $i,j\in A$ either $i\succ j$ or
$i=j$ or $j\succ i$. Moreover, $j\succ_{\rho}i$ if and only if $i\nsucc_{\rho
}j$.

Next we show that $\rho$ is transitive if and only if $\succ_{\rho}$ is.

If $\rho$ is transitive, and $\succ_{\rho}$ is not,$\ $then there exists $i$,
$j$, and $k$ such that $i\succ_{\rho}k\succ_{\rho}j$, but $i\nsucc_{\rho}j$.
In this case, $i$, $k$, and $j$ must be distinct and $j\succ_{\rho}i$.
Therefore, $i\succ_{\rho}k\succ_{\rho}j\succ_{\rho}i$ implying $\rho\left(
i\mid k\right)  =\rho\left(  k\mid j\right)  =\rho\left(  j\mid i\right)  =1$
and $\rho\left(  k\mid i\right)  =\rho\left(  j\mid k\right)  =\rho\left(
i\mid j\right)  =0$ contradicting (\ref{eq:trans-ddm}); thus must be
$\succ_{\rho}$ transitive.

Conversely, if $\rho$ is not transitive, then there are three distinct
alternatives $i$, $j$ and $k$ for which (\ref{eq:trans-ddm}) does not hold;
that is,
\[
\rho\left(  j\mid i\right)  \rho\left(  k\mid j\right)  \rho\left(  i\mid
k\right)  \neq\rho\left(  k\mid i\right)  \rho\left(  j\mid k\right)
\rho\left(  i\mid j\right)
\]
It is then impossible that both sides contain a zero factor. Since $\rho$ is
$0$-$1$ valued, then:

\begin{itemize}
\item either $\rho\left(  j\mid i\right)  \rho\left(  k\mid j\right)
\rho\left(  i\mid k\right)  =1$, then $i\succ_{\rho}k\succ_{\rho}j\succ_{\rho
}i$, so that $i\succ_{\rho}k\succ_{\rho}j$ and $i\nsucc_{\rho}j$, thus
$\succ_{\rho}$ is not transitive;

\item or $\rho\left(  k\mid i\right)  \rho\left(  j\mid k\right)  \rho\left(
i\mid j\right)  =1$, then $i\succ_{\rho}j\succ_{\rho}k\succ_{\rho}i$, so that
$i\succ_{\rho}j\succ_{\rho}k$ and $i\nsucc_{\rho}k$, thus $\succ_{\rho}$ is
not transitive.
\end{itemize}
\end{Hidden}


\newpage

\begin{thebibliography}{99}                                                                                               %


\bibitem {}S. P. Anderson, J. K. Goeree and C. A. Holt, Noisy directional
learning and the logit equilibrium, \emph{Scandinavian Journal of Economics},
106, 581-602, 2004.

\bibitem {}C. Baldassi, S. Cerreia-Vioglio, F. Maccheroni, M. Marinacci and M.
Pirazzini, A behavioral characterization of the drift diffusion model and its
multialternative extension for choice under time pressure, \emph{Management
Science}, 66, 5075-5093, 2020.

\bibitem {bm}H. D. Block and J. Marschak, Random orderings and stochastic
theories of response, in \emph{Contributions to probability and statistics}
(I. Olkin et al., eds), Stanford University Press, Stanford, 1960.

\bibitem {}R. Bogacz, E. Brown, J. Moehlis, P. Holmes and J. D. Cohen, The
physics of optimal decision making: a formal analysis of models of performance
in two-alternative forced-choice tasks, \emph{Psychological Review}, 113,
700-765, 2006.

\bibitem {}R. Bogacz, M. Usher, J. Zhang and J. L. McClelland, Extending a
biologically inspired model of choice: multi-alternatives, nonlinearity and
value-based multidimensional choice, \emph{Philosophical Transactions of the
Royal Society of London B: Biological Sciences}, 362, 1655-1670, 2007.

\bibitem {}S. Cerreia-Vioglio, F. Maccheroni, M. Marinacci and A. Rustichini,
Multinomial logit processes and preference discovery: inside and outside the
black box, \emph{Review of Economic Studies}, forthcoming.

\bibitem {}J. Ditterich, A comparison between mechanisms of multi-alternative
perceptual decision making: ability to explain human behavior, predictions for
neurophysiology, and relationship with decision theory, \emph{Frontiers in
Neuroscience}, 4, 2010.

\bibitem {feller}W. Feller, \emph{An introduction to probability theory and
its applications}, v. 1, Wiley, New York, 1968.

\bibitem {}P. C. Fishburn, \emph{Utility theory for decision making}, Wiley,
New York, 1970.

\bibitem {}N. J. Higham, \emph{Functions of matrices: theory and computation},
Society for Industrial and Applied Mathematics (SIAM), Philadelphia, 2008.

\bibitem {howard}R. A. Howard. \emph{Dynamic probabilistic systems}, v. 2,
Wiley, New York, 1971.

\bibitem {}N. L. Johnson,\ A.\ W.\ Kemp and S. Kotz,\ \emph{Univariate
discrete distributions}, Wiley, New York, 2005.

\bibitem {}F. P. Kelly, \emph{Reversibility and stochastic networks}, Wiley,
New York, 1979.

\bibitem {}A. N. Kolmogorov, Zur theorie der Markoffschen ketten,
\emph{Mathematische Annalen}, 112, 155-160, 1936 (trans. in \emph{Selected
Works of A. N. Kolmogorov}, Springer, 1992).

\bibitem {Krajbich et al. 2010}I. Krajbich, C. Armel and A. Rangel, Visual
fixations and the computation and comparison of value in simple choice,
\emph{Nature Neuroscience} 13, 1292-1298, 2010

\bibitem {}I. Krajbich and A. Rangel, Multialternative drift-diffusion model
predicts the relationship between visual fixations and choice in value-based
decisions, \emph{Proceedings of the National Academy of Sciences}, 108,
13852-13857, 2011.

\bibitem {}D. M. Kreps, \emph{Notes on the theory of choice}, Westview,
Boulder, 1988.

\bibitem {}R. D. Luce, \emph{Individual choice behavior: a theoretical
analysis}, Wiley, New York, 1959.

\bibitem {}R. D. Luce and P. Suppes, Preference, utility and subjective
probability, in \emph{Handbook of Mathematical Psychology }(R. D. Luce, R. R.
Bush and E. Galanter, eds.), v. 3, Wiley, New York, 1965.

\bibitem {mars}J. Marschak, Binary-choice constraints and random utility
indicators, in \emph{Mathematical Methods in the Social Sciences} (K. J.
Arrow, S. Karlin and P. Suppes, eds.), Stanford University Press, Stanford, 1960.

\bibitem {}T. McMillen and P. Holmes, The dynamics of choice among multiple
alternatives, \emph{Journal of Mathematical Psychology}, 50, 30-57, 2006.

\bibitem {}N. Metropolis, A. W. Rosenbluth, M. N. Rosenbluth, A. H. Teller and
E. Teller, Equation of state calculations by fast computing machines,
\emph{Journal of Chemical Physics}, 21, 1087-1092, 1953.

\bibitem {}M. Milosavljevic, J. Malmaud, A. Huth, C. Koch and A. Rangel, The
drift diffusion model can account for the accuracy and reaction time of
value-based choices under high and low time pressure, \emph{Judegment and
Decision Making}, 5, 437-449, 2010.

\bibitem {karlin}M. Pinsky and S. Karlin, \emph{An introduction to stochastic
modeling}, 4$^{\text{th}}$ ed., Academic Press, Burlington, 2011.

\bibitem {}R. Ratcliff, A theory of memory retrieval, \emph{Psychological
Review}, 85, 59-108, 1978.

\bibitem {}E. Reutskaja, R. Nagel, C. F. Camerer and A. Rangel, Search
dynamics in consumer choice under time pressure: an eye-tracking study,
\emph{American Economic Review}, 101, 900-926, 2011.

\bibitem {}R. F. Rinehart, The equivalence of definitions of a matric
function, \emph{American Mathematical Monthly}, 62, 395-414, 1955.

\bibitem {}R. M. Roe, J. R. Busemeyer and J. T. Townsend, Multialternative
decision field theory: a dynamic connectionst model of decision making,
\emph{Psychological Review}, 108, 370, 2001.

\bibitem {}J. E. Russo and L. D. Rosen, An eye fixation analysis of
multialternative choice, \emph{Memory and Cognition}, 3, 267-276, 1975.

\bibitem {}G. M. Stine, A. Zylberberg, J. Ditterich and M. N. Shadlen,
Differentiating between integration and non-integration strategies in
perceptual decision making, \emph{eLife}, 9, e55365, 2020.

\bibitem {}K. Valkanova, Markov stochastic choice, mimeo, 2020.
\end{thebibliography}
\end{document}